\documentclass[12pt]{article}%
\pdfoutput=1
\usepackage[nosort]{cite}
\usepackage{graphicx}
\usepackage{multicol}
\usepackage{amsfonts}
\usepackage{amssymb}
\usepackage{amsmath}
\usepackage{dsfont}
\usepackage{heck}
\usepackage{afterpage}
\usepackage{setspace}
\usepackage{verbatim}
\usepackage{color}
\usepackage{longtable}
\usepackage{float}
\usepackage{subcaption}
\usepackage{epsfig}
\usepackage{epstopdf}
\usepackage{adjustbox}
\usepackage{todonotes}

\usepackage{tikz}
\usepackage[margin=1in]{geometry}
\usepackage{titletoc}
\usepackage{hyperref}%
\usepackage{mathdots}
\providecommand{\U}[1]{\protect\rule{.1in}{.1in}}
\pdfoutput=1
\newsavebox{\mysavebox}

\usetikzlibrary[shapes.geometric]
\usetikzlibrary{positioning}
\usetikzlibrary{calc,intersections,through,backgrounds}
\tikzset{>=stealth}
\usetikzlibrary{decorations.pathmorphing}
\usetikzlibrary{decorations.markings}
\tikzset{>=stealth}
\hypersetup{colorlinks,citecolor=black,filecolor=black,linkcolor=black,urlcolor=black,pdftex}
\usetikzlibrary{decorations.markings}

\numberwithin{equation}{section}

\newcommand{\ba}{\begin{eqnarray}}
\newcommand{\ea}{\end{eqnarray}}

\newcommand{\Tr}{\, {\rm Tr}}

\newcommand{\be}{\begin{equation}}
\newcommand{\ee}{\end{equation}}

\tikzstyle{startstop} = [rectangle, rounded corners, minimum width=3cm, minimum height=1cm,text centered, draw=black, fill=blue!10]
\tikzstyle{startstop} = [rectangle, rounded corners, minimum width=3cm, minimum height=1cm,text centered, draw=black, fill=blue!10]
\tikzstyle{io} = [trapezium, trapezium left angle=70, trapezium right angle=110, minimum width=3cm, minimum height=1cm, text centered, draw=black, fill=blue!30]
\tikzstyle{process} = [rectangle, minimum width=3cm, minimum height=1cm, text centered, draw=black, fill=orange!30]
\tikzstyle{decision} = [diamond, minimum width=3cm, minimum height=1cm, text centered, draw=black, fill=green!30]
\tikzstyle{arrow} = [thick,->,>=stealth]
\begin{document}

\date{March 2018}

\title{4D Gauge Theories with Conformal Matter}

\institution{PENN}{\centerline{${}^{1}$Department of Physics and Astronomy, University of Pennsylvania, Philadelphia, PA 19104, USA}}

\institution{UNC}{\centerline{${}^{2}$Department of Physics, University of North Carolina, Chapel Hill, NC 27599, USA}}

\institution{UCSBmath}{\centerline{${}^{3}$Department of Mathematics, University of California Santa Barbara, CA 93106, USA}}

\institution{UCSBphys}{\centerline{${}^{4}$Department of Physics, University of California Santa Barbara, CA 93106, USA}}

\institution{Uppsala}{\centerline{${}^{5}$Department of Physics and Astronomy, Uppsala University, Box 516, SE-75120 Uppsala, Sweden}}

\authors{Fabio Apruzzi\worksat{\PENN, \UNC}\footnote{e-mail: {\tt fabio.apruzzi@unc.edu}},
Jonathan J. Heckman\worksat{\PENN}\footnote{e-mail: {\tt jheckman@sas.upenn.edu}},\\[4mm]
David R. Morrison\worksat{\UCSBmath, \UCSBphys}\footnote{e-mail: {\tt drm@physics.ucsb.edu}}, and
Luigi Tizzano\worksat{\Uppsala}\footnote{e-mail: {\tt luigi.tizzano@physics.uu.se}}}

\abstract{One of the hallmarks of 6D superconformal field theories (SCFTs) is that on
a partial tensor branch, all known theories resemble quiver gauge
theories with links comprised of 6D conformal matter, a generalization of weakly coupled
hypermultiplets. In this paper we construct 4D quiverlike gauge theories in which the
links are obtained from compactifications of 6D conformal matter on Riemann surfaces with
flavor symmetry fluxes. This includes generalizations of super QCD with
exceptional gauge groups and quarks replaced by 4D conformal matter.
Just as in super QCD, we find evidence for a conformal window as well as confining gauge group
factors depending on the total amount of matter. We also present F-theory realizations of
these field theories via elliptically fibered Calabi-Yau fourfolds. Gauge groups (and flavor symmetries)
come from 7-branes wrapped on surfaces, conformal matter localizes at the intersection of pairs of 7-branes,
and Yukawas between 4D conformal matter localize at points coming from triple intersections of 7-branes.
Quantum corrections can also modify the classical moduli space of the F-theory model,
matching expectations from effective field theory.}

\maketitle

\setcounter{tocdepth}{2}

\tableofcontents


\newpage

\section{Introduction}

In addition to providing the only consistent theory of quantum gravity, string
theory also predicts the existence of qualitatively new quantum field
theories. A notable example of this kind is the construction of conformal field
theories in six spacetime dimensions. These quantum field theories are
intrinsically strongly coupled, and the only known way to generate examples
is via compactifications of string theory backgrounds with six flat spacetime
dimensions \cite{Witten:1995ex, Witten:1995zh,
Strominger:1995ac, Seiberg:1996qx} (for a partial list of references,
see e.g. \cite{WittenSmall,
Ganor:1996mu,Morrison:1996pp,Seiberg:1996vs, Bershadsky:1996nu,
Brunner:1997gf, Blum:1997fw, Aspinwall:1997ye, Intriligator:1997dh, Hanany:1997gh} as
well as \cite{Heckman:2013pva, Gaiotto:2014lca, DelZotto:2014hpa,
DelZotto:2014fia, Heckman:2014qba, Bhardwaj:2015xxa, Chang:2017xmr,
Apruzzi:2013yva, Heckman:2015bfa, DelZotto:2015isa, Gaiotto:2015usa,
Ohmori:2015pua, Franco:2015jna, DelZotto:2015rca, Heckman:2015ola,
Louis:2015mka, Cordova:2015fha, Hanany:2015pfa, Aganagic:2015cta, Ohmori:2015pia,
Coman:2015bqq, Cremonesi:2015bld, Heckman:2016ssk, Cordova:2016xhm, Morrison:2016nrt,
Heckman:2016xdl, Cordova:2016emh, Kim:2016foj, Shimizu:2016lbw, Mekareeya:2016yal,
DelZotto:2016pvm, Apruzzi:2016nfr, Razamat:2016dpl, Bah:2017gph, Bah:2017wxp,
DelZotto:2017pti, Apruzzi:2017iqe, Heckman:2017uxe, Kim:2017toz,
Razamat:2017hda, Hassler:2017arf, Bourton:2017pee, Apruzzi:2017nck}).
Moreover, many impenetrable issues in strongly coupled phases of
lower-dimensional systems lift to transparent explanations in the
higher-dimensional setting.

This paradigm has had great success in theories with eight or more real
supercharges. Intuitively, the reason for this is that with so much
supersymmetry, the resulting quantum dynamics -- while still interesting -- is
tightly constrained. For example, in the
context of 4D $\mathcal{N}=2$ gauge theories, the metric on the Coulomb branch
moduli space is controlled by derivatives of the pre-potential, a holomorphic
quantity, as used for example, in \cite{Seiberg:1994rs, Seiberg:1994aj}.
Even with reduced supersymmetry, holomorphy is still an invaluable guide
\cite{Seiberg:1994pq}, but many quantities now receive quantum corrections.

One particularly powerful way to construct and study many features of strongly
coupled field theories in four and fewer dimensions is to first begin with a
higher-dimensional field theory, and compactify to lower dimensions. For
example, compactification of the $\mathcal{N}=(2,0)$ 6D SCFTs on a $T^{2}$
leads to a geometric characterization of S-duality in $\mathcal{N}=4$ Super
Yang-Mills theory \cite{Vafa:1997mh}, and analogous $\mathcal{N}=2$ dualities follow from
compactification on more general Riemann surfaces \cite{Witten:1997sc, Gaiotto:2009we}.
The extension to the vast
class of new $\mathcal{N}=(1,0)$ 6D SCFTs constructed and classified in
references \cite{Heckman:2013pva, DelZotto:2014hpa,
Heckman:2014qba, Heckman:2015bfa} (see also \cite{Apruzzi:2013yva, Apruzzi:2015wna}) has only
recently started to be investigated, but significant progress has already been
made in understanding the underlying 4D theories obtained from compactifying
special choices of $\mathcal{N} = (1,0)$ 6D SCFTs on various manifolds. For a partial list of references to compactifications of
6D SCFTs, see e.g. \cite{Gaiotto:2015usa, Ohmori:2015pua, DelZotto:2015rca, Ohmori:2015pia, Coman:2015bqq, Morrison:2016nrt, Apruzzi:2016nfr,
Razamat:2016dpl, Bah:2017gph,  DelZotto:2017pti, Kim:2017toz, Hassler:2017arf, Bourton:2017pee, Kim:2018bpg}.

In this paper we construct generalizations of 4D $\mathcal{N}=1$ quiver gauge
theories in which the role of the links are played by compactifications of 6D
conformal matter on a Riemann surface. To realize a quiver, we shall weakly
gauge flavor symmetries of these matter sectors by introducing corresponding
4D $\mathcal{N}=1$ vector multiplets. Additional interaction terms such as
generalized Yukawa couplings can also be introduced by gluing together
neighboring cylindrical neighborhoods of 6D conformal matter.
Depending on the number of matter fields, and the types of
interaction terms which have been switched on, we can expect a number of
different possibilities for the infrared dynamics of these generalized quivers.

Even the simplest theory of this kind, namely a generalization of SQCD with 4D
conformal matter turns out to have surprisingly rich dynamics. First of all,
we can consider a wide variety of 4D conformal matter sectors depending on the
choice of punctured Riemann surface and background flavor symmetry fluxes.
Weakly gauging a common flavor symmetry for many such sectors then leads us to
a generalization of SQCD with 4D conformal matter. It is natural to ask whether this SQCD-like
theory flows to an interacting fixed point. To address this, we first note that if
weakly gauge the flavor symmetry of conformal matter, we can compute
the beta function coefficient for the gauge coupling via anomalies. It is summarized schematically by the formula:
\begin{equation}
b_{G} = 3 h_{G}^{\vee} - b^{\mathrm{matter}}_{G},
\end{equation}
where $h_{G}^{\vee}$ is the dual Coxeter number of the gauge group $G$.
In the simplest case where we compactify 6D conformal
matter with flavor symmetry $G\times G$ on a $T^{2}$ with no fluxes, it is
well-known that we obtain a 4D $\mathcal{N}=2$ SCFT \cite{Ohmori:2015pua, DelZotto:2015rca, Ohmori:2015pia}.
Weakly gauging a common $G$ for $n$ such 4D conformal matter sectors produces a 4D $\mathcal{N}=1$
gauge theory with flavor symmetry $G^{n}$ in the ultraviolet. For $n=3$, we
are at the top of the conformal window, $n=2$ also produces an SCFT, and $n=1$
is outside the conformal window, instead producing a confining gauge theory
with a deformed quantum moduli space. Generalizations of this construction to
4D conformal matter on genus $g$ Riemann surfaces with fluxes from flavor
symmetries produce additional examples of generalized SQCD theories. Since we
can calculate the contribution from each such matter sector to the weakly gauged flavor symmetry, we
can see clear parallels with the conformal window of SQCD with gauge group $SU(N_c)$ and $N_f$ flavors \cite{Seiberg:1994pq}:
\begin{equation}
\frac{3}{2} N_c < N_f < 3 N_c.
\end{equation}
where in the present case we have:
\begin{equation}
\frac{3}{2} h_{G}^{\vee} \lesssim b^{\mathrm{matter}}_{G} \leq 3 h_{G}^{\vee}.
\end{equation}
The upper bound is sharp, since we can explicitly present examples which saturate the bound. The lower bound
appears to depend in a delicate way on the curvature of the punctured Riemann surface and flavor symmetry fluxes.
Note also that in contrast to ordinary SQCD, we expect an interacting fixed point at both the top and bottom of the window. The
reason is simply that the 4D conformal matter is itself an interacting fixed point.

Starting from this basic unit, we can construct elaborate networks of theories
by gauging additional flavor symmetry factors, producing a
tree-like graph of quiver gauge theories. We can also introduce the analog of
Yukawa couplings for conformal matter, though the field theory
interpretation of this case is particularly subtle.

To address this and related questions, it is helpful to use the UV\ complete framework
of string theory. Indeed, our other aim in this work will be to engineer examples of these theories
using F-theory on elliptically fibered Calabi-Yau fourfolds. The philosophy
here is rather similar to the approach taken in the F-theory GUT\ model building
literature \cite{Beasley:2008dc, Donagi:2008ca} (see \cite{Heckman:2010bq, Weigand:2010wm} for reviews),
namely we consider degenerations of the elliptic fibration over a complex surface
(codimension one in the base) to generate our gauge groups, degenerations over
complex curves (codimension two in the base) to generate 4D conformal matter,
and degenerations at points (codimension three in the base) to generate
interaction terms between conformal matter sectors. A common theme in this
respect is that the presence of singularities will mean that to make sense of
the Calabi-Yau geometry, we must perform blowups and small resolutions in the
base. In the context of 6D\ conformal matter from F-theory on an elliptically
fibered Calabi-Yau threefold, this is by now a standard story, namely we
introduce additional $\mathbb{P}^{1}$'s in the corresponding twofold base \cite{Bershadsky:1996nu, Heckman:2013pva,
Morrison:2012np}. For 4D models, the presence of codimension
three singularities in the base necessitates introducing compact collapsing surfaces.

Geometrically, then, we expect to obtain a rich set of canonical singularities
associated with the presence of collapsing surfaces and curves in the
threefold base. In higher dimensions, this is actually the main way to
generate examples of 6D SCFTs. In four dimensions, however, there can and will
be quantum corrections to the classical moduli space. This is quite clear in
the F-theory description since Euclidean D3-branes can wrap these collapsing
surfaces. This means there will be instanton corrections which mix the K\"ahler
and complex structure moduli of the compactification.

To track when we can expect such quantum corrections to be present, it is
helpful to combine top down and bottom up considerations. Doing so, we
show that in some cases, a putative 4D\ SCFT generated from a Calabi-Yau
fourfold on a canonical singularity instead flows to a confining phase in the
infrared. We also present some examples which realize 4D\ SCFTs.
All of this indicates a new arena for engineering strong
coupling phenomena in 4D theories.

The rest of this paper is organized as follows. In section \ref{sec:6Dcpct} we briefly
review some elements of 6D\ conformal matter and its compactification to 4D
conformal matter. We then introduce the general method of construction for
generating 4D gauge theories with conformal matter. Section \ref{sec:SQCD} focusses
on generalizations of SQCD in which the matter fields are replaced by 4D
conformal matter. In particular, we show there is a conformal window for our
gauge theories, and also analyze the dynamics of
these theories below the conformal window. We also present in section \ref{sec:QUIVER}
some straightforward generalizations based on networks of SQCD-like theories
obtained from gauging common flavor symmetries. With this field theoretic
analysis in hand, in section \ref{sec:FTH} we next turn to a top down construction of this and related
models via compactifications of F-theory. Proceeding by codimension, we
show how various configurations of 7-branes realize and extend
these field theoretic considerations. The field theory analysis also indicates
that the classical moduli space of the F-theory model is in many cases modified
by quantum corrections, which we also characterize. We present our conclusions
in section \ref{sec:CONC}. Some additional details on singularities in F-theory models
as well as details on conformal matter for various SQCD-like theories are presented in the Appendices.

\section{6D\ and 4D Conformal Matter \label{sec:6Dcpct}}

In this section we present a brief review of 6D conformal matter, and its
compactification on a complex curve. Our aim will be to use compactifications
of this theory as our basic building block in realizing a vast array of
strongly coupled 4D $\mathcal{N}=1$ quantum field theories.

From the perspective of string theory, there are various ways to engineer
6D\ SCFTs. For example, theories of class $\mathcal{S}_{\Gamma}$ admit a
description in both M-theory and F-theory \cite{DelZotto:2014hpa}.
In M-theory, they arise from a
stack of $N$ M5-branes probing an ADE\ singularity $\mathbb{C}^{2}%
/\Gamma_{ADE}$ in the transverse geometry $\mathbb{R}_{\bot}\times
\mathbb{C}^{2}/\Gamma_{ADE}$. We can move to a partial tensor branch by
keeping each M5-brane at the orbifold singularity and separating them along
the $\mathbb{R}_{\bot}$ direction. Doing so, we obtain a 7D Super Yang-Mills
theory of gauge group $G_{ADE}$ for each compact interval separating
neighboring M5-branes. The M5-branes have additional edge modes localized on
their worldvolume. The low energy limit is a 6D\ quiver gauge theory with
6D\ conformal matter localized on the M5-branes. In F-theory, each of these
gauge group factors is realized in F-theory by a compact $-2$ curve in a twofold base and
is wrapped by a 7-brane of gauge group $G_{ADE}$. Collisions of
7-branes at points of the geometry lead to additional singular behavior
for the elliptic fibration, namely the location of 6D conformal matter. The
F-theory picture is particularly helpful because it provides a systematic way
to determine the tensor branch moduli space. Essentially,
we keep performing blowups of collisions of 7-branes until all fibers on
curves are in Kodaira-Tate form. Said differently, in the Weierstrass model:%
\begin{equation}
y^{2}=x^{3}+fx+g,
\end{equation}
we seek out points where the multiplicity $f$, $g$ and
$\Delta=4f^{3}+27g^{2}$ is equal to or higher than $(4,6,12)$
(see Appendix~\ref{app:4612}). The presence of
such points is resolved by performing a sequence of blowups in the base
to proceed to the tensor branch. As
illustrative examples, the case of a single M5-brane probing an E-type
singularity is realized by the singular Weierstrass models:%
\begin{align}
(E_{6},E_{6})  &  \text{:}\text{ \ \ }y^{2}=x^{3}+u^{4}v^{4}\\
(E_{7},E_{7})  &  \text{:}\text{ \ \ }y^{2}=x^{3}+u^{3}v^{3}x\\
(E_{8},E_{8})  &  \text{:}\text{ \ \ }y^{2}=x^{3}+u^{5}v^{5}.
\end{align}

Another important feature of such constructions is that the anomaly
polynomial for background global symmetries can be computed for all such 6D
SCFTs \cite{Ohmori:2014kda}. The general structure of the anomaly polynomial is a formal
degree eight characteristic class:%
\begin{align}
I_{8}=  &  \alpha c_{2}(R_{6D})^{2}+\beta c_{2}(R_{6D})p_{1}(T)+\gamma
p_{1}(T)^{2}+\delta p_{2}(T)\\
&  +\sum_{i}\left[  \omega_{i}\,\frac{\mathrm{tr}_{\rm fund}F_{i}^{4}}{16}+\nu_{i}\left(\,\frac{\mathrm{Tr}%
F_{i}^{2}}{4}\right)^{2}+\,\frac{\mathrm{Tr}F_{i}^{2}}{4}\left(  \kappa_{i}p_{1}(T)+\xi_{i}%
c_{2}(R)+\sum_{j}\chi_{ij}\,\mathrm{Tr}F_{j}^{2}\right)  \right] \nonumber
\end{align}
where here, $R_{6D}$ refers to the $SU(2)$ R-symmetry bundle, $T$ the tangent
bundle, and the $F_{i}$ refer to possible flavor symmetries\footnote{${\rm Tr}$ refers to the normalized trace set by ${\rm tr}_{\rm adj}(F^2)=h^{\vee}_G{\rm Tr}(F^2)$, where $h^{\vee}_G$ is the dual Coxeter number. For convenience and similarity with \cite{Ohmori:2014kda} we keep instead the trace in the fundamental representation for higher powers (3 and 4) of the flavor symmetry curvatures.} of the 6D\ SCFTs.

Starting from these 6D theories, we reach a wide variety of lower-dimensional
systems by compactifying on a Riemann surface.\footnote{Punctures can also be
included, though this has only been studied for a few models.} This can also be
accompanied by activating various abelian background fluxes, as well as
holonomies of the non-abelian symmetries \cite{Morrison:2016nrt, Razamat:2016dpl, Kim:2018bpg}.

The F-theory realization of 4D field theories involves working with an elliptically fibered Calabi-Yau fourfold
with $\mathcal{B}$ a non-compact threefold base. In F-theory terms, we can engineer examples of
6D conformal matter on a curve by taking $\mathcal{B}$ to be
given by a complex curve $\Sigma_{g}$ of genus $g$ and a rank two vector
bundle $\mathcal{V}\rightarrow\Sigma_{g}$ so that the total space is a
threefold. A particularly tractable case to analyze is where $\mathcal{V}$ is a sum
of two line bundles $\mathcal{L}_{1}\oplus\mathcal{L}_{2}$ of respective
degrees $d_{1}$ and $d_{2}$ so that the base $\mathcal{B}$ is the total space $\mathcal{L}%
_{1}\oplus\mathcal{L}_{2}\rightarrow\Sigma_{g}$. Introducing local coordinates
$(u_{1},u_{2},v)$ for the line bundle directions and the Riemann surface, we
can expand the Weierstrass coefficients $f$ and $g$ as polynomials in these
local coordinates:%
\begin{equation}
f=\underset{i,j}{\sum}f_{ij}(v)(u_{1})^{i}(u_{2})^{j}\text{, \ \ }%
g=\underset{i,j}{\sum}g_{ij}(v)(u_{1})^{i}(u_{2})^{j}.
\end{equation}
In general, we can consider geometries in which there are various 7-brane
intersections over curves and points of the base. One way to further
constrain the profile of intersections so that the only intersection available
takes place over the curve $\Sigma_{g}$ is to enforce the condition that $\mathcal{B}$
is a local Calabi-Yau threefold, which in turn requires $d_{1}+d_{2}%
+(2-2g)=0$. Doing so, the coefficients $f_{ij}$ and $g_{ij}$ can be constant,
and we automatically engineer conformal matter compactified on a genus $g$
curve. Switching on background fluxes through the 7-branes then engineers
in F-theory the field theoretic constructions presented in the literature
\cite{Morrison:2016nrt, Razamat:2016dpl, Bah:2017gph, Kim:2017toz, Kim:2018bpg}.

Regardless of how we engineer these examples, it should be clear that even
this simple class of examples leads us to a rich class of 4D\ theories which
we shall refer to as 4D conformal matter. Indeed, in most cases there is
strong evidence that these theories flow to an interacting fixed point.

For example, we can, in many cases, calculate the anomalies of the 4D theory by
integrating the anomaly polynomial of the 6D theory over the Riemann surface \cite{Benini:2009mz}.
Then, the principle of a-maximization \cite{Intriligator:2003jj} yields a self-consistent
answer for the infrared R-symmetry for the putative SCFT. As standard in this
sort of analysis, we assume the absence of emergent $U(1)$ symmetries in the infrared.
In some limited cases, the procedure just indicated is inadequate for determining the
anomalies of the resulting 4D theory. When the
degree of the flavor flux is too low \cite{Razamat:2016dpl}, (typically when the Chern class is one),
or when the Riemann surface has genus one \cite{Ohmori:2015pua, Ohmori:2015pia},
then alternative methods must be used to determine the anomalies of the 4D theory.

A case of this type which will play a prominent role in our analysis of
SQCD-like theories is the theory obtained from compactification of rank one
$(G,G)$ conformal matter on a $T^{2}$ with no flavor fluxes. In this case, we
expect a 4D $\mathcal{N}=2$ SCFT with flavor symmetry $G\times G$. This 4D
$\mathcal{N}=2$ conformal matter is a natural generalization of a
hypermultiplet, but in which the \textquotedblleft matter
fields\textquotedblright\ are also an interacting fixed point.

\section{$\mathcal{N}=1$ SQCD\ with Conformal Matter \label{sec:SQCD}}

In the previous section we observed that there is a natural generalization of
ordinary matter obtained from compactifications of 6D conformal matter on
complex curves. In these theories, there is often a non-abelian flavor
symmetry. Our aim in this section will be to determine the field theory
obtained from weakly gauging this flavor symmetry. For simplicity, in this
section we focus on the special case of rank one $(G,G)$ 6D conformal matter
compactified on a $T^{2}$ with no fluxes, namely 4D $\mathcal{N}=2$ conformal
matter. We denote this theory by a link between two flavor symmetries:%
\begin{equation}
\lbrack G]\overset{CM}{-}[G].
\end{equation}
Before proceeding to the construction of SQCD-like theories, let us begin by
listing some properties of this theory. First of all, the anomaly polynomial
is given by (see Appendix \label{app:CONVENTIONS} for our conventions):
\begin{align}
\label{eq:I60}I_{6}=  &  \frac{k_{RRR}}{6} c_{1}(R)^{3}- \frac{k_{R}}{24}%
p_{1}(T)c_{1}(R)  + k_{RG_LG_L}\frac{\mathrm{Tr}(F^{2}_{G_{L}})}{4} c_{1}(R) + k_{RG_RG_R}%
\frac{\mathrm{Tr}(F^{2}_{G_{R}})}{4} c_{1}(R) + \ldots
\end{align}
where here $R$ is a $U(1)$ subalgebra of the R-symmetry of the 4D theory and $T$
is the formal tangent bundle, $F$ is the field strength of $G_L$ or $G_R$ flavor symmetries, and the dots indicate possible abelian flavor
symmetries and mixed contributions. From this, we read off both the conformal
anomalies $a$ and $c$,
\begin{align}
&  a=\frac{9}{32}k_{RRR}- \frac{3}{32}k_{R}\\
&  c=\frac{9}{32}k_{RRR}- \frac{5}{32}k_{R}.
\end{align}
If we weakly gauge the flavor symmetries,
the contribution to the beta function of
the gauge coupling is set by the term in the anomaly polynomial of the 4D theory involving an R-current
and two flavor currents, namely, the contribution as a matter sector is \cite{Anselmi:1997am}
(see also \cite{Benini:2009mz}):
\begin{equation}
b^{\mathrm{matter}}_{G} = \frac{3 k_{R G G}}{2},
\end{equation}
so that the numerator of the NSVZ beta function \cite{Novikov:1983uc, Novikov:1985rd, Shifman:1985fi, ArkaniHamed:1997mj}
is:
\begin{equation}
b_{G} = 3C_2(G) - b^{\mathrm{matter}}_{G}.
\end{equation}
In what follows, it will be helpful to recall that in our conventions,%
\begin{equation}
C_{2}(G)=h_{G}^{\vee},
\end{equation}
with $h_{G}^{\vee}$ the dual Coxeter number of the group $G$.

For the $(G,G)$ 6D conformal matter compactified on a $T^{2}$
with no fluxes, $a,c$ and the contribution to the beta function coefficients are
\cite{Ohmori:2015pua, Ohmori:2015pia}:
\begin{subequations}
\label{eq:chargesUV6Dr1}%
\begin{align}
&  a=24 \gamma- 12 \beta-18
\delta,\\
&  c=64 \gamma- 12\beta-8
\delta,\\
&  b^{\mathrm{matter}}_L = 24 \kappa_{L},\\
&  b^{\mathrm{matter}}_R = 24 \kappa_{R},
\end{align}
\end{subequations}
where, the coefficients $\beta, \gamma, \delta, \kappa_{L,R}$ can be read off from the 6D anomaly polynomial of rank one $(G,G)$ conformal matter theories
\begin{equation}
I_{8}= \alpha c_{2}(R_{6D})^{2}+\beta c_{2}(R_{6D}) p_{1}(T) + \gamma
p_{1}(T)^{2}+ \delta p_{2}(T) + \kappa_{L} p_{1}(T) \frac{\mathrm{Tr}(F_{L}^{2})}{4}+
\kappa_{R} p_{1}(T) \frac{\mathrm{Tr}(F_{R}^{2})}{4}+\ldots
\end{equation}
and the explicit values for the coefficients are listed in appendix \ref{app:4DCM} (where rank one means $Q=1$ in \eqref{eq:APGT}).

In addition to the anomalies, we also know the scaling dimension and
representation of some protected operators. Two such operators, which we denote by $M_{L}$
and $M_{R}$ transform in the adjoint representation of $G_{L}$ and $G_{R}$,
respectively. They have fixed R-charge of $4/3$ and have scaling dimension
$2$.\footnote{The existence of these operators follows from 
the appearance of a flavor symmetry $\mathfrak{g}_L \times \mathfrak{g}_R$, and a corresponding 
Higgs branch of moduli space. This moduli space is visible both in the 6D 
F-theory constructions (see e.g. \cite{Heckman:2014qba}) as well as in their 4D counterparts, 
as analyzed in reference \cite{Ohmori:2015pua}. The scaling dimension of these operators is 
fixed to be two because they parameterize the Higgs branch. For some discussion on this point, 
see reference \cite{Gaiotto:2008nz}.} These are a natural generalization of the mesons of SQCD. An outstanding
open problem is to determine the analog of the baryonic operators. We shall
return to this issue when we present our F-theory realization of various models.

Our plan in the remainder of this section will be to study various SQCD-like
theories in which the role of quarks and conjugate representation quarks are instead replaced
by 4D $\mathcal{N}=2$ conformal matter. This already leads to a wide variety
of new conformal fixed points and confining dynamics.
Along these lines, we start with $\mathcal{N}=2$ SQCD with 4D conformal
matter. Deformations which preserve $\mathcal{N}=1$ supersymmetry lead to a
new class of SQCD-like theories with $N=1$ supersymmetry. We then turn to an
analysis of $\mathcal{N}=1$ SQCD with 4D\ conformal matter.

\subsection{$\mathcal{N}=1$ Deformations of the $\mathcal{N}=2$ Case} \label{sec:N1defN2}

We obtain an $\mathcal{N}=2$ variant of SQCD by weakly gauging the left flavor
symmetry factor, namely replacing it with an $\mathcal{N}=2$ vector multiplet.
As explained in reference \cite{Ohmori:2015pua, Ohmori:2015pia}, the contribution to the beta function
coefficient of this gauge group is precisely $-h_{G}^{\vee}$, namely minus the
dual Coxeter number. In our conventions the beta function coefficient of the
$\mathcal{N}=2$ vector multiplet is $2h_{G}^{\vee}$. With this in mind, it is
now clear how we can engineer a 4D $\mathcal{N}=2$ SCFT: We can take two
copies of $(G,G)$ 4D $\mathcal{N}=2$ conformal matter, and weakly gauge a
diagonal subgroup. The resulting generalized quiver is then given by:%
\begin{equation}
\mathcal{N}=2\text{ Quiver: }[G]\overset{CM}{-}(G)\overset{CM}{-}[G].
\end{equation}
The beta function coefficient vanishes since we have: $b_{G}=2h_{G}^{\vee
}-h_{G}^{\vee}-h_{G}^{\vee}=0$.

Though we do not know the full operator content of this theory, there are some
protected operators we can still study. To set notation, we label the gauge
groups according to a superscript which runs from $1$ to $3$:%
\begin{equation}
\mathcal{N}=2\text{ Quiver: }[G^{(1)}]\overset{CM}{-}(G^{(2)})\overset{CM}{-}%
[G^{(3)}].
\end{equation}
The mesonic operators previously introduced now include $M_{L}^{(1,2)}$ and
$M_{R}^{(2,3)}$, in the obvious notation. The gauge invariant remnant of the
other mesonic operators is now replaced by the gauge invariant operators (in
weakly coupled notation):%
\begin{equation}
Y^{(1,2)}=M_{R}^{(1,2)}\cdot\varphi\text{ \ \ and \ \ }Y^{(2,3)}=\varphi\cdot
M_{L}^{(2,3)},
\end{equation}
namely, in $\mathcal{N}=1$ language, we now have a coupling between an adjoint
valued chiral superfield (from the $\mathcal{N}=2$ vector multiplet)\ and the
respective mesonic operators from the left and right conformal matter. These
operators have R-charge $+2$, namely scaling dimension $3$.

Having identified some operators of interest, we can now catalog their
symmetry properties. Of particular interest are the $G_{L}\times G_{R}$ flavor
symmetries as well as the $SU(2)\times U(1)$ R-symmetry of the 4D
$\mathcal{N}=2$ SCFT. Since we shall ultimately be interested in
$\mathcal{N}=1$ SCFTs where at most the Cartan generator $I_{3}$ of the
$SU(2)$ factor remains, we simply list the charges under these abelian
symmetries in what follows. Here then, are the relevant symmetry assignments:%
\begin{equation}%
\begin{tabular}
[c]{|c|c|c|c|c|c|}\hline
& $M_{L}^{(1,2)}$ & $M_{R}^{(2,3)}$ & $u$ & $Y^{(1,2)}$ & $Y^{(2,3)}$\\\hline
$R_{UV}$ & $4/3$ & $4/3$ & $4/3$ & $2$ & $2$\\\hline
$J_{\mathcal{N}=2}$ & $-2$ & $-2$ & $+4$ & $0$ & $0$\\\hline
$G_{L}$ & adj$(G_{L})$ & $1$ & $1$ & $1$ & $1$\\\hline
$G_{R}$ & $1$ & adj$(G_{L})$ & $1$ & $1$ & $1$\\\hline
\end{tabular}
\ \ \ ,
\end{equation}
where viewed as a 4D $\mathcal{N}=1$ SCFT, the linear combination of $U(1)$'s
corresponding to the 4D $\mathcal{N}=1$ $U(1)$ R-symmetry and the associated
global symmetry $J_{\mathcal{N}=2}$ is (see e.g. \cite{Benini:2009mz, Tachikawa:2009tt}):
\begin{align}
R_{UV}  &  =\frac{4}{3}I_{3}+\frac{1}{3}R_{\mathcal{N}=2} \label{eq:RUVRIR}\\
J_{\mathcal{N} = 2}  &  =R_{\mathcal{N}=2}-2I_{3}  \label{eq:JN=2RIR}
\end{align}
Note also that for a superconformal scalar primary, we can read off the
scaling dimension $\Delta$ from the R-charge $R$ via the relation:%
\begin{equation}
\Delta=\frac{3}{2}R_{\mathcal{N}=1},
\end{equation}
so we see that the various scalar operators have dimensions:%
\begin{equation}%
\begin{tabular}
[c]{|c|c|c|c|c|c|}\hline
& $M_{L}^{(1,2)}$ & $M_{R}^{(2,3)}$ & $u$ & $Y^{(1,2)}$ & $Y^{(2,3)}$\\\hline
$\Delta$ & $2$ & $2$ & $2$ & $3$ & $3$\\\hline
\end{tabular}
\ \ \ .
\end{equation}

From the perspective of an $\mathcal{N}=1$ theory, we can activate some
marginal couplings such as the $Y^{(1,2)}$ and $Y^{(2,3)}$. We can also
entertain relevant deformations, as specified by deforming by the dimension
two primaries and their superconformal descendants:%
\begin{equation}
\delta W=m_{L}\cdot M_{L}+m_{R}\cdot M_{R}+m_{\text{mid}}u.
\end{equation}
Here, $m_{L}$ and $m_{R}$ are dimension one mass parameters which respectively
transform in the adjoint representation of $G_{L}$ and $G_{R}$, and
$m_{\text{mid}}$ is a dimension one mass parameter which in weakly coupled
terms gives a mass to the Coulomb branch scalar.

Each of these deformations leads to a class of conformal fixed points.
Similar deformations of $\mathcal{N}=2$ theories have been studied previously.
For example, mass deformations of the adjoint valued scalar were considered in
\cite{Tachikawa:2009tt}, and in the context of compactifications of class $\mathcal{S}$
theories in \cite{Benini:2009mz}. In both cases, there is strong evidence that the resulting theory
is a 4D $\mathcal{N}=1$ SCFT.\footnote{References \cite{Benini:2009mz,Tachikawa:2009tt} consider both 
Lagrangian and non-Lagrangian theories with such a relevant deformation added, and in both 
cases present strong evidence that this yields an interacting fixed point. Some of the Lagrangian cases include 
$\mathcal N=2$ $SU(N)$ gauge theory with $2N$ flavors and its deformation to $SU(N)$ SQCD with $2N$ flavors and a 
non-trivial quartic interaction between the quarks superfields, upon adjoint mass deformation. The analysis of these paper does not require a Lagrangian description, and this case is analyzed as well. By assuming that there are no emergent $U(1)$'s in the IR, the conformal 
anomalies $a$ and $c$ as well as the dimension of some protected operators have been computed for the IR theories. These do not show any pathologies, thus providing evidence for the existence of these interacting fixed points. Moreover, they provide examples where the leading order conformal anomalies $a$ and $c$ match the ones computed from the AdS duals. It would be interesting to provide further checks on this self-consistent proposal.} In the case of deformations by the mesonic
operators, there is a further distinction to be made between the case of
having a diagonalizable or nilpotent deformation. In the former case where
$[m_{L},m_{L}^{\dag}]=0$, we actually retain $\mathcal{N}=2$ supersymmetry and
therefore we expect the flow to not generate an SCFT in the IR since the
contribution to the beta function will necessarily decrease. When we instead
have a nilpotent deformation, only $\mathcal{N}=1$ supersymmetry is preserved,
and there can still be an SCFT in the IR. This is referred to as a T-brane
deformation in the literature (see e.g. \cite{Heckman:2010qv, McGrane:2014pma,
Maruyoshi:2016aim, Collinucci:2016hpz, Collinucci:2017bwv}).

As we will need it in our analysis of $\mathcal{N} = 1$ SQCD-like theories, here we mainly
focus on the case of deformations specified by $m_{\text{mid}}u$:%
\begin{equation}
\delta W=m_{\text{mid}}u.
\end{equation}
Assuming there are no emergent $U(1)$'s in the infrared, the result of
reference \cite{Tachikawa:2009tt} completely fix the R-charge assignments of operators
in the infrared theory. For example, the scaling dimensions for our parent theory
operators are now:
\begin{equation}%
\begin{tabular}
[c]{|c|c|c|c|c|c|}\hline
& $M_{L}$ & $M_{R}$ & $u$ & $Y_{L}$ & $Y_{R}$\\\hline
$R_{IR}$ & 1 & 1 &2  &2  &2 \\\hline
$\Delta$ & 3/2 & 3/2 &3  &3  & 3\\\hline
$G_{L}$ & adj$(G_{L})$ & $1$ & $1$ & $1$ & $1$\\\hline
$G_{R}$ & $1$ & adj$(G_{L})$ & $1$ & $1$ & $1$\\\hline
\end{tabular}
\ .
\end{equation}
We can also calculate the values of $a$, $c$, and the contribution to the weakly gauged beta
function coefficients $b^{\mathrm{matter}}_{L}$ and $b^{\mathrm{matter}}_{R}$ in terms of the original UV theory which read
\begin{subequations}\label{eq:UVgen}
\begin{align}
&  a_{UV}=\frac{5d_G}{24} + 2(24 \gamma- 12 \beta-18\delta)\\
&  c_{UV}=\frac{d_G}{6}+ 2(64 \gamma- 12 \beta-8\delta) ,\\
&  b^{\mathrm{matter}}_{L,\,UV}=24 \kappa_L,\\
&  b^{\mathrm{matter}}_{R,\,UV}=24 \kappa_R,
\end{align}
\end{subequations}
where $\beta,\gamma,\delta$ are coefficient of the rank one 6D conformal matter theories (compared with
\cite{Ohmori:2015pua, Ohmori:2015pia} we have stripped off the contribution from the reduction of the
6D tensor multiplet). We now need to plug these into the general expressions for the
IR central charges and beta function coefficients, which (as in \cite{Tachikawa:2009tt}) are
\begin{subequations}
\begin{align} \label{eq:IRgen}
&  a_{IR}=\frac{9}{32}\left(  4a_{UV}-c_{UV}\right)  ,\\
&  c_{IR}=\frac{1}{32}\left(  -12a_{UV}+39c_{UV}\right)  ,\\
&  b^{\mathrm{matter}}_{L,\,IR}=\frac{3}{2}\times b^{\mathrm{matter}}_{L,\,UV},\\
&  b^{\mathrm{matter}}_{R,\,IR}=\frac{3}{2}\times b^{\mathrm{matter}}_{R,\,UV}.
\end{align}
\end{subequations}
The end result of this analysis for $SU(k), SO(2k), E_{6,7,8}$ rank one conformal matter is given in table \ref{tab:IRres} where we used the explicit expression for the coefficients of the 6D anomaly polynomial in \eqref{eq:APGT} (with $Q=1$) and with the explicit group theory data in table \ref{tab:groupconst}. The central charges and beta function coefficients of $SU(k)$ SQCD with $2k$ flavors matches the one in computed in \cite{Agarwal:2014rua}. In fact our construction can be thought as generalizations of \cite{Benini:2009mz,Bah:2012dg,Bah:2013aha,Agarwal:2014rua}, with conformal matter instead of standard matter coupled to $\mathcal N=1$ vectors. Perhaps these theories can be obtained in a similar way by $\mathcal{N}=1$ deformation of class $\mathcal{S}$ theories in \cite{Chacaltana:2013oka,Chacaltana:2015bna,Chacaltana:2017boe, Chcaltana:2018zag}.
\begin{table}[t!]
\centering
\begin{tabular}
[c]{|c||c|c|c|}\hline
$G$ & $a_{IR}$ & $c_{IR}$ & $b^{\mathrm{matter}}_{L,IR}=b^{\mathrm{matter}}_{R,IR}$\\\hline
$SU(k)$ & $\frac{3}{64} \left(5 k^2-4\right)$ & $\frac{1}{64} \left(19 k^2-8\right)$ & $\frac{3}{2}k$\\\hline
$SO(2k)$ & $\frac{3}{16} ( 7k^2 - 8k - 20)$ & $\frac{1}{16} (23k^2 - 25k -58)$ & $3 (k-1)$\\\hline
$E_{6}$ & $\frac{447}{8}$ & $\frac{487}{8}$ & $18$\\\hline
$E_{7}$ & $\frac{2187}{16}$ & $\frac{1161}{8}$ & $27$\\\hline
$E_{8}$ & $\frac{1635}{4}$ & $\frac{3395}{8}$ & $45
$\\\hline
\end{tabular}
\caption{Values of $a_{IR},c_{IR},b^{\mathrm{matter}}_{L,IR}=b^{\mathrm{matter}}_{R,IR}$
for $\mathcal N=1$ deformation of $\mathcal{N}=2$ theories with two G-type
conformal matter sectors coupled by gauging the diagonal of two (out of the four) flavor groups.}%
\label{tab:IRres}
\end{table}

\subsection{Conformal Window and a Confining Gauge Theory}

Rather than starting from $\mathcal{N}=2$ SQCD with 4D\ conformal matter and
performing $\mathcal{N}=1$ deformations, we can instead ask what happens if we
gauge a common flavor symmetry of multiple 4D conformal matter theories by
introducing $\mathcal{N}=1$ vector multiplets. In this case, the contribution
to the beta function coefficient from this vector multiplet is $3h_{G}^{\vee}%
$. Given $n$ copies of $(G,G)$ 4D $\mathcal{N}=2$ conformal matter, we can
weakly gauge a diagonal subgroup by introducing a corresponding $\mathcal{N}%
=1$ vector multiplet. This produces a theory of SQCD with 4D conformal matter.
By inspection, we see that when $n=3$, the beta function coefficient vanishes
since $b_{G}=3h_{G}^{\vee}-h_{G}^{\vee}-h_{G}^{\vee}-h_{G}^{\vee}=0$. This
strongly suggests that we have successfully engineered a 4D $\mathcal{N}=1$
SCFT. We denote the quiver by:%
\begin{equation}
\mathcal{N}=1\text{ Quiver: }[G]\overset{CM}{-}\overset{[G]}{\overset{|}{(G)}%
}\overset{CM}{-}[G].
\end{equation}
This is a close analog to the case of ordinary SQCD with gauge group $SU(N)$
and $N_{f}=3N_{c}$ flavors.

Now, much as in ordinary SQCD, we know that with fewer flavors, it is still
possible to have a conformal fixed point. With this in mind,
we can consider varying the total number of flavors in large jumps of $n$
4D\ conformal matter sectors.%
\begin{align}
n  &  =3\text{: \ \ }[G]\overset{CM}{-}\overset{[G]}{\overset{|}{(G)}}\overset{CM}{-}[G]\\
n  &  =2\text{: \ \ }[G]\overset{CM}{-}(G)\overset{CM}{-}[G]\\
n  &  =1\text{: \ \ }(G)\overset{CM}{-}[G].
\end{align}
Performing a similar computation of the beta function for the weakly gauged
flavor symmetry reveals that at least in the limit of weak coupling, the
theory will flow to strong coupling in the infrared. To determine whether this
flow terminates at a fixed point or a confining phase, we
shall adapt some of the standard methods from SQCD \cite{Seiberg:1994pq} to the present case.

Consider first the case of $n=2$ conformal matter sectors, namely the quiver:
\begin{equation}
[G]\overset{CM}{-}(G)\overset{CM}{-}[G].
\end{equation}
We have already encountered a
variant of this theory in the previous section, namely we can start from an
$\mathcal{N} = 2$ gauging of a flavor symmetry and then add a mass term to the adjoint
valued chiral multiplet. In $\mathcal{N}=1$ language, there is a
superpotential coupling to the conformal matter sectors on the left and right
through R-charge two operators:%
\begin{equation}
W\supset\sqrt{2}\text{Tr}_{G}\left(  M_{R}^{(1,2)}\cdot\varphi\right)
+\sqrt{2}\text{Tr}_{G}\left(\varphi \cdot  M_{L}^{(2,3)}\right)  ,
\end{equation}
where here, we have indicated by a subscript $(1,2)$ that the conformal matter
connects (reading from left to right) gauge groups $1$ and $2$ with similar
notation for $(2,3)$. Taking our cue from references \cite{Benini:2009mz, Tachikawa:2009tt},
we introduce a mass term for the adjoint valued
scalar, initiating a relevant deformation to a new conformal fixed point:%
\begin{equation}
\delta W=\frac{1}{2}m\text{Tr}_{G} \varphi^{2}  .
\end{equation}
Integrating out the chiral multiplet, we learn that the scaling dimensions of
operators have shifted. For example, by inspection of the interaction terms,
we see that $M_{R}^{(1,2)}$ and $M_{L}^{(2,3)}$ both have
R-charge $+1$, and scaling dimension $3/2$. This can also be seen by
integrating out $\varphi$, resulting in a marginal operator:%
\begin{equation}
\frac{2}{m}\text{Tr}_{G}\left(  M_{R}^{(1,2)}\cdot M_{L}^{(2,3)}\right)  ,
\label{OOmarg}%
\end{equation}
with R-charge $+2$. This is almost the same as the $n=2$ theory, aside from
the presence of this additional constraint on the moduli space of vacua. In
the limit where we tune this constraint to zero (formally by sending the
coefficient of this superpotential deformation to zero), we arrive at our
$n=2$ theory. No correlation functions or global anomaly exhibit singularities as a function of the parameter at either the putative fixed point (with the deformation switched off) or in deformations by the marginal operator. 
For these reasons, we actually expect that the conformal anomalies and 
flavor symmetry correlators will be the same. We thus conclude
that both $n=2$ and $n=3$ lead to conformal fixed points. Returning to the
case of ordinary SQCD, the $n=2$ theory is analogous to having $SU(N_{c})$
gauge group with $N_{f}=2N_{c}$ flavors, namely it is in the middle of the
conformal window \cite{Seiberg:1994pq}.

Consider next the case of a single conformal matter sector, namely $n=1$. In
ordinary SQCD with $N_{f}=N_{c}$ flavors, we expect a confining gauge theory
with chiral symmetry breaking. Moreover, the classical moduli space receives
quantum corrections. We now argue that a similar line of reasoning applies for
the quiver gauge theory:%
\begin{equation}
(G)\overset{CM}{-}[G].
\end{equation}
To see why, consider the UV\ limit of this gauge theory, namely where the
gauge coupling is still perturbative. In this regime, we can approximate the
dynamics in terms of 4D conformal matter and a weakly gauged flavor symmetry.
The mesonic operator $M_{R}$ has scaling dimension $2$, and we can form degree
$i$ Casimir invariants of $M_{R}$, Cas$_{i}(M_{R})$ of classical scaling dimension $2i$.
The specific degrees of these Casimir invariants depend on the gauge group in
question, but we observe that the highest degree invariant has $i_{\text{max}%
}=h_{G}^{\vee}$, the dual Coxeter number of the group. We denote this special
case by Cas$_{\text{max}}(M_{R})$.

Now, as proceed from the UV\ to the IR, the coupling constant will flow to
strong coupling. In the UV, the beta function coefficient for the weakly
gauged flavor symmetry is:%
\begin{equation}
b_{G}=3h_{G}^{\vee}-h_{G}^{\vee}=2h_{G}^{\vee}%
\end{equation}
Given the scale of strong coupling $\Lambda$, instanton corrections will scale
as $\exp(-S_{\text{inst}})\sim\Lambda^{b_{G}}$. So, based on scaling
arguments, and much as in ordinary SQCD \cite{Seiberg:1994pq},
we see that nothing forbids a quantum correction to the moduli
space which lifts the origin of the mesonic branch. Indeed, starting from the
classical chiral ring relations of 4D conformal matter (whatever they may be)
namely Cas$_{\text{max}}(M_{R})-$ (Baryons) $=0$ is now modified to (see also \cite{Maruyoshi:2013hja}):%
\begin{equation}
\text{Cas}_{\text{max}}(M_{R})-(\text{Baryons})=\Lambda^{b_{G}}\text{.}
\label{deformo}%
\end{equation}
The quantum correction to the classical chiral ring relations are expected by scaling arguments, even if 
we do not know the precise form of the baryonic operators.\footnote{As we will see from the top-down approach, string theory predicts the existence of T-brane deformations, which preserve the geometry of the F-theory compactification, whereas mesonic deformation do not. For this reason we expect these T-branes to be natural candidates for baryon operators in the physical theory. We keep them in the non-perturbative constraint on the moduli space, even if we do not explicitly discuss the lift of the baryonic branch.}
This strongly suggests that the mesonic branch is lifted, and moreover, that
our theory confines in the IR rather than leading to an interacting fixed point. As we will explain later in section \ref{subsec:WGF4DM}, further evidence for confinement of these theories is provided by the F-theory Calabi-Yau geometry. A mesonic vev generates a recombination mode in the geometry, e.g. $y^2 = x^3 + (uv - r)^5$. The only singularity remaining is on a non-compact divisor, so there is nothing left to make an interacting CFT.

We can obtain even further variants on SQCD-like theories by wrapping 6D conformal matter on more general
Riemann surfaces in the presence of background fluxes. The analysis of Appendix \ref{app:4DCM}
determines the contribution to the running of the gauge couplings of the weakly gauged flavor symmetries.
This also suggests that much as in SQCD with classical gauge groups and matter fields,
there will be a non-trivial conformal window. We have already established the ``top of the conformal window,''
though the bottom of the window is more difficult to analyze with these methods. It nevertheless seems plausible
that if we denote the contribution from the conformal matter sectors in the UV as $b^{\mathrm{matter}}_{G}$,
that the conformal window is given by the relation:
\begin{equation}
\frac{3}{2} h_{G}^{\vee} \lesssim b^{\mathrm{matter}}_{G} \leq 3 h_{G}^{\vee}.
\end{equation}
Note that in contrast to SQCD, the matter fields are themselves an
interacting fixed point so we expect the theories at the upper and lower bounds to also
be interacting fixed points.

\section{Quivers with 4D\ Conformal Matter \label{sec:QUIVER}}

Having seen some of the basic avatars of SQCD-like theories with conformal
matter, it is now clear how to generalize these constructions to a wide
variety of additional fixed points. First of all, we can generalize our notion
of 4D conformal matter to consider a broader class of 6D\ SCFTs compactified
on Riemann surfaces with flavor fluxes. This already leads to new fixed points
in four dimensions with large flavor symmetries. Additionally, we can consider
gauging common flavor symmetries of these 4D $\mathcal{N}=1$ conformal matter
sectors. For example, the case of $n=3$ 4D conformal matter sectors for
$G$-type SQCD provides a \textquotedblleft trinion\textquotedblright\ which we
can then use to glue to many such theories. Note that we can also produce
generalized quivers which form closed loops. In the context of quivers with
classical gauge groups and matter, this usually signals the possibility of
additional superpotential interactions. These are likely also present here,
but purely bottom up considerations provide (with currently known methods)
little help in determining how such interaction terms modify the chiral ring.

\begin{figure}[t!]
\begin{center}
\scalebox{1}[1]{
\includegraphics[clip,scale=0.45]{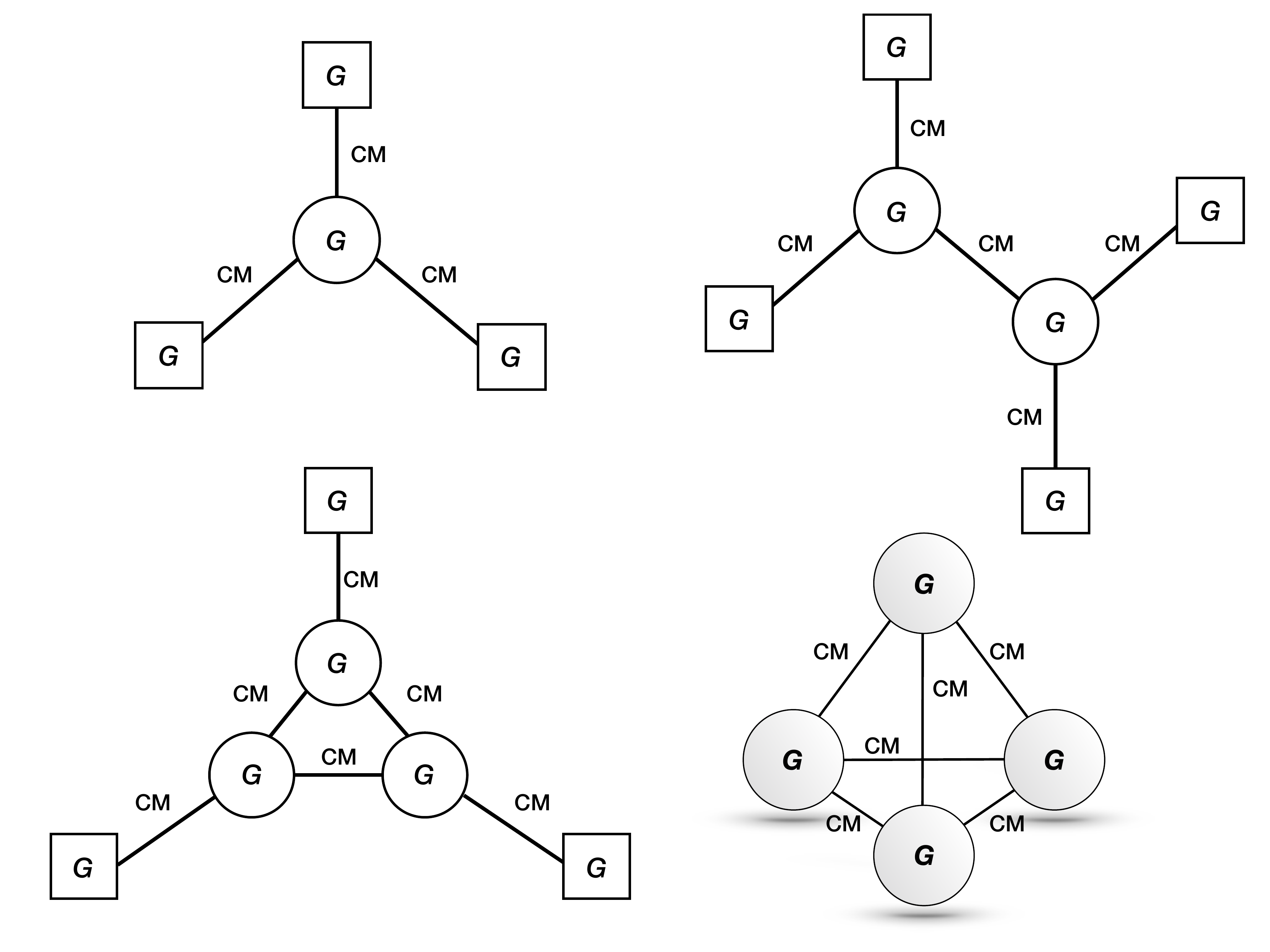}}
\end{center}
\caption{Examples of quiver gauge theories with conformal matter. Here, each line with ``CM'' indicates
4D $\mathcal{N} = 2$ $G \times G$ conformal matter, namely 6D $G \times G$ conformal
matter compactified on a $T^2$.}
\label{fig:quivnet}
\end{figure}

Let us give a few examples which illustrate these general points. Consider the
$\mathcal{N}=2$ quiver with conformal matter:%
\begin{equation}
\mathcal{N}=2\text{ Quiver: }(G)%
\genfrac{}{}{0pt}{}{\overset{CM}{\frown}}{\underset{CM}{\smile}}%
(G).
\end{equation}
Switching on a mass deformation for each adjoint valued chiral multiplet, we
obtain a $G$-type generalization of the conifold:%
\begin{equation}
G\text{-type Conifold: }(G)%
\genfrac{}{}{0pt}{}{\overset{CM}{\frown}}{\underset{CM}{\smile}}%
(G).
\end{equation}
Indeed, we also see that there is a natural superpotential relation dictated
by the mesonic operators of our 4D conformal matter sectors, as follows from
the extension of our discussion near line (\ref{OOmarg}). As another simple
example, we can consider a tree-like pattern of $n=3$ SQCD-like theories which
spread out to produce $\mathcal{N}=1$ theories with large flavor symmetry
factors $G^{N}$. We obtain an even larger class of theories by using genuinely
$\mathcal{N}=1$ 4D conformal matter. In Appendix \ref{app:4DCMAP} we use the anomaly
polynomial of 6D conformal matter to extract properties of these 4D conformal
matter sectors. Lastly, we can also construct quiver networks connected
by conformal matter as in figure \ref{fig:quivnet}.

\section{F-theory Embedding \label{sec:FTH}}

In the previous sections we studied 4D theories with conformal matter from a
\textquotedblleft bottom up\textquotedblright\ perspective in the sense that
we took the 6D SCFT as a starting point for our field theory analysis. In this
section we turn to a \textquotedblleft top down\textquotedblright\ analysis.
One reason for doing so is that the 6D\ SCFTs considered thus far all have an
F-theoretic origin. Besides this, the top down construction can also point the
way to structures which would otherwise be mysterious from a purely
field theoretic approach. Of course, the arrow of implication runs both way.
In some cases we will encounter classical geometric structures which can
receive quantum corrections. The field theory analysis presented in the
previous sections will then help to indicate when we should expect such effects to
be present.

With this picture in mind, let us now turn to the F-theory realization of
quiver gauge theories with 4D conformal matter. Recall that in F-theory, the
structure of the gauge theory sector, matter sectors, and interaction terms
organize according to intersections of components of the discriminant
locus:\footnote{Here we neglect the possibility of T-brane phenomena \cite{Cecotti:2010bp, Anderson:2013rka,
Collinucci:2014qfa, Bena:2016oqr, Marchesano:2016cqg, Anderson:2017rpr, Bena:2017jhm, Marchesano:2017kke}.
It is quite likely that such deformations are associated with the ``baryonic branch'' of
the 4D conformal matter sector.  We also neglect the ``frozen phase'' of
F-theory \cite{Witten:1997bs,deBoer:2001px,Tachikawa:2015wka,frozen-phase}.}
\begin{equation}
\Delta=4f^{3}+27g^{2}\text{,}%
\end{equation}
where $f$ and $g$ are coefficients of the minimal Weierstrass model describing
the elliptically fibered Calabi-Yau fourfold:%
\begin{equation}
y^{2}=x^{3}+fx+g.
\end{equation}
Said differently, gauge theory, matter and interactions organize respectively
on codimension one, two and three subspaces of the threefold base.

The analysis of the gauge theory sector follows a by now standard story for
7-branes wrapped on K\"{a}hler surfaces, and we refer the interested reader to
\cite{Beasley:2008dc, Donagi:2008ca} for additional details on this aspect of the construction. One
important distinction from the purely field theoretic construction is that
even in the limit where gravity is decoupled, the volume modulus of the K\"ahler
surface is a dynamical mode.\footnote{Recall the general rule of thumb is that for a cycle
of middle dimension or higher, the corresponding volume modulus is
normalizable even in limits where gravity is decoupled.} The modulus is
naturally complexified since we can also integrate the RR\ four-form potential
over the K\"ahler surface, so we write the complexified combination (in
dimensionless units)\ as:%
\begin{equation}
T=\frac{4\pi i}{g^{2}}+\frac{\theta}{2\pi},
\end{equation}
in the obvious notation. Instanton corrections will then be organized in terms
of a power series in $\exp(2\pi i T)$.\ Indeed, we should generically expect
quantum corrections to the classical F-theory moduli space:
Euclidean D3-branes can wrap compact surfaces, and they will
mix K\"ahler and complex structure moduli. This also depends on the
details of the geometry as well as background fluxes.

In the case of matter, we must distinguish between the case of
\textquotedblleft ordinary matter\textquotedblright\ in which the
multiplicities
of $(f,g,\Delta)$ are less than $(4,6,12)$, and where the vanishing
is more singular, in which case we have \textquotedblleft conformal
matter.\textquotedblright\ The effective theory associated with
\textquotedblleft ordinary matter\textquotedblright\ has been extensively
studied in the F-theory literature, but the case of 4D conformal matter is, at
the time of this writing, still a rather new structure. Since this takes place
over a complex curve, the resulting 4D theory ought to be thought of as 6D
conformal matter on a curve. The procedure for handling this case follows
already from the algorithmic procedure outlined in reference \cite{Heckman:2013pva}, namely
we keep blowing up collisions of the discriminant locus until all elliptic
fibers are in Kodaira-Tate form. Since this blowing up procedure treats one of
the coordinates as a spectator, we obtain a collection of compact
$\mathbb{P}^{1}$'s with local geometry:%
\begin{equation}
\mathcal{O}\oplus\mathcal{O}(-n_{i})\rightarrow\mathbb{P}^{1},
\end{equation}
where the $n_{i}$ are the sequence of integers appearing in the algorithmic
blowup procedure of reference \cite{Heckman:2013pva}. In models on a threefold base, it
can also happen that we need to perform additional blowups with respect to a different pair of coordinates.
This leads to a further shift in the degrees of the line bundle assignments, so in general, the local geometry of
these $\mathbb{P}^{1}$'s will have the form:
\begin{equation}
\mathcal{O}(-m_i) \oplus \mathcal{O}(-m_i^{\prime}) \rightarrow \mathbb{P}^{1}.
\end{equation}

Much as in the case of \textquotedblleft ordinary matter\textquotedblright\ we
find that compactification on a complex curve with curvature and 7-brane
flux produces a 4D $\mathcal{N}=1$ quantum field theory at low energies. In
fact, the analysis of compactification on various curves illustrates that
these theories are typically 4D $\mathcal{N}=1$ SCFTs. We have already
presented an F-theory construction of such theories in terms of the local
threefold base given by the total space $\mathcal{L}_{1}\oplus\mathcal{L}%
_{2}\rightarrow\Sigma_{g}$. Weakly gauging the flavor symmetry in this
construction means that we compactify one of these line bundle factors, on
which we have wrapped a 7-brane.

Continuing on to codimension three singularities in the base, we encounter
Yukawa couplings between matter fields. In the case of three \textquotedblleft
ordinary\textquotedblright\ matter fields this leads to gauge invariant cubic
couplings between $\mathcal{N}=1$ chiral multiplets. If any of these terms are
replaced by conformal matter, we obtain a generalization of this
situation. Again, we distinguish between the case of \textquotedblleft
ordinary\textquotedblright\ Yukawas in which the multiplicities of
$(f,g,\Delta)$ are less than $(8,12,24)$, and where the vanishing is more
singular, in which case we have a \textquotedblleft Yukawa for conformal
matter.\textquotedblright\ The distinction comes down to whether we need to
perform a blowup in the base to again place all elliptic fibers over surfaces
in Kodaira-Tate form. An example of this kind is the triple intersection of
three non-compact 7-branes with $E_{8}$ gauge group:%
\begin{equation}
y^{2}=x^{3}+(u v w)^{5},
\end{equation}
with $(u,v,w)$ local coordinates of the base. This leads to an intricate
sequence of blowups, which in turn introduces a number of additional compact
collapsing surfaces into the F-theory background. This in turn suggests a
natural role for non-perturbative corrections to the classical moduli space.

Our plan in this section will be to focus on the geometric realization of 4D
theories similar to the ones considered from a bottom up perspective in the
previous section. Since we anticipate a wide variety of new phenomena in the
construction of 4D theories, our aim will be to instead focus on some of the
main building blocks present in such F-theory constructions. We
first explain how to weakly gauge a flavor symmetry of 4D conformal
matter. After this, we turn to the construction of \textquotedblleft conformal
Yukawas.\textquotedblright\ Due to the fact that we should expect quantum
corrections to the geometry, we begin with the construction of the classical
geometries of each case. We then analyze quantum corrections.

\subsection{Weakly Gauging Flavor Symmetries of 4D Conformal Matter} \label{subsec:WGF4DM}

Recall that to realize 6D conformal matter, we can consider a non-compact
elliptically fibered Calabi-Yau threefold with the collision of two components
of the discriminant locus such that the multiplicities of $(f,g,\Delta)$
at the intersection points are at least $(4,6,12)$. An example of this type is the
collision of two $E_{6}$ 7-branes, namely the collision of two $IV^{\ast}$
fibers:%
\begin{equation}
y^{2}=x^{3}+(u_{1}u_{2})^{4}\text{.}%
\end{equation}
The 6D conformal matter sector has a manifest $E_{6}\times E_{6}$ flavor
symmetry. We can extend this to 4D conformal matter by taking a threefold base
given by the total space of a sum of two line bundles over a complex curve,
i.e. $\mathcal{B}=\mathcal{L}_{1}\oplus\mathcal{L}_{2}\rightarrow\Sigma_{g}$.
Then, the $u_{i}$ specify non-compact divisors in the threefold base.

By a similar token, we can also compactify one of these directions, leaving
the other non-compact. For example, we can weakly gauge an $E_{6}$ factor by
wrapping one of the 7-branes over a K\"ahler surface $\mathcal{S}$. Letting
$v$ denote a local coordinate normal to the surface so that $v=0$ indicates
the locus wrapped by the 7-brane, the local presentation of the F-theory
model is:
\begin{equation}
y^{2}=x^{3}+v^{4}(\widetilde{g}_{\Sigma})^{4}\text{,} \label{EEgauge}%
\end{equation}
where $\widetilde{g}_{\Sigma}$ is a section of a bundle on our surface which vanishes
along $\Sigma$, a complex curve in $\mathcal{S}$. The assignment of
this section depends, on the details of the geometry, and in particular the
normal geometry of the surface $\mathcal{S}$ inside the threefold base
$\mathcal{B}$.

To keep our discussion general, suppose that we expand $f$ and $g$ of the
Weierstrass model as power series in the local normal coordinate $v$:%
\begin{equation}
f=\underset{i}{\sum}v^{i}f_{\Sigma(i)}\text{ \ \ and \ \ }g=\underset{j}{\sum
}v^{j}g_{\Sigma(j)},
\end{equation}
where here, the coefficients
$f_{\Sigma(i)}$ and $g_{\Sigma(j)}$ are sections of bundles
defined over the surface. Our aim will be to determine the divisor
class dictated by where these sections vanish. Recall that
$f$ and $g$ transform as sections of $\mathcal{O}(-4K_{\mathcal{B}})$ and
$\mathcal{O}(-6K_{\mathcal{B}})$, so in the restriction to $\mathcal{S}$, we
have:%
\begin{align}
f_{\Sigma(i)}  &  \in\mathcal{O}_{\mathcal{B}}(-4K_{\mathcal{B}}%
-i\mathcal{S})|_{\mathcal{S}}=\mathcal{O}_{S}(-4K_{\mathcal{S}}%
+(4-i)\mathcal{S}\cdot\mathcal{S})\\
g_{\Sigma(j)}  &  \in\mathcal{O}_{\mathcal{B}}(-6K_{\mathcal{B}}%
-j\mathcal{S})|_{\mathcal{S}}=\mathcal{O}_{\mathcal{S}}(-6K_{\mathcal{S}%
}+(6-i)\mathcal{S}\cdot\mathcal{S})
\end{align}
where in the rightmost equalities of the top and bottom lines we used the
adjunction formula.

The multiplicities of $f$ and $g$ along a divisor on $S$ will depend
on the order of vanishing of the coefficient sections, and
we can now see that it
is indeed possible to engineer conformal matter, in which we also weakly gauge
the flavor symmetry of the 7-brane.

To illustrate, consider the case of $E_{6}\times E_{6}$ conformal matter in which we weakly gauge one
of these flavor symmetry factors. Then, we can specialize the form of the
Weierstrass model to be as in line (\ref{EEgauge}), and in which we also take
$g_{\Sigma(4)}=(\widetilde{g}_{\Sigma})^{4}$. Provided our answer makes sense
over the integers, we can then determine the divisor class on which we have
wrapped our 6D conformal matter:%
\begin{equation}
\widetilde{g}_{\Sigma}\in\mathcal{O}_{\mathcal{S}}\left(  \frac
{-6K_{\mathcal{S}}+ 2 \mathcal{S}\cdot\mathcal{S}}{4}\right)  . \label{gtilde}%
\end{equation}
For example, if we take $\mathcal{B}$ to a local Calabi-Yau threefold,
then $\mathcal{S}\cdot\mathcal{S=}K_{\mathcal{S}}$, and we learn that the
divisor class is $-K_{\mathcal{S}}$, so this corresponds to 6D
conformal matter on an elliptic curve (namely, a $T^{2}$).
To realize an SQCD-like theory, we specialize
$\mathcal{S}$ to a surface which does not contain any additional matter from
the bulk 8D vector multiplet (reduced on the surface). One such choice is $\mathcal{S}$
a del Pezzo surface with no gauge field fluxes switched on. The quiver has the form:
\begin{equation}
(E_{6})\overset{CM}{-}[E_{6}].
\end{equation}
Similar considerations clearly apply for other gauge group assignments.

It is also possible to engineer higher genus curves. Again, it is helpful to
work with illustrative examples. We take
$\mathcal{S}$ to be a $\mathbb{P}^{2}$ so that $K_{S}=-3H$, with $H$ the
hyperplane class. Setting $\mathcal{S}\cdot\mathcal{S=}nH$, line
(\ref{gtilde}) reduces to:%
\begin{equation}
\widetilde{g}_{\Sigma}\in\mathcal{O}_{\mathcal{S}}\left(  \frac{18H+2nH}%
{4}\right)  .
\end{equation}
So, for $n=-3$ we have a genus one curve, and for $n=-1$ we have a genus three
curve. The case of a genus three
curve is particularly interesting, because as explained in Appendix \ref{app:4DCM}, this
contributes just enough to the $E_{6}$ gauge theory beta function to realize a
conformal fixed point at the top of the conformal window.

\subsection{Yukawas for Conformal Matter}

Having introduced a systematic way to build 7-brane gauge theories coupled to
conformal matter, we now turn to interactions between conformal matter
sectors. Much as in the case of ordinary matter, such interaction terms are
localized along codimension three subspaces of the threefold base, namely
points. The local geometry of the Calabi-Yau fourfold will involve the triple
intersection of three components of the discriminant locus. Depending on the
multiplicities  of $f$ and $g$ along each curve, this can lead to
interactions between three ordinary matter sectors, two ordinary matter
sectors and one conformal matter sector, one ordinary matter sector and two
conformal matter sectors, and three conformal matter sectors.

At a broad level, we can interpret such interaction terms as a deformation of
the related system defined by three decoupled 4D matter sectors. Let us label
these three matter sectors as theories $\mathcal{T}_{i,i+1}$, with index
$i=1,2,3$ defined mod three. Each matter sector is specified by the pairwise
intersection of two 7-branes, so there is also a corresponding flavor symmetry
$G_{i}\times G_{i+1}$ for each one. Provided we know the operator content of
these sectors, we can introduce a superpotential deformation, which we
interpret as the presence of a Yukawa coupling. This will in many cases
generate a flow to a new 4D theory which a priori could either be a
conformal fixed point or a gapped phase.

So, let us posit the
existence of \textquotedblleft bifundamental\textquotedblright\ operators
$\Psi_{i,i+1}$ such that the product $\Psi_{1,2}\cdot\Psi_{2,3}\cdot\Psi
_{3,1}$ is invariant under all flavor symmetries. In the case of ordinary 4D
matter, we are at weak coupling so these operators each have scaling dimension
one, and the superpotential deformation has dimension three, i.e. it is
marginal. Depending on the details of the weakly gauged flavor symmetries, it
could end up being marginal relevant, marginal irrelevant or exactly marginal.

Now, for strongly coupled 4D conformal matter, we expect on general grounds
that such 4D Yukawas will be relevant operator deformations. The reason is
that the dimension of the mesonic fields tends to decrease after gauging a
flavor symmetry, so since the mesons are \textquotedblleft
composites\textquotedblright\ of bifundamental operators such as the
$\Psi_{i,j}$, we should expect (at least at a formal level)\ the corresponding
Yukawas to now be relevant deformations. We expect this to happen provided
there is at least one 4D conformal matter sector present at a Yukawa point.

Even so, in practice we do not have such detailed information on the operator
content of the 4D conformal matter sector. Because of this, we will resort to
a combination of top down and bottom up analyses to trace the effects of such
Yukawas on the 4D effective field theory.

The plan of this subsection will be to analyze the classical
F-theory geometry defined by a codimension three singularity involving a
collision of three components of the discriminant locus. Provided each 7-brane
carries gauge group $G_{i}$, this can be visualized as three 6D conformal
matter theories with respective flavor symmetries $G_{i}\times G_{i+1}$ which
we then compactify on a semi-infinite cylinder with a metric which narrows at
one end, namely the \textquotedblleft tip of a cigar.\textquotedblright\ What
we are doing when we introduce a codimension three singularity is joining the
three theories together at the tip of each cigar.

According to the classical geometry, then, we expect to realize a
field theory with flavor symmetry:%
\begin{equation}
G_{\text{classical}}=G_{1}\times G_{2}\times G_{3}.
\end{equation}
We emphasize that this is only the classical answer, and that the quantum
theory may end up having a smaller flavor symmetry. To present evidence that
there could be a symmetry breaking effect due to non-perturbative effects, we
need to analyze the geometry of these codimension three singularities. In
particular, it is valid to ask whether such singularities are permissible in
F-theory at all.

In the remainder of this subsection we perform
explicit resolutions of the threefold base so that all fibers over surfaces
and curves can be put into Kodaira-Tate form.

As discussed in Appendix~\ref{app:4612}, the possibility of blowing up
the base of an F-theory model in codimension two or codimension three
is determined by the multiplicities of  $f$, $g$, and $\Delta$
along the codimension two and codimension three loci in question.  We wish to consider
three divisors on which F-theory 7-branes are wrapped which meet pairwise in conformal
matter curves, with all three meeting
at a common point. We refer to this as a Yukawa for conformal matter.
The divisors and curves in our setup are generally
non-compact, but the point is compact.  Our strategy will be to
blowup only compact points and curves, achieving a partial resolution
of singularities in which conformal matter is still present along
noncompact curves.  In the following subsections, we will see how to put these
local constructions together to form quivers.

\subsubsection{Warmup $SO(8) \times SO(8) \times SO(8)$}

To start, let us consider the intersection of three divisors $D_1$, $D_2$,
$D_3$, on each of which there is an $SO(8)$ global symmetry group.
At the pairwise intersections we get the familiar $SO(8)$--$SO(8)$
conformal matter (which is just an instance of the $E$-string).
What happens at the point of intersection?

To be concrete, we are considering a Weierstrass equation of the form
\begin{equation}
y^2 = x^3 + f_0 \, (uvw)^2 x+ g_0\, (uvw)^3.
\end{equation}
with discriminant $\Delta = (4f_0^3+27g_0^2)(uvw)^6$.
Along the curves of pairwise intersection, we find multiplicities
$(4,6,12)$ so these are the usual conformal matter curves.  At the
origin, where all three divisors meet, the multiplities are $(6,9,18)$.
This is not enough to support a blowup at the origin.  We thus conclude
that these Yukawa points do not have any degrees of freedom in their
Coulomb branch beyond those implied by the conformal matter curves.

\begin{figure}[t!]
\begin{center}
\includegraphics[trim={6cm 5cm 6cm 5cm},clip,scale=0.5]{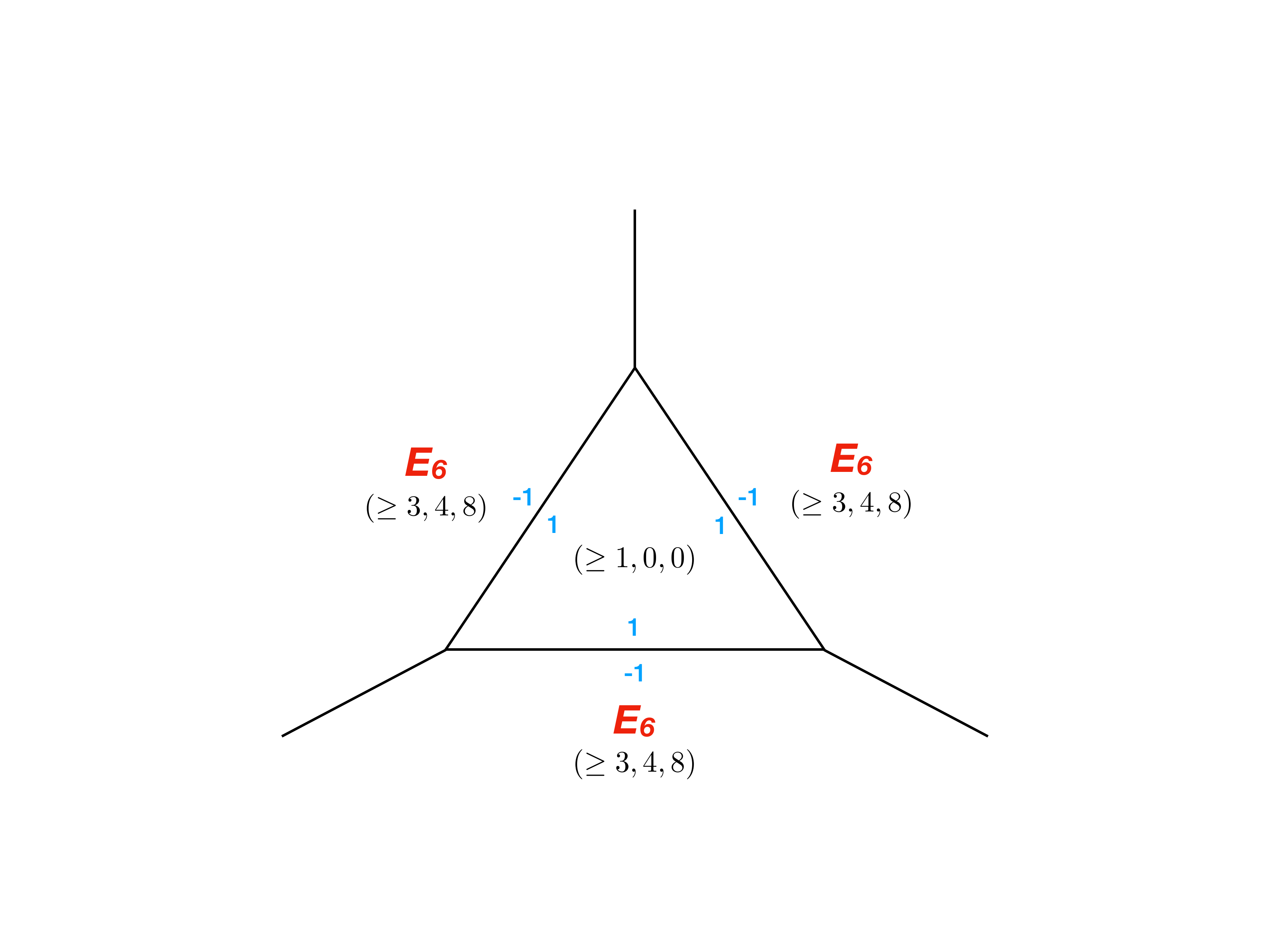}
\end{center}
\caption{The initally blown up $E_6$--$E_6$--$E_6$ Yukawa point.}
\label{fig:E6E6E6}
\end{figure}

\subsubsection{$E_6 \times E_6 \times E_6$}

Turning to the case in which each divisor has type $E_6$, we can represent
this by the equation
\begin{equation}
y^2 = x^3 + (uvw)^4
\end{equation}
with $\Delta = 27(uvw)^8$.  (In this case, $f$ is not relevant for the
computations and we may as well set it equal to $0$.)  Along curves
of intersection such  as $u=v=0$, we find  $E_6$--$E_6$ conformal
matter, and those non-compact curves could be blown up.  Rather than doing
so, however, we examine the Yukawa point.

The multiplicities of $(f,g,\Delta)$ at the origin are $(\ge 9,12,24)$
which means that the origin may be blown up.  The residual vanishing
of $(f_0,g_0,\Delta_0)$ (after reducing the orders of vanishing by
$(8,12,24)$) are $(\ge1,0,0)$.  Thus, we have Kodaira type $I_0$
(nonsingular elliptic fibers) over the exceptional divisor $E$.
There are three new Yukawa points introduced by this blowup,
but they each have multiplicities $(\ge6, 8, 16)$ which does not
allow a blowup.  In addition, no new curves of conformal matter were
introduced by this blowup, but of course we still have the original three
noncompact conformal matter curves.
The exceptional divisor is $\mathbb{P}^2$ and it meets the other
exceptional divisors in lines (which have self-intersection $1$).  These
same lines are exceptional curves of self-intersection $-1$ within
the blown up divisor.
All of this is illustrated in figure~\ref{fig:E6E6E6},
in which we give both the
gauge or flavor group and the orders of vanishing of $(f,g,\Delta)$ for each
divisor.  (When the divisor is unlabeled, there is no gauge symmetry
or flavor symmetry
associated to that divisor.)

\begin{figure}[t!]
\begin{center}
\includegraphics[trim={6cm 2cm 6cm 2cm},clip,scale=0.5]{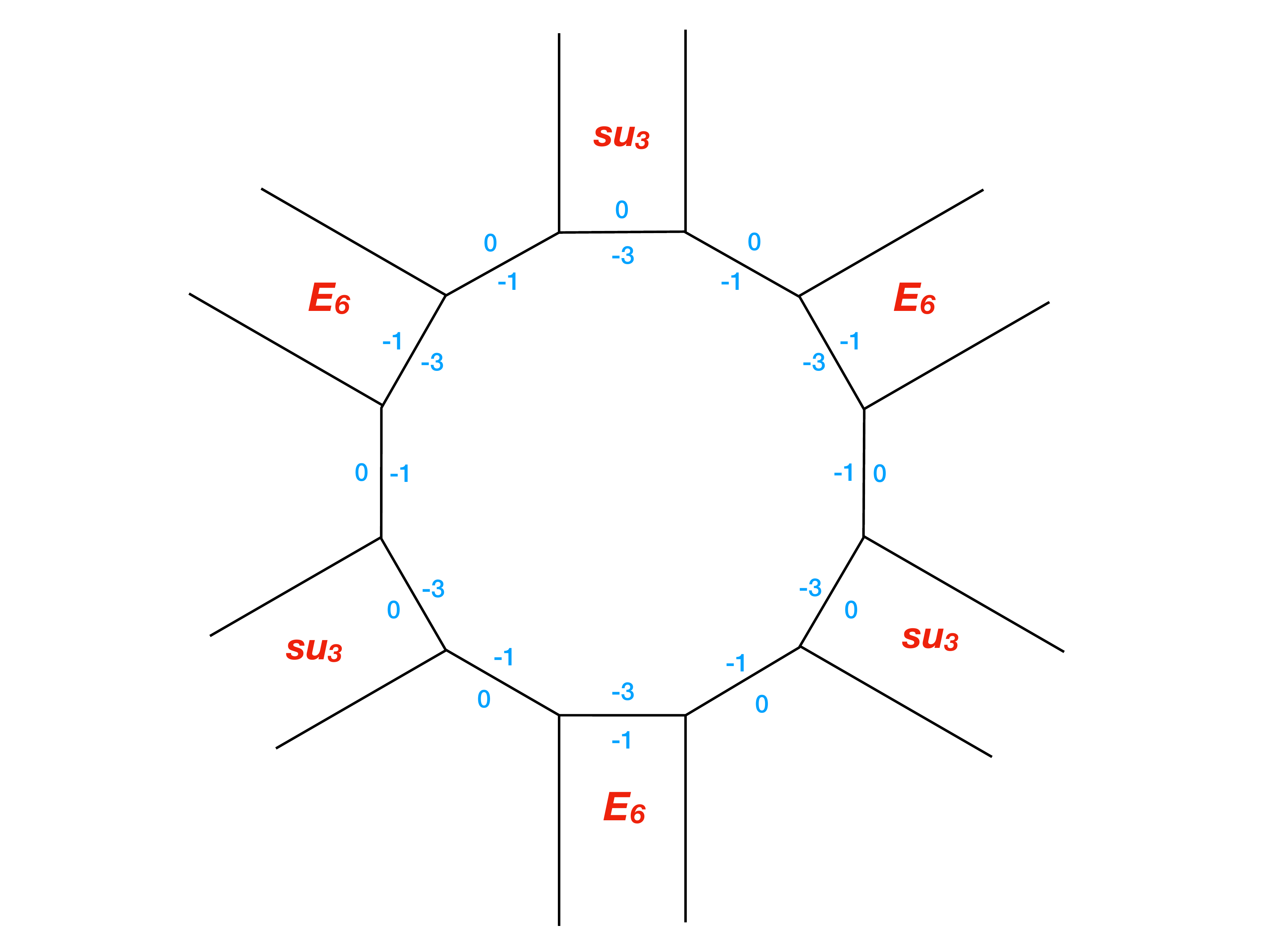}
\end{center}
\caption{The fully blown up $E_6$--$E_6$--$E_6$ Yukawa point.}
\label{fig:E6E6E6-Complete}
\end{figure}

We next blow up the non-compact conformal matter curves (see
 figure~\ref{fig:E6E6E6-Complete}).  The pattern of the blowups
is determined by the $E_6$--$E_6$ collision (known as the $IV^*$--$IV^*$
collision in Kodaira notation) whose sequence of blowups was determined
long ago \cite{Aspinwall:1997ye,Morrison:2012np}.

Note that when blowing up a non-compact conformal matter curve $\Gamma$
we automatically
blowup the point of intersection of $\Gamma$ with any divisor $D$,
creating
an exceptional curve $C$ on the blow up of $D$.  The self-intersection of $C$
is $-1$ on the blown up divisor $\widetilde{D}$
and is $0$ on the non-compact exceptional
divisor.  Moreover, any curve on $D$ which passes through the point
being blown up will have its self-intersections lowered on $\widetilde{D}$.
All of these properties are visible in figure~\ref{fig:E6E6E6-Complete},
which shows the results of an iterated sequence of blowups.

\subsubsection{$E_7 \times E_7 \times E_7$}

The next case to consider is
the one in which each divisor has type $E_7$.  We can represent
this by the equation
\begin{equation}
y^2 = x^3 + (uvw)^3x
\end{equation}
with $\Delta = 4(uvw)^9$.  (In this case, it is $g$ which
is not relevant for the
computations and which we set  to $0$.)  Along curves
of intersection such  as $u=v=0$, we find  $E_7$--$E_7$ conformal
matter.  We first examine the Yukawa point without blowing up
the conformal matter curves.

\begin{figure}[t!]
\begin{center}
\includegraphics[trim={6cm 4cm 6cm 5cm},clip,scale=0.5]{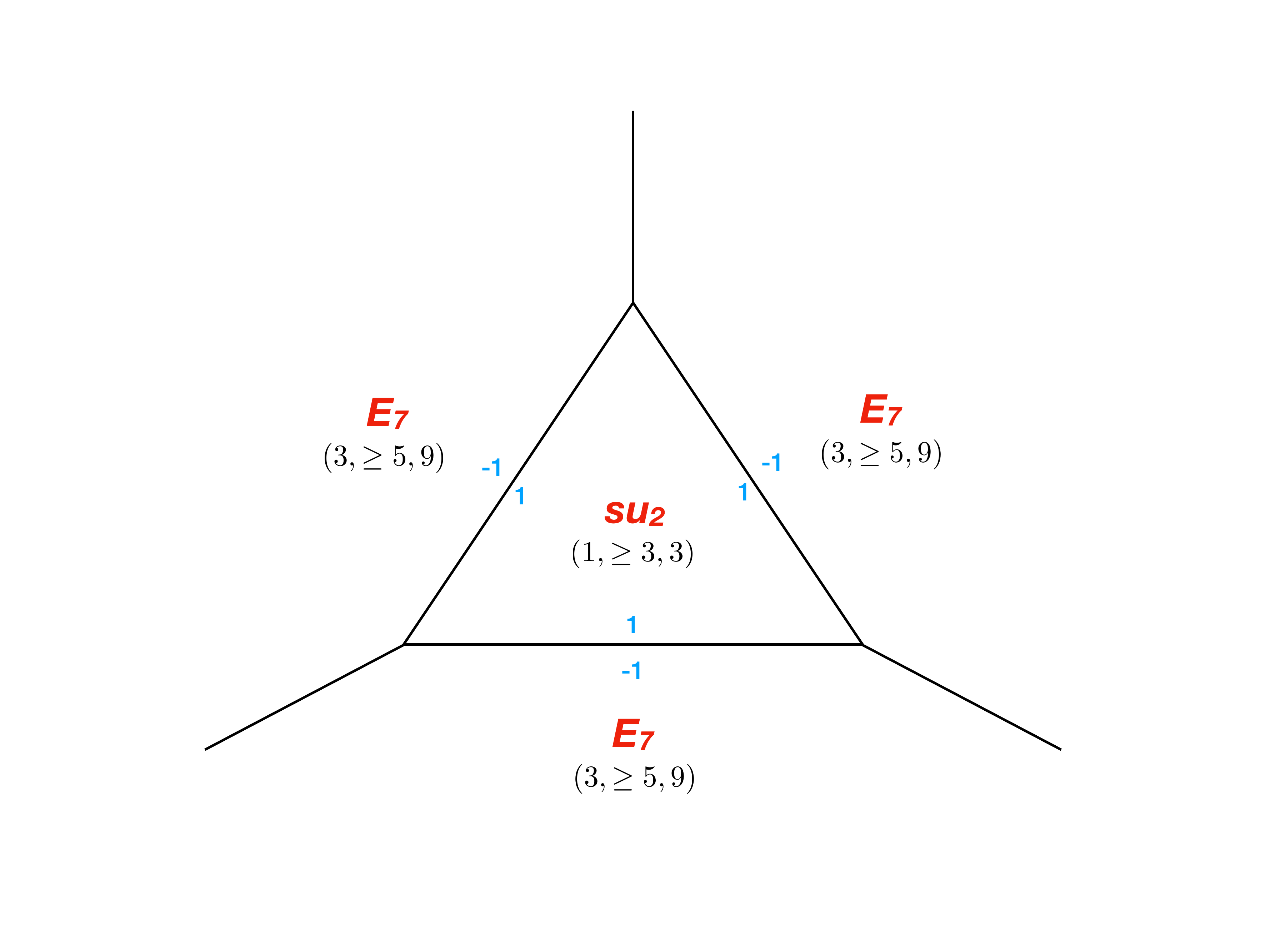}
\end{center}
\caption{The initially blown up $E_7$--$E_7$--$E_7$ Yukawa point.}
\label{fig:E7E7E7}
\end{figure}

The multiplicities of $(f,g,\Delta)$ at the origin are $(9,\ge15,27)$
which means that the origin should be blown up.  The residual vanishing
of $(f_0,g_0,\Delta_0)$ (after reducing the orders of vanishing by
$(8,12,24)$) are $(1,\ge3,3)$, which is Kodaira type $III$ with gauge
group $SU(2)$ over
the exceptional divisor $E$.
There are three new Yukawa points introduced by this blowup,
but they each have multiplicities $(7,\ge13, 21)$ which does not
allow a blowup.  There are also three new compact conformal matter curves
introduced by this blowup, each having $E_7$--$SU(2)$ conformal matter.
These could also be blown up if desired, but we shall postpone doing so.
The illustration of this initial blowup
is in figure~\ref{fig:E7E7E7}, in which we again give both the
gauge group and the orders of vanishing of $(f,g,\Delta)$ for each
divisor.

\begin{figure}[t!]
\begin{center}
\includegraphics[trim={6cm 4cm 6cm 5cm},clip,scale=0.7]{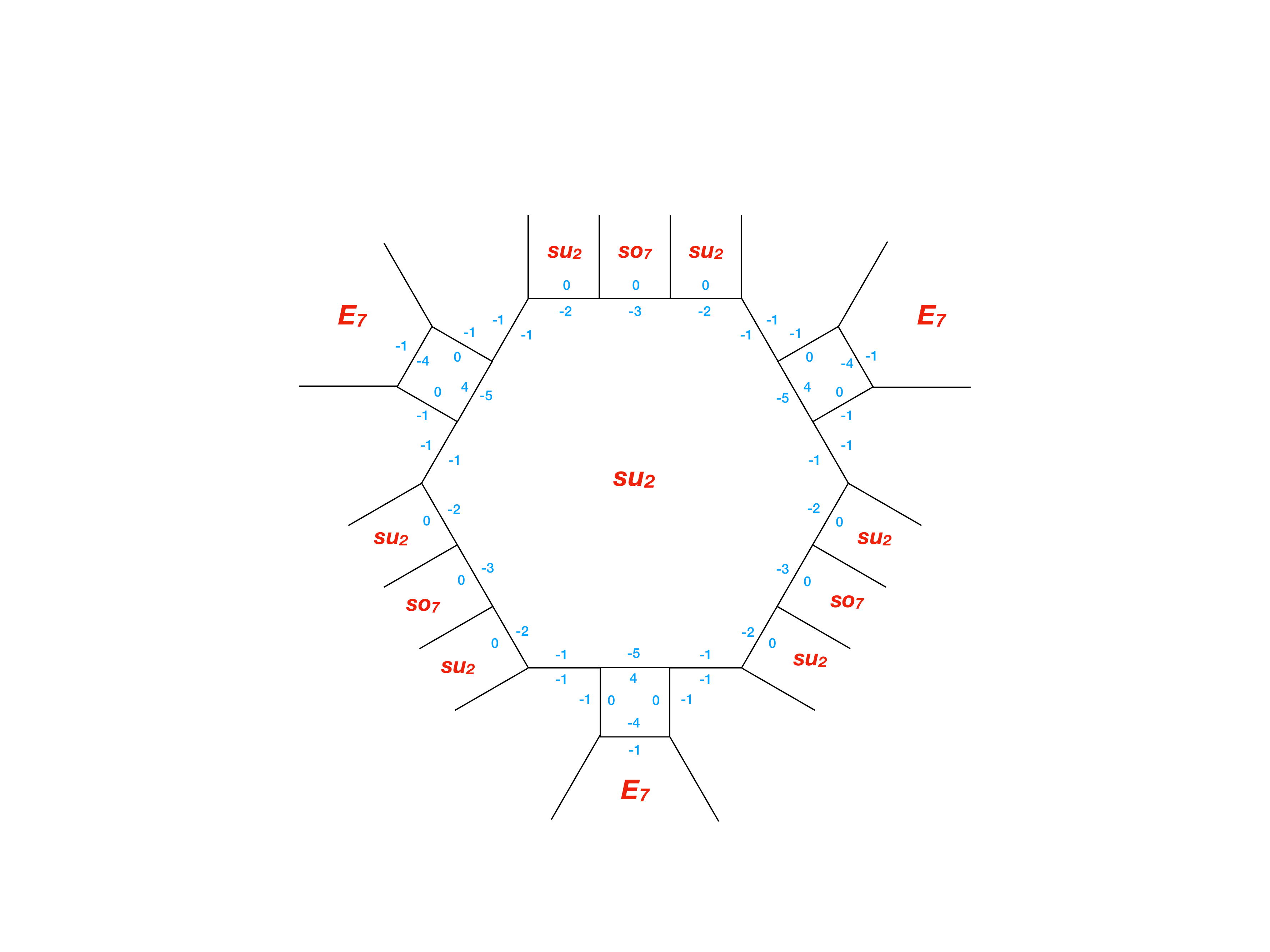}
\end{center}
\caption{The fully blown up $E_7$--$E_7$--$E_7$ Yukawa point.}
\label{fig:E7E7E7-Complete}
\end{figure}

We now blow up the non-compact conformal matter curves, this time relying on
the known sequence of blowups for the $E_7$--$E_7$ collision
(also known as the $III^*$--$III^*$ collision).  At the end of
this process, there are still three compact curves supporting conformal
matter of $E_7$--$SU(2)$ type (also known as $III^*$--$III$ type).
Blowing up those compact curves one time each
completes the resolution, illustrated
in figure~\ref{fig:E7E7E7-Complete}.

In general, when we blow up a compact curve $\Gamma$ which is the intersection
of two divisors in which the self-intersections are $-a$ and $-b$, the
normal bundle of the curve in the threefold is $\mathcal{O}(-a) \oplus
\mathcal{O}(-b)$.  Blowing up the curve creates the Hirzebruch surface
$\mathbb{F}_{a-b}$.  That surface is ruled with curves of self-intersection $0$
(which may appear as exceptional curves of self-intersection $-1$ on
other divisors).  Moreover, there are two disjoint
sections, with self-intersection $a-b$ and $b-a$.

In particular, there are two new compact curves:  one with normal
bundle $\mathcal{O}(-a)\oplus \mathcal{O}(a-b)$ and the other with
normal bundle $\mathcal{O}(b-a)\oplus \mathcal{O}(-b)$.  In the present
example, the last three blowups are on curves with normal bundle
$\mathcal{O}(-1)\oplus\mathcal{O}(-5)$.  Blowing each of them up creates
a Hirzebruch surface $\mathbb{F}_4$ and two curves on it:  one with
normal bundle $\mathcal{O}(-1)\oplus\mathcal{O}(-4)$ and the other
with normal bundle $\mathcal{O}(4)\oplus\mathcal{O}(-5)$.

All of these features are visible in figure~\ref{fig:E7E7E7-Complete}.

\subsubsection{$E_8 \times E_8 \times E_8$}

As our last example with constant $J$-invariant, we let
 each divisor have type $E_8$ which we can represent
 by the equation
\begin{equation}
y^2 = x^3 + (uvw)^5
\end{equation}
with $\Delta = 27(uvw)^{10}$.  (Once again, $f$
is not relevant for the
computations and which we set  to $0$.)  Along curves
of intersection of pairs of divisors, we find  $E_8$--$E_8$ conformal
matter.  We will first examine the Yukawa point without blowing up
the conformal matter curves.

\begin{figure}[t!]
\begin{center}
\includegraphics[trim={4cm 1cm 4cm 3cm},clip,scale=0.5]{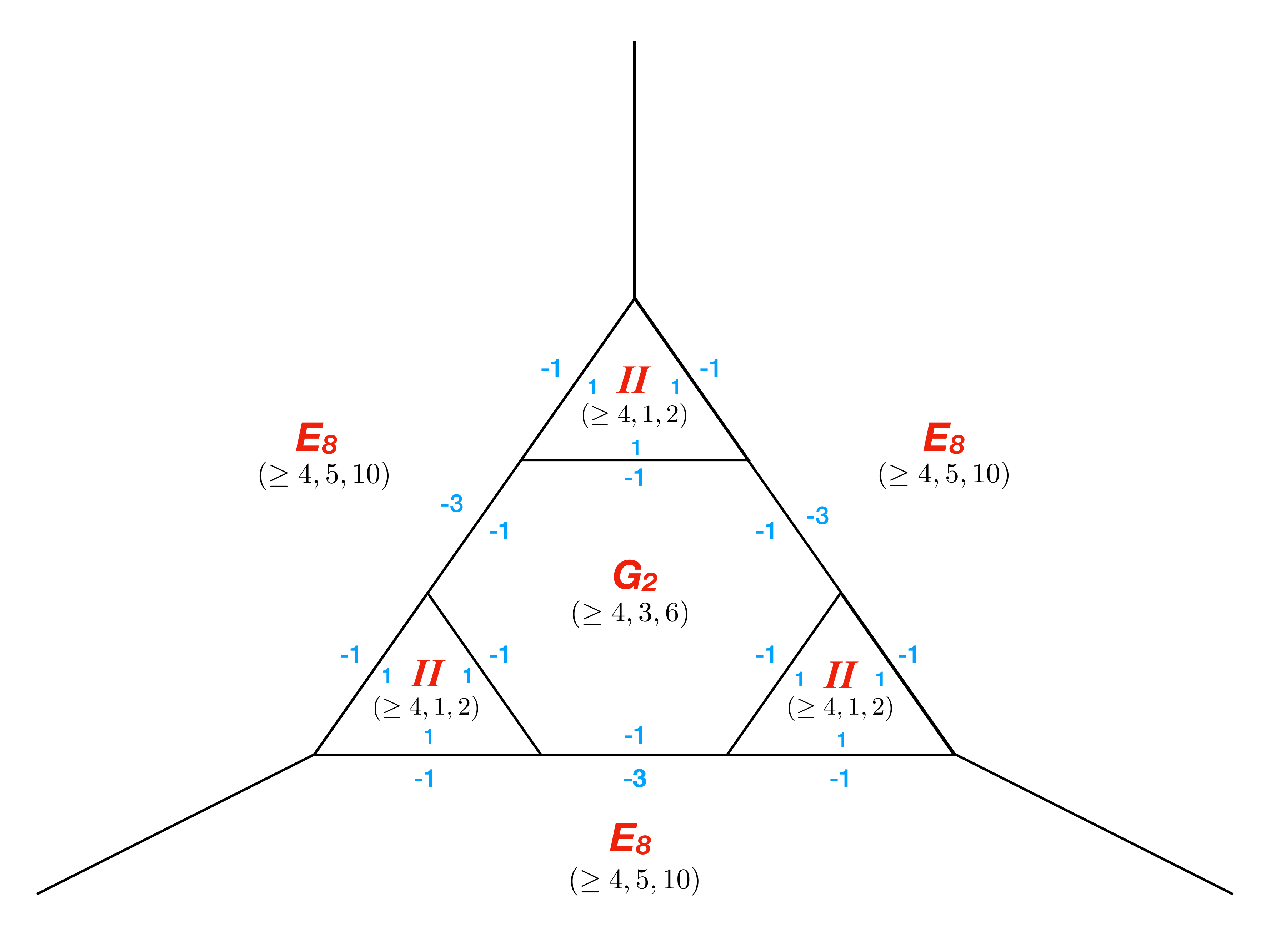}
\end{center}
\caption{The initially blown up $E_8$--$E_8$--$E_8$ Yukawa point.}
\label{fig:E8E8E8}
\end{figure}

The multiplicities of $(f,g,\Delta)$ at the origin are $(\ge12,15,30)$
which means that the origin should be blown up.  The residual vanishing
of $(f_0,g_0,\Delta_0)$ (after reducing the orders of vanishing by
$(8,12,24)$) are $(\ge4,3,6)$, which is Kodaira type $I_0^*$ with gauge
group\footnote{The fact that the gauge group must be $G_2$ rather than
$SO(7)$ or $SO(8)$ is implied by the behavior of the gauge groups
in $E_8$--$E_8$ conformal matter, in which $G_2$ appears
\cite{Aspinwall:1997ye}.} $G_2$ over
the exceptional divisor $E$.
There are three new Yukawa points introduced by this blowup.
Each has multiplicities $(\ge12,13,26)$ so that they can be
blown up.  Blowing them up will generate three more exceptional
divisors $E_1$, $E_2$, $E_3$, each of which has residual multiplicities
$(\ge4,1,2)$ and hence Kodaira type $II$.  This is the Kodaira type
which does not have any gauge symmetry, and yet for which the elliptic
fibers in the total space are singular (with cusps).
An intersection curve between a divisor of Kodaira type $II$ and an $E_8$
divisor has conformal matter with global symmetry $E_8$ and that must
be considered in this situation.  For this reason, we are careful to
label the type $II$ divisors even though there is no gauge or flavor
symmetry associated to them.
The configuration of divisors and gauge groups after the initial blowups
of Yukawa points is illustrated in
figure~\ref{fig:E8E8E8}.

After the four blowups at points, we are left with the original three noncompact
$E_8$--$E_8$ (or $II^*$--$II^*$) conformal matter curves, supplemented by three compact
$G_2$--$E_8$ (or $I_0^*$--$II^*$)
conformal matter curves, and six $II$--$E_8$ (or $II$--$II^*$) conformal matter
curves.

\begin{figure}[t!]
\begin{center}
\includegraphics[trim={4cm 2cm 4cm 2cm},clip,scale=0.5]{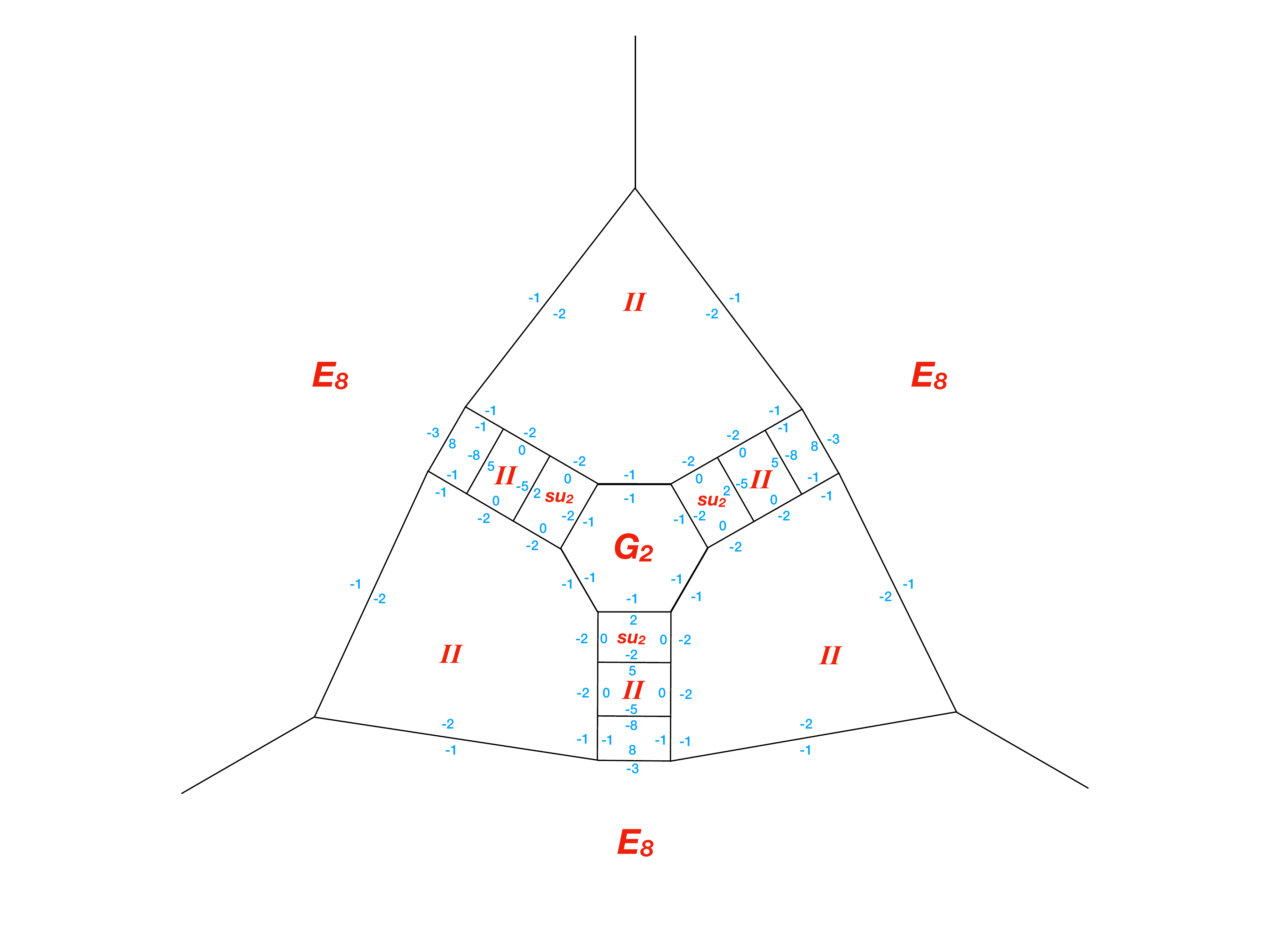}
\end{center}
\caption{The partially blown up $E_8$--$E_8$--$E_8$ Yukawa point.}
\label{fig:E8E8E8-Intermediate}
\end{figure}

We blow the conformal matter curves up in two steps.  First, we use the
known sequence of blowups for the $G_2$--$E_8$ collision to resolve
the conformal matter there.  The known sequence of blowups includes
information of about gauge algebras, and the rules articulated above allow the
determination of the self-intersection of each compact curve in
the diagram.
The results are illustrated in figure~\ref{fig:E8E8E8-Intermediate}.

\begin{figure}[t!]
\begin{center}
\includegraphics[trim={4cm 1cm 4cm 1cm},clip,scale=0.5]{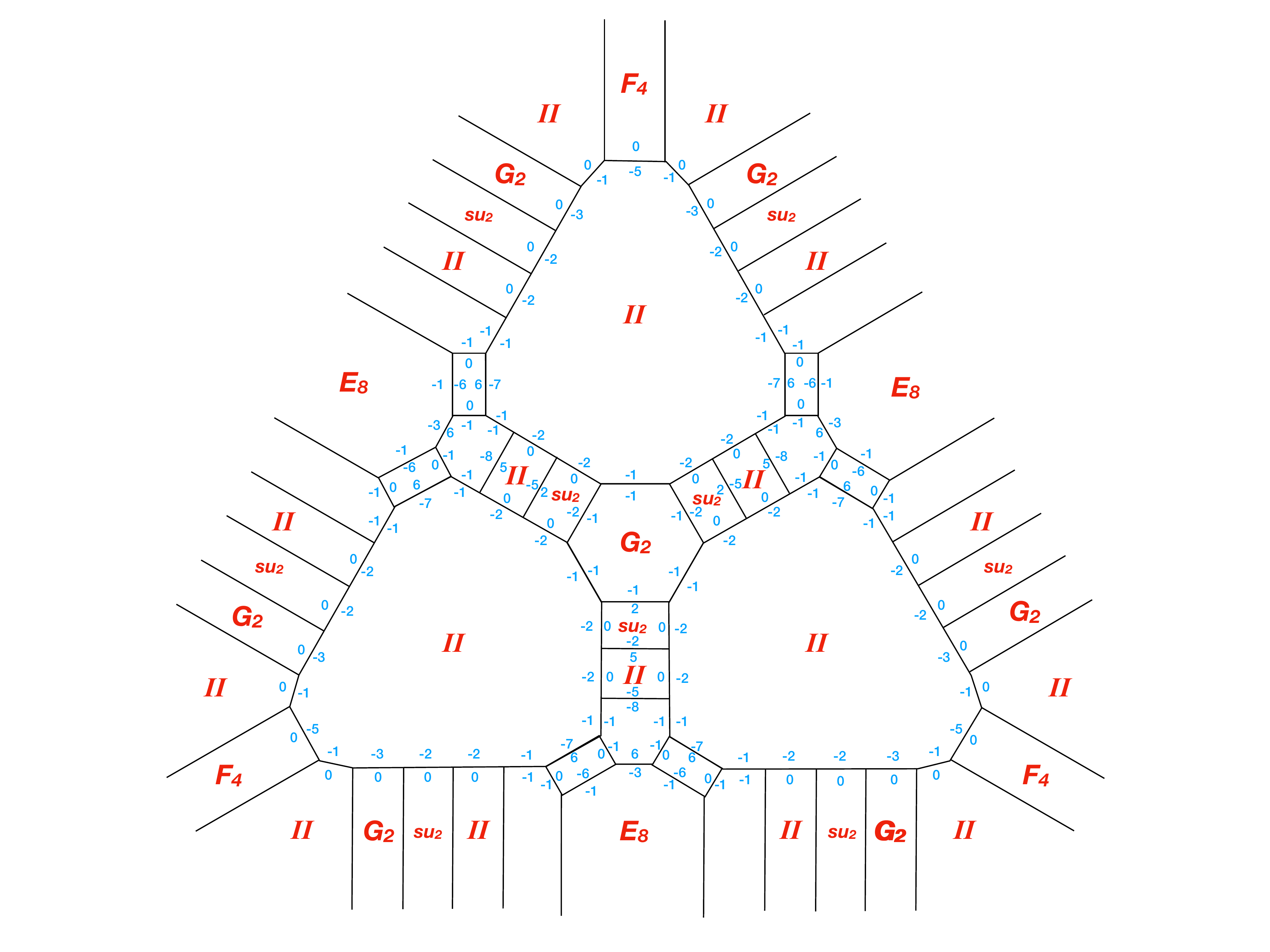}
\end{center}
\caption{The completely blown up $E_8$--$E_8$--$E_8$ Yukawa point.}
\label{fig:E8E8E8-Complete}
\end{figure}

In the final step, just as in the $E_7$--$E_7$--$E_7$ case we blow up
the non-compact conformal matter curves using the known blowup
sequence and flavor groups for the $II^*$--$II^*$ collision.
Finally, we blow up the six remaining $II$--$II^*$ collisions,
obtaining a Hirzebruch $\mathbb{F}_6$ in each case.
The result is illustrated in
figure~\ref{fig:E8E8E8-Complete}.

\subsubsection{Mixed G's with Non-Constant J-function}

As a final example, we treat a case with non-constant $J$-function.
The computations are dependent on both $f$ and $g$, and in fact we
can get variable answers depending on apparently subtle details of
the equation.  We allow the three divisors to have gauge symmetry
$E_6$, $E_7$ and $E_8$ which we can realize in a variety of ways
with the Weierstrass equation
\begin{equation}
y^2 = x^3 + u^{3+k}v^3w^{4+m}x + u^4v^{5+\ell}w^5
\end{equation}
for various choices of nonnegative integers $k\ge0$, $\ell\ge0$, and
$m\ge0$.  The discriminant takes the form
\begin{equation}
4u^{9+3k}v^9w^{12+3m} + 27 u^8v^{10+2\ell}w^{10}
= u^8v^9w^{10}(4u^{3k+1}w^{3m+2} + 27v^{2\ell+1})
\end{equation}
and has multiplicity at the origin $27 + \operatorname{min}(3k+3m+3,2\ell+1)$.
In particular, the multiplicity is $28$ if $k=\ell=m=0$.  The multiplicities
of $f$ and $g$ at the origin are
$(10+k+m,14+\ell)$.

The origin can be blown up, leaving residual orders of vanishing
\begin{equation}
(2+k+m,2+\ell, 4+\operatorname{min}(3k+3m+2,2\ell));
\end{equation}
the minimum value is $(2,2,4)$.  As can be seen, the minimum value has
Kodaira type $IV$, but other values are possible.  For example, $\ell=1$
implies Kodaira type $I_0^*$.  As another example, if $\ell=2$ and $k+m=1$
we get Kodaira type $IV^*$ (and gauge group $E_6$).

\begin{figure}[t!]
\begin{center}
\includegraphics[trim={4cm 3cm 4cm 3cm},clip,scale=0.5]{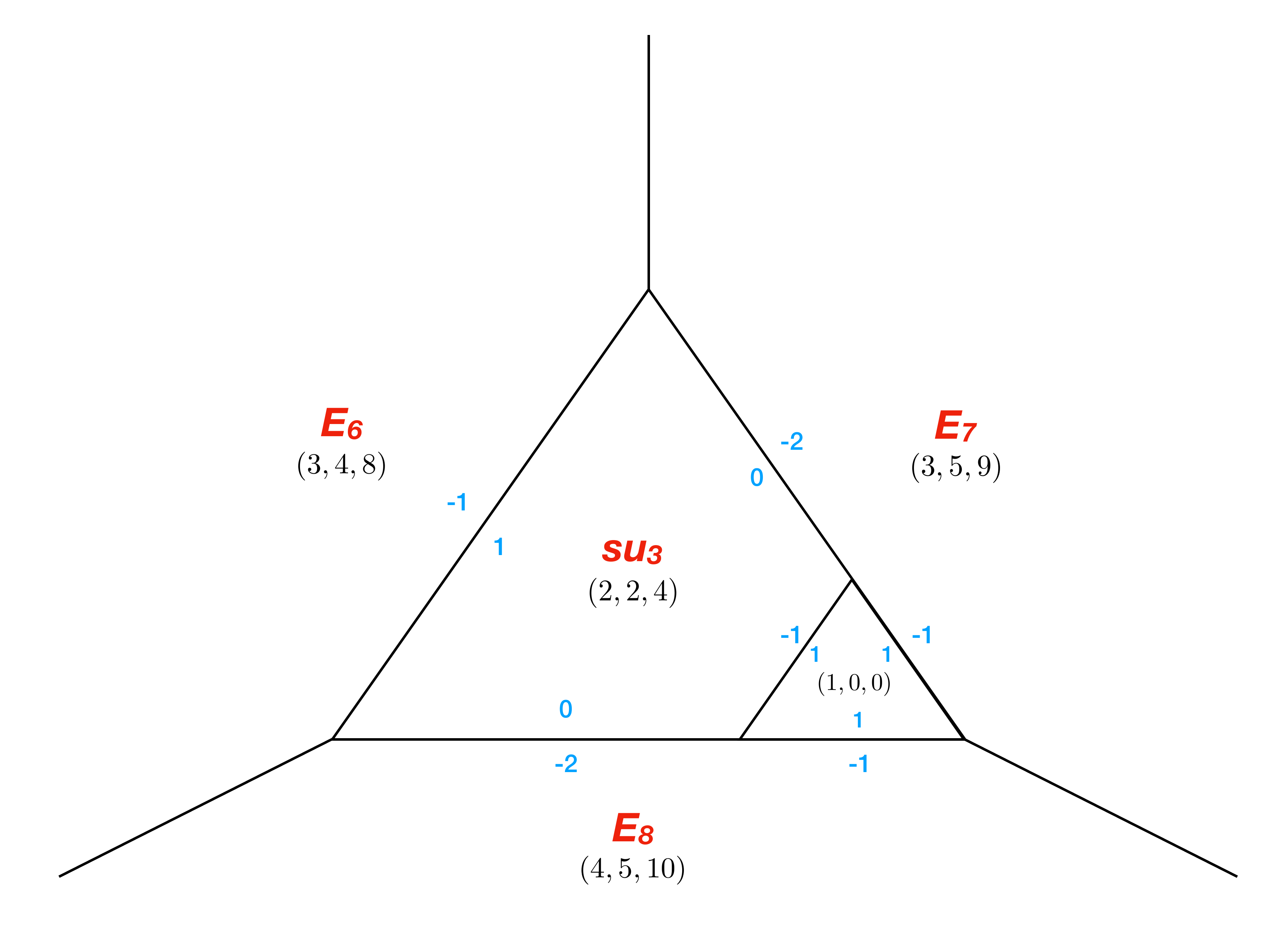}
\end{center}
\caption{The blowup of the minimal  $E_6 - E_7 - E_8$ Yukawa point.}
\label{fig:E6E7E8}
\end{figure}

Let us analyze the minimal case $k=\ell=m=0$.  The Kodaira type after the
first blowup is $IV$ with gauge group $SU(3)$.  There are three new
Yukawa points created after the first blowup, and again, assessing
the multiplicity of $\Delta$ is tricky.  If $\{t=0\}$
is the new exceptional divisor, then the Weierstrass equation after
the first blowup can be written
\begin{equation}
y^2 = x^3 + t^2u^3v^3w^4 x + t^2 u^4v^5w^5
\end{equation}
with discriminant $t^4u^8v^9w^{10}(4t^2uw^2+27v)$.  It follows that
the discriminant has multiplicity higher than naively expected
if $t=u=v=0$ or $t=v=w=0$.  Thus, the multiplicities at the three
new Yukawa points are $(8,11,22)$, $(9,11,22)$, and $(9,12,24)$.  Only
the last one can be blown up, and it gives an exceptional divisor
of Kodaira type $I_0$ with a nonsingular elliptic fibration over it.
This is illustrated in figure~\ref{fig:E6E7E8}.

We omit the description of the complete blowup in this case.

\subsection{Quiver Networks in F-theory}

Clearly, there are many ways we can piece together the codimension one, two
and three singularities of the threefold base to engineer 4D quantum field
theories. Indeed, in addition to these geometric ingredients there will also
be fluxes from 7-branes which allow us to induce a \textquotedblleft chiral
spectrum\textquotedblright\ on each conformal matter curve.

\begin{figure}[t!]
\begin{center}
\scalebox{1}[1]{
\includegraphics[clip,scale=0.45]{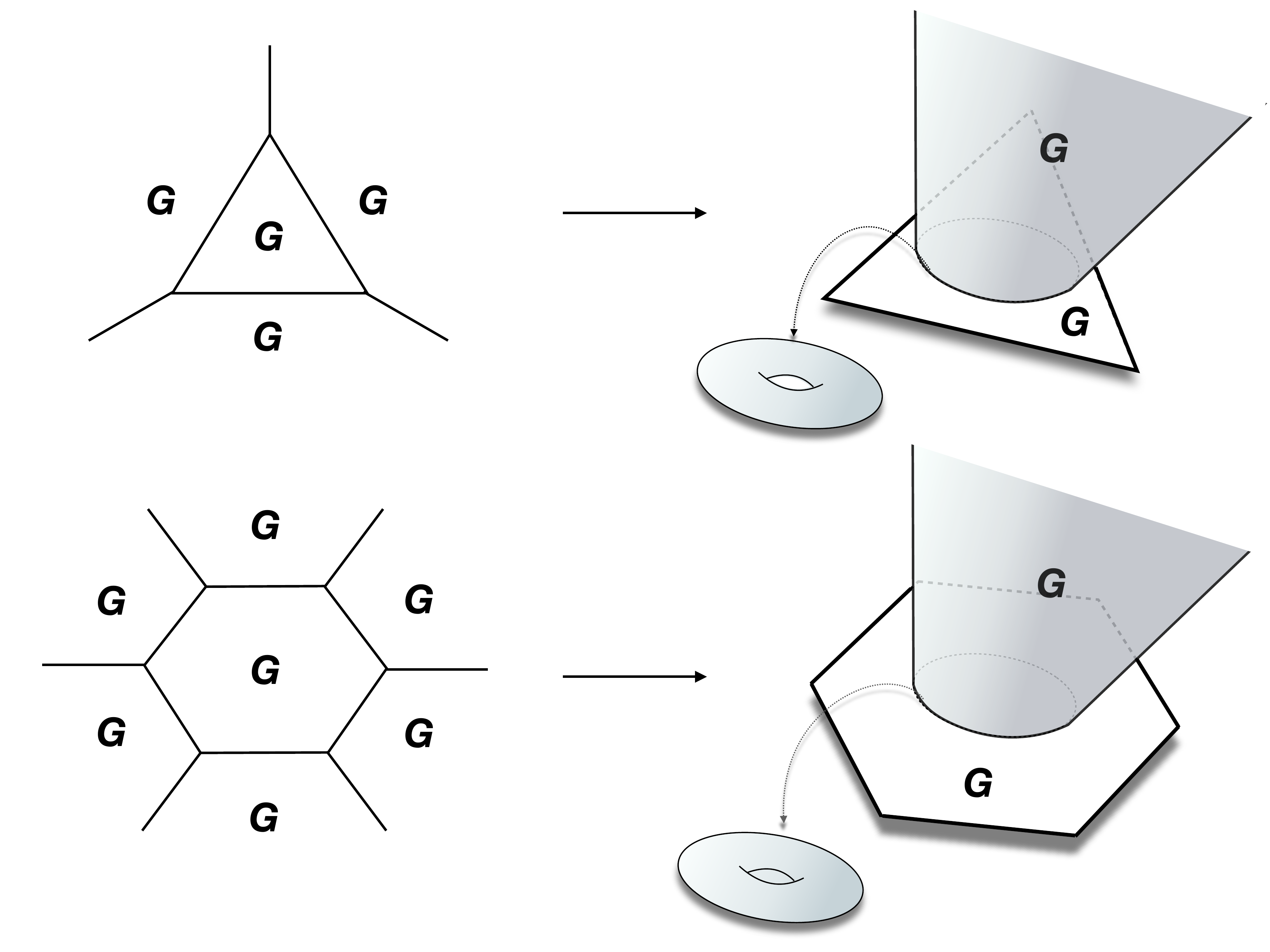}}
\end{center}
\caption{On the left mutually meeting non-compact surfaces with group $G$ and all meeting at a compact $\mathbb P^2$ (toric diagram: triangle) on top, or $dP_3$ (toric diagram, hexagon) on the bottom with gauge group $G$ on $\mathbb P^1$ curves. On the right the smoothing where a recombined non-compact surfaces intersects the $dP_3$ or $\mathbb P^2$ on a $T^2$.}
\label{fig:smoothdp3}
\end{figure}

From the perspective of F-theory, a particularly simple class of examples
involve taking the base to be a non-compact toric Calabi-Yau threefold. In
this case, the analysis of the previous sections illustrates that for any
associated $(p,q)$ web diagram describing a toric Calabi-Yau threefold, we can
decorate each face (be it compact or non-compact) by wrapping a 7-brane
over it. This clearly produces a quiver-like structure, in which the
bifundamentals are 4D conformal matter compactified on compact legs of the
geometry, and with Yukawa couplings between the 4D conformal matter. The
analysis of resolutions of singularities presented in the previous subsection
illustrates that this also leads to a well-defined elliptically fibered
Calabi-Yau fourfold, albeit one with many canonical singularities. An
interesting feature of this classical geometry is the presence of a large
flavor symmetry group. By inspection,
there is a complex structure deformation which takes the \textquotedblleft
pinched\textquotedblright\ complex curves meeting at conformal Yukawas to the
case of a single smooth curve of 4D conformal matter compactified on a high
genus curve (see for example figure \ref{fig:smoothdp3}, for a single $dP_3$
intersecting six non-compact surfaces all with the same group $G$.)
As an illustrative example, consider a
7-brane with gauge group $E_{8}$ wrapped on a compact $\mathbb{P}^{2}$ which
intersects a non-compact $E_{8}$ flavor 7-brane. The local Weierstrass model
for this geometry has already been given on line (\ref{EEgauge}) which we
reproduce for the convenience of the reader:%
\begin{equation}
y^{2}=x^{3}+v^{5}(\widetilde{g}_{\Sigma})^{5}\text{,} \label{EEagain}%
\end{equation}
In this equation, $\widetilde{g}_{\Sigma}$ is a section of the bundle
$\mathcal{O}(3H)$ with $H$ the hyperplane class divisor, namely a degree three
homogeneous polynomial on $\mathbb{P}^{2}$ with vanishing locus an elliptic
curve. We can further specialize the form of this polynomial by factoring the
cubic into three linear terms:
\begin{equation}
\widetilde{g}_{\Sigma}=\underset{i=1}{\overset{3}{%
{\displaystyle\prod}
}}\widetilde{g}_{i}.
\end{equation}
Plugging back in to our minimal Weierstrass model, the geometry now includes
the appearance of Yukawas between conformal matter sectors:%
\begin{equation}
y^{2}=x^{3}+v^{5}(\widetilde{g}_{1}\widetilde{g}_{2}\widetilde{g}_{3}%
)^{5}\text{,} \label{P2example}%
\end{equation}
so the classical geometry now has three $E_{8}$ flavor symmetries!
Additionally, there are now codimension three singularities in the threefold
base along which three $E_{8}$ factors meet.

\begin{figure}[t!]
\begin{center}
\scalebox{1}[1]{
\includegraphics[trim={2cm 2cm 2cm 2cm},clip,scale=0.5]{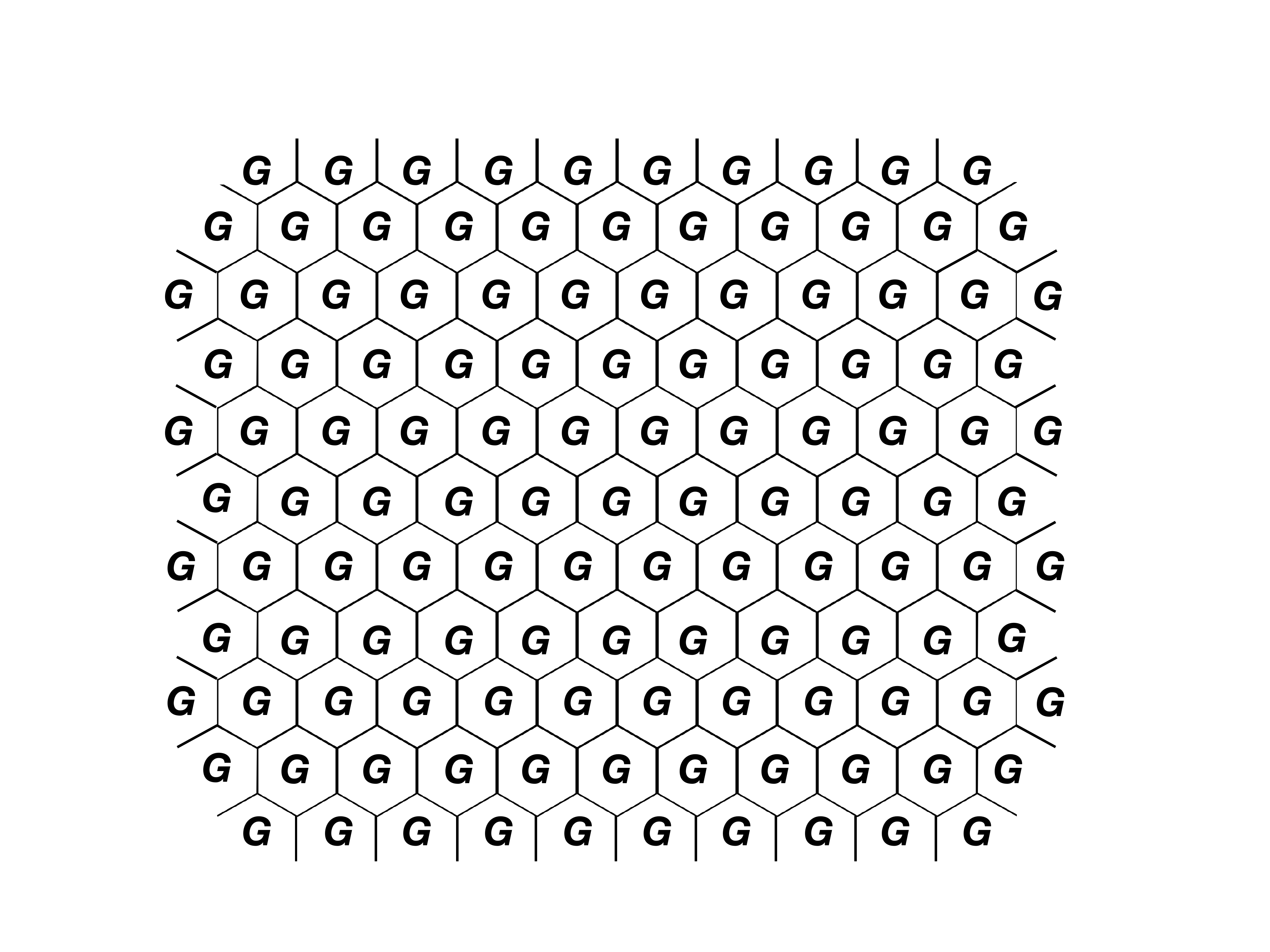}}
\end{center}
\caption{Toric diagram for a threefold base $\mathcal{B}$. Here, each compact face of the honeycomb represents
a $dP_3$ surface. By suitable tuning of the elliptic fiber, we can wrap 7-branes with gauge group $G$ over each
face of the geometry.}
\label{fig:honeycomb}
\end{figure}

Similar considerations hold for 7-branes wrapped on other compact surfaces. As
a more involved example, we can consider a toric threefold base tiled by a honeycomb
lattice as in figure \ref{fig:honeycomb}. Each hexagon in this lattice describes a local del Pezzo three
geometry $dP_{3}$, namely $\mathbb{P}^{2}$ blown up at three points in general
position. Observe that in the case of a single hexagon, the local geometry is
again given by the same sort of Weierstrass models described previously. For
example, in the collision of two $E_{8}$ 7-branes, we again have a complex
equation as in line (\ref{EEagain}), but where now, $\widetilde{g}_{\Sigma}$
is a section of $\mathcal{O}(-K_{\mathcal{S}})$, namely the vanishing locus is
the anti-canonical divisor. Recall that the ring of divisors for the surface
$dP_{3}$ has generators $H$, the hyperplane class, and $E_{i}$ three
exceptional divisors. In terms of these generators, we have:%
\begin{equation}
\mathcal{O}(-K_{\mathcal{S}})=\mathcal{O(}3H-E_{1}-E_{2}-E_{3}\mathcal{)}%
\text{.}%
\end{equation}
By suitable tuning, we can further factorize $\widetilde{g}_{\Sigma}$ so that
the elliptic curve degenerates into a necklace of six lines:%
\begin{equation}
\widetilde{g}_{\Sigma}=\widetilde{g}_{1}\widetilde{g}_{12}\widetilde{g}%
_{2}\widetilde{g}_{23}\widetilde{g}_{3}\widetilde{g}_{31},
\end{equation}
where these polynomials are sections of the following bundles:%
\begin{equation}
\widetilde{g}_{i}\in\mathcal{O(}E_{i}\mathcal{)}\text{ \ \ and \ \ }%
\widetilde{g}_{ij}\in\mathcal{O(}H-E_{i}-E_{j}\mathcal{)}\text{.}%
\end{equation}
Doing so, we see that the entire honeycomb lattice can be filled with gauge
groups of $E_{8}$ type with non-compact flavor branes on the outside of the
picture. Similar considerations clearly hold for other choices of gauge
groups, and lead to a vast array of quantum field theories.

Note that in the construction of these networks of quiver gauge theories, we
can also see the presence of pinched off curves which meet at trivalent
junctions. Applying a smoothing deformation, we see that generically, these
curves can fatten up into a high genus curve, and applying a further smoothing
deformation eliminates all the compact gauge group factors.

It is natural to ask whether this tuning operation has a
counterpart in the field theory. To properly address this issue, we will need
to study quantum corrections to the geometry. In particular, we shall present
evidence that quantum corrections smooth out such tunings, lifting such
enhancements points from the quantum moduli space.

\subsection{Quantum Corrected Geometry}

So far, we have focussed on the classical geometry specified by our F-theory
background. We have also seen that generically, the quiver gauge theories
constructed have compact complex surfaces. This includes the 7-branes wrapped
over the gauge theory divisors of the F-theory model, but also includes
additional collapsing surfaces associated with the codimension three
singularities present in the base.

Now, it is well known from the analysis of \cite{Witten:1996bn} that the
presence of such surfaces can generate quantum corrections which generically
mix the complex structure and K\"ahler moduli of an F-theory background. In
F-theory terms, this comes about from Euclidean D3-branes wrapped over compact
surfaces of the model. In the M-theory background defined by reduction
on a circle, these instanton corrections are captured by Euclidean M5-branes
wrapped on divisors of the Calabi-Yau fourfold.

In either the F-theory description or its dimensional reduction in M-theory,
the assessing the presence (or absence) of such instanton corrections amounts to a
multi-step process. First, we must determine the spectrum of light states
stretched between our Euclidean brane and the other branes present in the
background (now treated as fixed objects). Next, integrating over the zero
mode moduli space, and also possible flux sectors from the Euclidean brane
then leads to instanton corrections to the F- and D-terms of the 4D quantum field theory.

The first aspect of determining whether instanton corrections can be generated
should be clear. In the F-theory models considered, we clearly have compact
surfaces which can be wrapped, so we ought to generically expect the presence
of instanton corrections. Note that even in the case of the $SO(n)^{3}$
codimension three singularities, we should expect instanton corrections. The
reason is that even though the resolved geometry contains no compact complex
surfaces, it contains compact curves in an F-theory background with constant
axio-dilaton. In such configurations, Euclidean D1- and F1-strings can wrap
the complex curves, again generating a quantum deformation of the classical moduli space.

The second aspect, where we actually attempt to extract the zero mode spectrum
for states stretched between the Euclidean brane and the background branes of
the geometry is more challenging in general due to the presence of exceptional
7-branes. Simliar systems with D3-branes in the presence of exceptional
7-branes often lead to strongly interacting SCFTs \cite{Minahan:1996fg, Minahan:1996cj}.
In our case, these exceptional 7-branes can either share four common directions with
the Euclidean D3-brane, or intersect along a complex curve.

Though it would clearly be interesting to directly calculate the form of these
superpotential deformations from this perspective, our primary aim in this
work will be to determine when to expect such corrections to the classical geometry.
So, we shall instead piece together our bottom up
and top down considerations to analyze the quantum corrected moduli space.

The first situation where we can track the effects of an instanton correction
comes from the SQCD-like theory with generalized quiver:%
\begin{equation}
(G)\overset{CM}{-}[G].
\end{equation}
From our field theory considerations, we expect an instanton correction to
contribute which deforms the moduli space of vacua, namely the origin of the
mesonic branch of moduli space will be lifted. By inspection of the F-theory
geometry where we have a 7-brane wrapped on a compact surface, we can see that
a Euclidean D3-brane could indeed wrap this surface. In the F-theory
construction, however, the gauge coupling is promoted to a dynamical field, so
in contrast to equation (\ref{deformo}), we now get the relation (see also \cite{Maruyoshi:2013hja}):
\begin{equation}
\text{Cas}_{\text{max}}(M_{R})-(\text{Baryons})=\left(  \Lambda_{UV} \times\exp(2\pi
iT)\right)  ^{b_{G}}\text{,}%
\end{equation}
where $T$ is the complexified volume modulus associated with the
compact complex surface in the geometry. From the perspective of the effective
field theory, we can introduce a chiral
superfield:%
\begin{equation}
X=\Lambda_{UV} \times\exp(2\pi iT).
\end{equation}
The non-trivial element in this identification is that the origin of the $X$
moduli space corresponds to decompactifying the surface $\mathcal{S}$.

This sort of correction term mixes the complex structure moduli with the
K\"{a}hler moduli. Additionally, it triggers a brane recombination. For
example, in the case where $G=E_{8}$, the deformation is of the form:%
\begin{equation}
y^{2}=x^{3}+v^{5}(\widetilde{g}_{\Sigma})^{5}\rightarrow y^{2}=x^{3}+\left(
v\widetilde{g}_{\Sigma}-r\right)  ^{5}, \label{mesondeform}%
\end{equation}
with $r$ a recombination mode coming from the vevs of the meson fields. It would be interesting
to perform a direct calculation of this effect in string theory, perhaps along the lines
of references \cite{Witten:1996bn, Katz:1996th}.

More generally, we can see that complex structure deformations of the F-theory model translate to ``mesonic deformations'' of the
field theory. T-brane deformations which retain the form of the Weierstrass model are thus natural candidates for ``baryonic deformations.''

Consider next the F-theory model defined by a triple intersection of three
$G$-type 7-branes. For ease of exposition, we focus on the special case where
$G=E_{8}$, which as we have already remarked, is described by a minimal
Weierstrass model of the form:%
\begin{equation}
y^{2}=x^{3} + (u v w)^{5}.
\end{equation}
The resolution of this codimension three singularity introduces compact
K\"{a}hler surfaces, so there is the possibility of an instanton correction.

In this case, we see that the classical geometry consists of three
semi-infinite tubes of 6D conformal matter which are being joined together at
a singular point of the geometry. It identifies the flavor symmetries of each
tube so that we have a $G_{\text{classical}}=E_{8}\times E_{8}\times E_{8}$
flavor symmetry classically.

Let us compare this with compactifications of class $\mathcal{S}_{\Gamma}$
theories. In the present case, the F-theory geometry suggests that we look at
a theory on a thrice punctured sphere which retains a $G_{\text{classical}}$
flavor symmetry. By inspection of the family of metrics for the thrice
punctured sphere, however, we see that there is no degeneration in the family
of metrics which will take us from this smooth geometry to that in which three
tubes degenerate. Moreover, there does not appear to be a point in the moduli
space of theories in which an enhancement to $G_{\text{classical}}$ occurs,
even taking into account data near the punctures. Indeed, in the present case
the F-theory model suggests that all boundary conditions at these marked
points are trivial, so the best we could hope for from the M-theory
construction is an $E_{8}\times E_{8}$ flavor symmetry (though even this is
typically broken to a smaller flavor symmetry).

Another awkward feature of this construction from the perspective of class $\mathcal{S}_\Gamma$ theories
is that the volume of the tubes becomes larger as we proceed away from the trivalent junction. This is precisely
the opposite situation to what one expects to encounter for a punctured Riemann surface.

\begin{figure}[t!]
\begin{center}
\scalebox{1}[1]{
\includegraphics[trim={2cm 5cm 1cm 5cm},clip,scale=0.5]{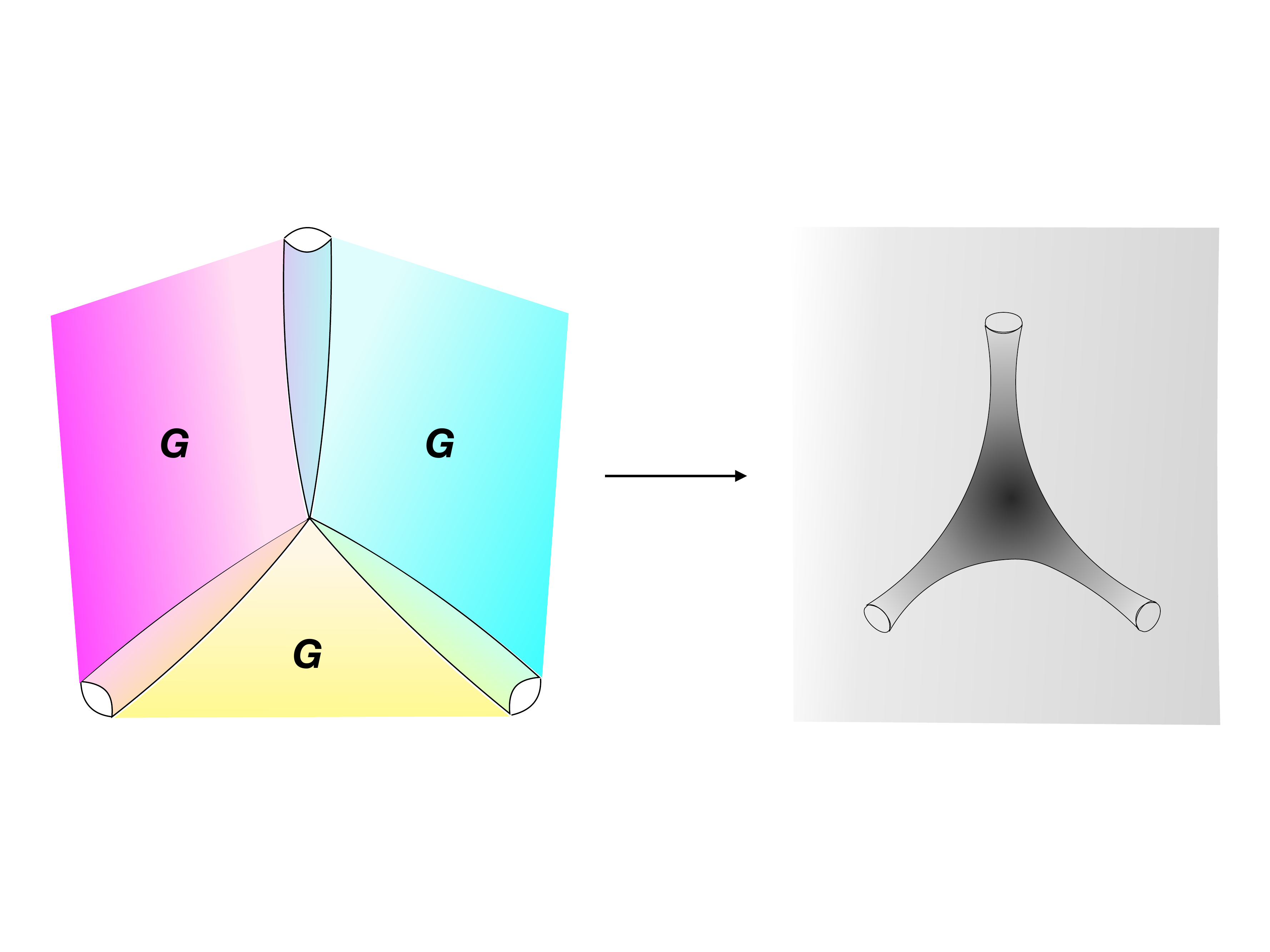}}
\end{center}
\caption{On the left three non-compact surfaces with group $G$ mutually meeting on cigar geometries, which intersect at a point. On the right the smoothing of the configuration to a single surface (grey) where only a subgroup $H\subset G$ may survive.}
\label{fig:smoothyuk}
\end{figure}

This strongly suggests that the classical F-theory picture cannot be realized
in the class $\mathcal{S}_\Gamma$ construction. We thus have two options: Either the F-theory picture receives
no quantum corrections -- in spite of the presence of collapsing divisors in the base --, or instead the F-theory
geometry receives quantum corrections which smooth out some of these singularities. The latter scenario
seems far more plausible, and presents a self-consistent picture. See figure \ref{fig:smoothyuk} for a depiction
of this smoothing process.

Hence, we expect a smoothing deformation to rejoin the three complex lines $u = 0$, $v = 0$, $w = 0$
into a single cubic polynomial in the variables $(u,v,w)$ which need not factorize. Denoting this as $\widetilde{g}_{3}(u_i)$, we see that
the presence of such codimension three singularities will be accompanied by a smoothing deformation:
\begin{equation}
y^{2}=x^{3}+ (u v w)^{5}\rightarrow y^{2}=x^{3}+ (\widetilde{g}_{3}(u,v,w))^{5},
\end{equation}
so that only a single $E_8$ survives.

More generally, we expect that Yukawas for 4D conformal matter lead to smoothings from factorized curves to more generic
curves.

These two sorts of instanton corrections can also appear simultaneously in a
given F-theory background. As an example of this sort, consider again the
model specified by line (\ref{EEagain}), with Weierstrass model:%
\begin{equation}
y^{2}=x^{3}+v^{5}(\widetilde{g}_{1}\widetilde{g}_{2}\widetilde{g}_{3}%
)^{5}\text{.}%
\end{equation}
By inspection, we see three codimension three enhancement points at the torus
fixed points of the $\mathbb{P}^{2}$. We thus expect a quantum deformation
which eliminates this tuned factorization, leading to smoothings of $v$ and a pair $g_i$ and $g_j$.
If, however, we assume that this smoothing leaves intact the presence of the $E_8$ gauge group on $v = 0$,
then the form of this smoothing is further constrained to take the form:
\begin{equation}
y^{2}=x^{3}+v^{5}(\widetilde{g}_{1}\widetilde{g}_{2}\widetilde{g}_{3}%
)^{5}\rightarrow y^{2}=x^{3}+v^{5}(\widetilde{g}_{3H})^{5}.
\end{equation}
But this is the model $(E_{8})\overset{CM}{-}[E_{8}]$, which also confines, so
the net result of the deformations is:%
\begin{equation}
y^{2}=x^{3}+v^{5}(\widetilde{g}_{1}\widetilde{g}_{2}\widetilde{g}_{3}%
)^{5}\rightarrow y^{2}=x^{3}+v^{5}(\widetilde{g}_{3H})^{5}\rightarrow
y^{2}=x^{3}+(v\widetilde{g}_{3H}-r)^{5}%
\end{equation}
in the obvious notation. Namely, the mesonic fields pick up a non-zero vev and
break to the diagonal $E_{8}$ which is a flavor symmetry.

Similar considerations clearly apply in the network defined by the honeycomb
lattice of figure \ref{fig:honeycomb}. Even though the classical geometry contains a large
flavor symmetry group, quantum deformations to the moduli space lead in the
infrared to a confining phase.

\subsubsection{Conformal Fixed Points}

In some cases, instantons do not produce such strong deformations of the
classical moduli space. An example of this type is the intersection of two
$E_{6}$ 7-branes along a genus three curve. We know from our field theory
analysis that weakly gauging one of the $E_{6}$ factors leads to a conformal
fixed point at the top of the conformal window for $E_{6}$ gauge theory. This
involves \textquotedblleft genus three\textquotedblright\ conformal matter.
Based on the analysis of Appendix \ref{app:4DCM}, we also see that once we incorporate
fluxes from the 7-branes, we can again engineer conformal fixed points.

It is also possible to engineer examples of conformal fixed points without
resorting to the presence of higher genus or 7-brane fluxes. To give an
example of this type, we now engineer the 4D model:%
\begin{equation}
\lbrack E_{8}]\overset{CM}{-}\overset{[E_{8}]}{\overset{|}{(E_{8})}%
}\overset{CM}{-}[E_{8}].
\end{equation}

We take our gauge theory surface to be a $dP_{9}$. This surface can be viewed
as a $\mathbb{P}^{2}$ blown up at nine points in general position, but can
also be viewed as an elliptically fibered surface over a base $\mathbb{P}^{1}%
$. In the latter description, we can mark three points, each of which is
associated with a $T^{2}$. The local Weierstrass model is then:%
\begin{equation}
y^{2}=x^{3}+v^{5}(\widetilde{g}_{1}\widetilde{g}_{2}\widetilde{g}_{3})^{5},
\end{equation}
where the divisor class of each $\widetilde{g}_{i}$ is $3H-E_{1}-...-E_{9}$,
with $E_{1},...,E_{9}$ the nine exceptional divisors of the surface. The three matter
curves do not intersect, so there are no conformal Yukawas present in the
model. Note that in this case, the local threefold cannot be a Calabi-Yau,
since as we have already remarked, that would have corresponded to having a
single genus one 4D conformal matter sector. Instead, we learn that
since the product $\widetilde{g}_{1}\widetilde{g}%
_{2}\widetilde{g}_{3}$ is a section of the bundle $\mathcal{O}_{\mathcal{S}%
}\left(  -3K_{\mathcal{S}}\right)  $, we require the divisor relation:%

\begin{equation}
\frac{-6K_{\mathcal{S}}+\mathcal{S}\cdot\mathcal{S}}{5}=-3K_{\mathcal{S}}.
\end{equation}
So, the self-intersection of $\mathcal{S}$ in the threefold base is fixed to
be:%
\begin{equation}
\mathcal{S}\cdot\mathcal{S}=-9K_{\mathcal{S}}.
\end{equation}
In other words, the threefold base is locally given by the total space
$\mathcal{O}(-9K_{\mathcal{S}})\rightarrow\mathcal{S}$. Even though
the gauge theory surface is not a contractible cycle in the threefold base, the
4D gauge theory appears to make sense in its own right.

\section{Conclusions \label{sec:CONC}}

Compactifications of 6D conformal matter on curves lead to novel
building blocks in the construction and study of strongly coupled quantum field
theories in four dimensions. The main idea in our construction is that by
weakly gauging the flavor symmetries of such theories by 4D $\mathcal{N}=1$
vector multiplets, we can obtain a broad class of quiver-like gauge theories
in which the link fields are themselves strongly coupled sectors. We have
presented evidence from a bottom up perspective that much as in ordinary
SQCD with classical gauge groups, there is a notion of a conformal window with 4D
conformal matter, and that the analog of SQCD with gauge group $SU(N_{c})$ and
$N_{f}=N_{c}$ flavors leads to a confining gauge theory. We have also
presented a top down construction of this and related quantum field theories
using F-theory on elliptically fibered Calabi-Yau fourfolds in the presence of
canonical singularities. An additional ingredient suggested by the F-theory
models is the presence of Yukawa couplings between 4D conformal matter, which
in geometric terms is associated with codimension three points in the base
where the multiplicities of the Weierstrass model coefficients $f$, $g$
and the discriminant $\Delta$ are at least $(8,12,24)$, respectively.
Combining our bottom up and top down analyses, we have also argued that the
presence of collapsing four-cycles in these geometries generically indicates
the presence of Euclidean D3-brane instanton corrections which mix the
K\"{a}hler and complex structure moduli. Moreover, they can often smooth out
some singularities present in the classical geometries. This suggests a wide
variety of applications within both field theory and F-theory. In the rest of this
section we indicate some potential avenues for future investigation.

We have pieced together evidence for instanton corrections to
the classical moduli space, based mainly on consistency with both bottom up
and top down considerations. Given that we have the explicit F-theory geometry
for these 4D theories, it should in principle be possible to carefully track
the worldvolume theory of Euclidean D3-branes to calculate such superpotential
corrections. Though the worldvolume theory of these D3-branes may have some
non-Lagrangian building blocks, one of the guiding philosophies of this work
has been that such effects can often still be analyzed, so it would seem
worthwhile to see whether a direct stringy calculation could indeed be performed, perhaps along the lines of
references \cite{Heckman:2008es, Donagi:2010pd, Cvetic:2012ts}.

One of the most intriguing features of SQCD\ with classical gauge groups and
matter content is the notion of Seiberg duality. It is tempting to extend such
considerations to the case of exceptional gauge groups with 4D conformal
matter. To carry this out geometrically, we would need to understand the
structure of flop transitions in the models just engineered. Alternatively, we
could attempt to make an educated guess as to the nature of Seiberg duals for
gauge theories with 4D conformal matter. This would provide additional insight
into the structure of strongly coupled 4D field theories.

Much as in other cases, engineering these 4D field theories turns out to be
simplest in cases where we can leverage the full power of holomorphic
geometry. This in particular is the reason we have chosen F-theory to analyze
the string theory lift of the resulting 4D\ quantum field theories. Even so,
it is tempting to directly engineer these systems using M-theory on a singular
manifold with $G_{2}$ metric holonomy. In this description, weakly gauging a
flavor symmetry amounts to introducing compact three-cycles (presumably
calibrated with respect to the associative three-form). There are in principle
two ways that one could introduce 4D\ conformal matter in this setting. One
way is to simply wrap M5-branes on two-cycles of the geometry. Another way
would be to consider specialized intersections of three-cycles along Riemann
surfaces. Turning the discussion around, the successful realization of these
structures in F-theory strongly suggests there is also an M-theory avatar for
constructing such geometries. This would likely provide further insight into
the construction of $G_2$ manifolds.

Another natural direction to consider is the extension of these results to the
construction of 2D quantum field theories. References \cite{Apruzzi:2016nfr,
Schafer-Nameki:2016cfr, Apruzzi:2016iac, Lawrie:2016rqe} have initiated a general program for engineering 2D
quantum field theories via compactifications of F-theory on elliptically
fibered Calabi-Yau fivefolds. Continuing in the same vein as in the 4D case,
we can clearly see a similar quiver-like structure with exceptional gauge
groups and 2D conformal matter will be present in this lower dimensional case.
Note that in this case it is also natural to consider quartic intersections of
7-branes when the multiplicities of the Weierstrass model coefficients
$f$, $g$ and the discriminant $\Delta$ are at least  $(12,18,36)$,
respectively. It would be quite interesting to study the resulting field
theories engineered from this starting point.

\section*{Acknowledgements}

We thank F. Benini, M. Del Zotto and S. Razamat for helpful discussions.
FA, JJH\ and DRM\ thank the 2018 Summer Workshop at
the Simons Center for Geometry for hospitality during
part of this work. FA, JJH\ and DRM also thank the Banff International
Resesearch Station for hospitality during workshop 18w5190 on the Geometry and
Physics of F-theory.\ LT\ thanks the high energy theory group at the University of
Pennsylvania for hospitality. The work of FA\ and JJH is supported by NSF
CAREER grant PHY-1756996. The work of FA\ is also supported by NSF\ grant
PHY-1620311. The work of DRM\ is supported by NSF\ grant PHY-1620842.
The work of LT\ is supported by VR grant \#2014-5517 and by the ``Geometry
and Physics'' grant from Knut and Alice Wallenberg Foundation.

\appendix

\section{$(4,6,12)$ and all that} \label{app:4612}

As has been discussed
since the earliest days of F-theory \cite{Morrison:1996pp,Bershadsky:1996nh},
in a Weierstrass model
\begin{equation}
 y^2 = x^3 + fx + g
\end{equation}
over a base $\mathcal{B}$,
if the orders of vanishing of $f$, $g$, and $\Delta:=4f^3+27g^2$ exceed
$(4,6,12)$ along some divisor, then the singularities are bad enough to destroy the Calabi--Yau
property of the total space.  For example, in the simplest case
\begin{equation} \label{eq:46}
 y^2 = x^3 + f_0\,t^4x + g_0\,t^6
\end{equation}
(with $f_0$ and $g_0$ constant),
the corresponding surface
singularity can be resolved by a weighted blowup
\cite{MR1710753} with the weights of $t$, $x$, and $y$ being $1$, $2$, and
$3$, respectively.
The exceptional divisor $E$ of the
weighted blowup is an elliptic curve of self-intersection $-1$
 of weighted degree $6$ in
$\mathbb{P}^{1,2,3}$
(see
Example~2.5 in \cite{MR0291170} for details).
  The original  fiber has its cusp singularity
resolved, and becomes
a nonsingular rational curve $C$ of self-intersection $-1$
meeting $E$ transversally.  Moreover,
the holomorphic $2$-form on the Weierstrass model acquires a pole
along $E$ when we blow up and the blowup no longer has trivial canonical
bundle.  By the Hayakawa-Wang criterion \cite{hayakawa,MR1432818},
in type IIA or M-theory on the corresponding Calabi--Yau moduli space,
such points are at infinite distance from the interior of moduli (and so
they should be in F-theory as well).

The curve $C$ can be blown down to a smooth point, and doing so
increases the self-intersection of $E$ to $0$; $E$ is now a fiber
of an elliptic fibration.  But the total space no longer has trivial
canonical bundle.  The easiest way to see this is to divide
\eqref{eq:46} by $t^6$ and rewrite  as
\begin{equation} \label{eq:Weierstrasschange}
(y/t^3)^2 = (x/t^2)^3 + f_0\, (x/t^2) + g_0,
\end{equation}
or, introducing new variables $X=x/t^2$, $Y=y/t^3$,
\begin{equation}
Y^2 = X^3 + f_0\,X + g_0.
\end{equation}

If $x$, $y$, $f$, and $g$ were sections of
$\mathcal{O}(-2K_B)$,
$\mathcal{O}(-3K_B)$,
$\mathcal{O}(-4K_B)$,
and
$\mathcal{O}(-6K_B)$,
respectively (as needed for a Calabi--Yau total space), then
$X$, $Y$, $f_0$, and $g_0$ are sections of
$\mathcal{O}(-2K_B-2D)$,
$\mathcal{O}(-3K_B-3D)$,
$\mathcal{O}(-4K_B-4D)$,
and
$\mathcal{O}(-6K_B-6D)$,
respectively, where $D$ is the divisor $\{t=0\}$.  Thus, the ``wrong''
line bundle $\mathcal{O}(K_B+D)$
is being used for the Weierstrass model, and the canonical
bundle of the total space is not trivial.

The argument we have just given applies in any dimension, with $\{t=0\}$
a local equation for a divisor $D$ along which $f$ vanishes to order at
least $4$ and $g$ vanishes to order at least $6$.  More generally,
if we had started with a Weierstress model which was based on a line
 bundle $\mathcal{L}$
(not necessarily $\mathcal{O}(K_{\mathcal B})$) then this construction changes it
to the bundle $\mathcal{L}\otimes \mathcal{O}(D)$.

Suppose now that we replace the divisor $D$ by a locus $\Gamma$
of higher codimension.
(For example, $\Gamma$ could be a point when $\mathcal{B}$ is a surface, $\Gamma$
could be a point or a curve when $\mathcal{B}$ is a threefold, and so on.)
In this case, if $f$ and $g$ have sufficiently
large multiplicity\footnote{It is common in
algebraic geometry to use the term ``multiplicity'' rather than ``order
of vanishing'' in higher codimension.  This is because it is not the
vanishing of a single function which is being described, but rather
the degrees of all monomials in a local expression for $f$ or $g$ near the
locus $\Gamma$.}  along $\Gamma$, it signals the possibility of blowing
up the base $\mathcal{B}$ without disturbing the Calabi--Yau condition.  We now
explain how this works.

Let $k\ge2$ denote the codimension of $\Gamma$ and suppose that
$\Gamma$ is locally defined by the vanishing of
$k$ functions $\{x_1=x_2=\cdots=x_k=0\}$
which are part of a local coordinate system (i.e., the other coordinates
are $x_{k+1}$, \dots, $x_n$ where $n$ is the dimension of $\mathcal{B}$).
A sample coordinate chart
on the blowup has coordinates $u_j=x_j/x_k$ for $j=1, \dots , k-1$ and then
$x_k$, $x_{k+1}$, \dots, $x_n$ with $\{x_k=0\}$ defining the
exceptional divisor $E$ of the blowup.  Then we have
\begin{multline} \label{eq:canonblowup}
\pi^*(f(x_1,\dots,x_n)\, dx_1\wedge dx_2 \wedge \cdots
\wedge dx_{k-1} \wedge dx_k  \wedge \cdots \wedge dx_n)
=\\
x_k^{k-1}\, f(u_1x_k,\dots,u_1x_{k-1},x_k,\dots,x_n)\,
du_1\wedge du_2 \wedge \cdots \wedge du_{k-1} \wedge dx_k \wedge \cdots
\wedge dx_n .
\end{multline}
If $\pi:\widetilde{\mathcal{B}}\to \mathcal{B}$ denotes the blowup map, then \eqref{eq:canonblowup}
can be described as
\begin{equation} \label{eq:pullback}
K_{\widetilde{\mathcal{B}}} = \pi^*(K_{\mathcal{B}}) + (k{-}1)E
\end{equation}
since the canonical divisor is the divisor of zeros of a holomorphic
form of top degree.
For $f$ and $g$ generic, the pullbacks $\pi^*(f)$ and $\pi^*(g)$
define a Weierstrass model based on the line bundle
$\mathcal{L}=\mathcal{O}(\pi^*(K_{\mathcal B}))\ne \mathcal{O}(K_{\widetilde{\mathcal{B}}})$
for which the canonical bundle of
the total space is not trivial.  But if $\pi^*(f)$ and $\pi^*(g)$
vanish to  orders $4$ and $6$ along a divisor $D$,
we can combine \eqref{eq:pullback} with the
effect in \eqref{eq:Weierstrasschange} to obtain the
Weierstrass line bundle
\begin{equation}
\mathcal{L}\otimes \mathcal{O}(D) =
\mathcal{O}(\pi^*(K_{ B})+D) = \mathcal{O}(K_{\widetilde B} - (k{-}1)E+D).
\end{equation}
We will  get Weierstrass line bundle $\mathcal{O}(K_{\widetilde{\mathcal{B}}})$
(and hence trivial canonical
bundle in the total space) if we  choose $D=(k{-}1)E$.

Since the multiplicities of $f$ and $g$ along $\Gamma$ coincide with
the orders of vanishing of $\pi^*(f)$ and $\pi^*(g)$ along $E$,
we see that in codimension two we need $f$ and $g$ to have multiplicities
 $4$ and $6$ along $\Gamma$ in order for a blowup to be possible.
(In that case, we automatically get that $\Delta$ has multiplicity at least $12$).

When the blowup is possible, the total space of the new Weierstrass
model determined from \eqref{eq:Weierstrasschange}
again has trivial canonical bundle.  This means that the original
Weierstrass model had so-called ``canonical singularities'' so that
these points are at finite distance from the interior of the moduli
space \cite{hayakawa,MR1432818}.  This phenomenon in codimension two
has been studied since the early days of F-theory, and leads to
conformal field theories in 6D, including the conformal matter studied
in this paper.  After the blowup, the orders of vanishing of $f$, $g$, and
$\Delta$ along $E$ have been reduced by $4$ $6$, and $12$, respectively,
and this allows us
to determine the type of gauge symmetry along $E$.

Note that in codimension two, if the multiplicities of $(f,g,\Delta)$
are at least $(8,12,24)$ then the blowup will not be canonical:  after
removing one power of $x_k$ (the local equation of $E$) we will still
have orders of vanishing at least $4$ and $6$ and so there will be
singularities which destroy the Calabi--Yau property.

Turning to the case of codimension three, since the canonical divisor
changes by $2E$ under the blowup, we need to compensate with higher
multiplicities.  If $\Gamma$ has codimension three (i.e., a point
on a threefold base, or a curve on a fourfold base), then the condition
for the blowup to have a Calabi--Yau total space is that $(f,g,\Delta)$
have multiplicity at least $(8,12,24)$ but not higher than $(12,16,36)$.
This time, we will use the ``change of Weierstrass model''
\eqref{eq:Weierstrasschange} with $t=x_k^2$ so that we remove $D=2E$
from the canonical bundle.  After the blowup, the orders of vanishing
of $f$, $g$,  and $\Delta$ along $E$ have been reduced by
$8$, $12$,  and $24$, respectively,
and this allow us to determine the type of gauge symmetry along $E$.

\section{4D Conformal Matter Contributions} \label{app:4DCM}
In this Appendix we study the matter building blocks of the
4D generalized quivers, that will be coupled to $\mathcal{N}=1$ vector multiplets.
To do so we need to compute useful quantities by reducing the 6D anomaly polynomial of
conformal matter theories on a Riemann surface, $\Sigma$. First, recall
the anomaly polynomial of a generic 6D conformal matter theory in the case of $Q$
M5-branes probing a $\mathbb{C}^{2}/\Gamma_{ADE}$ singularity with $\Gamma_{ADE}$
a finite subgroup of $SU(2)$ (see reference \cite{Ohmori:2014kda} for details):
\begin{align}
\label{eq:6Dap}I_{8}=  &  \alpha c_{2}(R_{6D})^{2}+ \beta c_{2}(R_{6D})
p_{1}(T) + \gamma p_{1}(T)^{2} + \delta p_{2}(T) + \nu_{L} \frac
{\mathrm{Tr}(F_{L}^{2})^{2}}{16}+\nu_{R} \frac{\mathrm{Tr}(F_{R}^{2})^{2}}%
{16}\\
&  +\xi_{L} \frac{\mathrm{Tr}(F_{L}^{2})}{4} c_{2}(R_{6D}) +\xi_{R}
\frac{\mathrm{Tr}(F_{R}^{2})}{4} c_{2}(R_{6D}) +\kappa_{L} \frac
{\mathrm{Tr}(F_{L}^{2})}{4} p_{1}(T) +\kappa_{R} \frac{\mathrm{Tr}(F_{R}^{2}%
)}{4} p_{1}(T)\nonumber\\
&  + \omega_{L} \frac{\mathrm{tr_{fund}}(F_{L}^{4})}{16} + \omega_{R}
\frac{\mathrm{tr_{fund}}(F_{R}^{4})}{16} + \chi\frac{\mathrm{Tr}(F_{R}%
^{2})\mathrm{Tr}(F_{L}^{2})}{16} ,\nonumber
\end{align}
where $R_{6D}$ is the 6D R-symmetry, $T$ is the tangent bundle, and $F_{L,R}$
are the field strength of the left and right flavor symmetries respectively,
where we assume that $G_{L}=G_{R}=G$ and is dictated by the ADE singularity.
The coefficients in front of the monomials have all been determined in reference \cite{Ohmori:2014kda}:
\begin{subequations}
\label{eq:APGT}%
\begin{align}
&  \alpha=\frac{1}{24}\left(  |\Gamma|^{2}Q^{3}- 2Q(|\Gamma|(r_{G}+1)-1) +
d_{G}-1\right) \\
&  \beta=\frac{1}{48}\left(  Q-Q(|\Gamma|(r_{G}+1)-1)+ d_{G}-1\right) \\
&  \gamma=\frac{1}{5760}\left(  7d_{G}+30 Q-23\right) \\
&  \delta=-\frac{1}{5760}\left(  4d_{G}+120Q - 116 \right) \\
&  \nu_{L}=\nu_{R}=\frac{1}{4}\left(  u_{G}-\frac{2}{Q}\right) \\
&  \xi_{L}=\xi_{R}=\frac{1}{2}\left(  h^{\vee}_{G}-Q|\Gamma| \right) \\
&  \kappa_{L}=\kappa_{R}=\frac{h^{\vee}_{G}}{24}\\
&  \omega_{L}=\omega_{R}=\frac{t_{G}}{3}\\
&  \chi=\frac{1}{Q},
\end{align}
\end{subequations}
and the group theory data are given in table \ref{tab:groupconst}.
\begin{table}[t!]
\centering
\begin{tabular}
[c]{|c||c|c|c|c|c|c|c|c|}\hline
$G$ & $SU(k) $ & $SO(k) $ & $Sp(k)$ & $G_{2} $ & $F_{4} $ & $E_{6}$ &
$E_{7}$ & $E_{8}$\\\hline\hline
$r_{G}$ & $k-1$ & $\lfloor k/2 \rfloor$ & $k$ & 2 & 4 & 6 & 7 & 8\\\hline
$h^{\vee}_{G}$ & $k$ & $k-2$ & $k+1$ & 4 & 9 & 12 & 18 & 30\\\hline
$d_{G}$ & $k^{2}-1$ & $k(k-1)/2$ & $k(2k+1)$ & 14 & 52 & 78 & 133 &
248\\\hline
$d_{\mathrm{fnd}}$ & $k$ & $k$ & $2k$ & 7 & 26 & 27 & 56 & 248\\\hline
$s_{G}$ & $\frac12$ & $1$ & $\frac12$ & 1 & 3 & 3 & 6 & 30\\\hline
$t_{G}$ & $2k$ & $k-8$ & $2k+8$ & 0 & 0 & 0 & 0 & 0\\\hline
$u_{G}$ & $2$ & $4$ & $1$ & $\frac{10}{3} $ & $5 $ & $6 $ & $8$ & $12$\\\hline
$|\Gamma|$ & $k$ & $2k-8$ with even $k>7$ & no ADE & no ADE & no ADE & $24 $ &
$48$ & $120$\\\hline
\end{tabular}
\caption{Group theory constants defined for all $G$. }%
\label{tab:groupconst}
\end{table}

In particular, since we will need to weakly gauge a flavor symmetry (which we choose to be the left one in conformal matter theories) we focus
on the UV beta function contribution for these 4D generalized matter. First, we set our conventions for
anomalies of 4D theories. Then, we consider 4D $\mathcal{N} = 2$
 and $\mathcal{N} = 1$ conformal matter.

\subsection{4D Anomalies}

To set our conventions, let us briefly recall the general expression for the 4D anomaly polynomial.
Indeed, even though we are dealing with a non-Lagrangian theory, it is still possible
to extract some calculable quantities such as the anomalies of the system.
The anomaly polynomial for a 4D theory with a simple non-abelian
flavor symmetry with field strength $F$ will contain the following terms:%
\begin{equation}
I_{6}=k_{RRR}\frac{c_{1}(R)^{3}}{6}-k_{RTT}\frac{p_{1}(T)c_{1}(R)}{24}%
+k_{RFF}c_{1}(R)\frac{\text{Tr}F^{2}}{4} + ....
\end{equation}
Here, we recall that just as in \cite{Ohmori:2014kda}, our normalization of Tr$F^{2}$ is
specified so that an instanton one configuration has Tr$F^{2}/4$ integrate to
one over a compact four-manifold. It is related to the trace over the adjoint
representation as:%
\begin{equation}
\text{Tr}F^{2}=\frac{1}{h_{G}^{\vee}}\text{tr}_{\text{adj}}F^{2}.
\label{normalization}%
\end{equation}
Reading off the coefficients for the various anomalies, we have:%
\begin{equation}
\text{Tr}(R^{3})  = k_{RRR}, \,\,\, \text{Tr}(R)  = k_{RTT}, \,\,\, \text{Tr}(RF^{A}F^{B}) = \frac{k_{RFF}}{2}\delta^{AB},
\end{equation}
where the indices $A$ and $B$ label generators of the non-abelian algebra, and
the \textquotedblleft Tr\textquotedblright\ over the gauge algebra generators
is normalized as in line (\ref{normalization}). From these quantities, we can
also extract various physical combinations, including the conformal anomalies
$a$ and $c$, as well as the contribution to the numerator of the NSVZ\ beta
function from weakly gauging this flavor symmetry:%
\begin{equation}
a=\frac{3}{32}\left(  3k_{RRR}-k_{RTT}\right)  \text{, \ \ }c=\frac{1}%
{32}(9k_{RRR}-5k_{RTT})\text{, \ \ }b_{G}^{\text{matter}}=\frac{3k_{RFF}}{2}.
\label{physquant}%
\end{equation}
Here, $b_{G}^{\text{matter}}$ indicates the contribution to the weakly gauged
flavor symmetry; The net contribution including an $\mathcal{N}=1$ vector
multiplet is:%
\begin{equation}
b_{G} = 3C_{2}(G)-b_{G}^{\text{matter}}.
\end{equation}

Let us now confirm that all quantities have been properly normalized. Recall
the anomaly polynomial for a Weyl fermion of R-charge $q$ in a representation
$\rho$ of the simple non-abelian gauge group:%
\begin{equation}
I_{6}(\text{Weyl,}q,\rho)=\left(  \text{ch}(qR)\right)  \left(  \text{tr}%
_{\rho}e^{iF}\right)  \widehat{A}(T)|_{6}.
\end{equation}
Using the expansions:%
\begin{align}
\text{ch}(qR)  &  =1+qc_{1}(R)+\frac{q^{2}}{2!}c_{1}(R)^{2}+\frac{q^{3}}%
{3!}c_{1}(R)^{3}+...\\
\text{tr}_{\rho}e^{iF}  &  =d_{\rho}-\frac{1}{2}\text{tr}_{\rho}F^{2}-\frac
{i}{3!}\text{tr}_{\rho}F^{3}+...\\
\widehat{A}(T)  &  =1-\frac{1}{24}p_{1}(T)+...,
\end{align}
with $d_{\rho}$ the dimension of the representation, this expands to:%
\begin{equation}
I_{6}(\text{Weyl,}q,\rho)=\frac{d_{\rho}q^{3}}{3!}c_{1}(R)^{3}-\frac{d_{\rho
}q}{24}c_{1}(R)p_{1}(T)-\frac{q}{2}c_{1}(R)\text{tr}_{\rho}F^{2}-\frac{i}%
{3!}\text{tr}_{\rho}F^{3}.
\end{equation}
Converting the trace in the representation $\rho$ to a trace in the adjoint
representation introduces the index of the representation, namely:%
\begin{equation}
\frac{1}{\text{Ind}(\rho)}\text{tr}_{\rho}F^{2}=\frac{1}{\text{Ind}%
(\text{adj)}}\text{tr}_{\text{adj}}F^{2}=\text{Tr}F^{2},
\end{equation}
so the anomaly polynomial takes the form:%
\begin{equation}
I_{6}(\text{Weyl,}q,\rho)=\frac{d_{\rho}q^{3}}{3!}c_{1}(R)^{3}-\frac{d_{\rho
}q}{24}c_{1}(R)p_{1}(T)-2qC_{2}\left(  \rho\right)  c_{1}(R)\frac
{\text{Tr}F^{2}}{4},
\end{equation}
and the corresponding anomalies are, from line (\ref{physquant}) given by:%
\begin{equation}
a=\frac{3}{32}\left(  3d_{\rho}q^{3}-d_{\rho}q\right)  \text{, \ \ }c=\frac
{1}{32}(9d_{\rho}q^{3}-5d_{\rho}q)\text{, \ \ }b_{G}^{\text{matter}}%
=-3q\times\text{Ind}\left(  \rho\right)  .
\end{equation}
Note that to weakly gauge this flavor symmetry in a self-consistent way,
additional matter must be present. This is necessary to cancel off the
tr$_{\rho}F^{3}$ term in the anomaly polynomial. Having performed this
calculation, we can now verify that we indeed obtain a consistent expression
for the physical quantities of line (\ref{physquant}). For example, for a
collection of weakly coupled chiral multiplets in the fundamental
representation of $SU(N)$, the Weyl fermions have $q=-1/3$ and $C_{2}\left(
\rho\right)  =1/2$, so these quantities reduce to:%
\begin{equation}
a=N\times\frac{1}{48}\text{, \ \ }c=N\times\frac{1}{24}\text{, \ \ }%
b_{G}^{\text{matter}}=1/2,
\end{equation}
which is correct.

Note that in the above, we implicitly made reference to the infrared R-symmetry of the 4D theory.
In practice, we tend to start at a UV fixed point and flow to a new infrared fixed point. The infrared R-symmetry
will then be given by a linear combination of all candidate $U(1)$ symmetries. The principle of
a-maximization tells us that the conformal anomaly $a$ is maximized for the true IR R-symmetry \cite{Intriligator:2003jj}.
Provided there are no emergent $U(1)$'s in the infrared, we can then use 't Hooft anomaly matching to extract
this quantity.

\subsection{4D $\mathcal{N}=2$ Conformal Matter}

In this subsection we discuss a few additional details on the structure of 4D $\mathcal{N} = 2$ conformal matter,
namely the special case of 6D conformal matter compactified on a $T^2$ in the absence of fluxes.

From reference \cite{Ohmori:2015pia}, we know that the contribution to the flavor symmetry two point
function of the 4D $\mathcal{N} = 2$ theory descends from the coefficient of the monomial $p_{1}(T)
\mathrm{Tr}(F_{L}^{2})$. In fact, in order to couple the matter with the $\mathcal{N}%
=1$ vector multiplet, we need to gauge one of the two flavor symmetry factors, which
for convenience we pick to be $G_{L}$. The contribution to the
coefficient of the beta function for the gauge group $G$ of $\mathcal{N}=2$ 4D
conformal matter is
\begin{equation}
b^{\mathrm{matter}}_{G}=24\kappa_{L}= h^{\vee}_{G}%
\end{equation}
where $h^{\vee}_{G}$ is the dual Coxeter number, which for ADE groups are
listed in \ref{tab:groupconst}. In principle, we can add an arbitrary
number of conformal matter sectors. However, the theory may be conformal in the IR only for certain
values of $b^{\mathrm{matter}}_{G}$, and the conjectured conformal window is given
by
\begin{equation}
\frac{3}{2} h_{G}^{\vee} \lesssim b^{\mathrm{matter}}_{G} \leq 3 h_{G}^{\vee} ,
\end{equation}
where we assumed the value of the lower bound based on the $SU(N)$ SQCD, and on the fact that $b_L=2h^{\vee}_G$ leads to an
IR fixed point and $b_L=h^{\vee}_G$ gives confinement in the IR, whereas we have strong evidence for the upper bound of the conformal window.
This means that we can add a minimum of two conformal matter sectors and a
maximum of three for any $G$-type conformal matter.

We can also add conformal matter sector compactified on more general Riemann
surfaces with also abelian flavor fluxes, these will be strictly
$\mathcal{N}=1$ 4D conformal matter. The conformal constraints gives more
complicated bounds on the genus and flavor fluxes configurations, and we
study this next.

\subsection{More General 4D Conformal Matter} \label{app:4DCMAP}
In order to study the contribution to the UV beta function coefficients from
$\mathcal{N}=1$ 4D conformal matter,
we need to reduce the anomaly polynomial of the 6D conformal matter theories
\eqref{eq:6Dap}, taking into the fact that there is a background value for the
6D R-symmetry bundle, in accord with the presence of a partial twist on the Riemann surface.
The 4D conformal matter will be labeled by the genus, $g$, of the Riemann surface, $\Sigma$, and a
normalized unit of $U(1)$ flavor flux on $\Sigma$, $l$. The $SU(2)_{R_{6D}}$
bundle gets twisted with the $U(1)$ holonomy of $\Sigma$ as follows
\begin{equation}
R_{6D}=(R_{4D} \otimes K_{\Sigma}^{1/2}) \oplus(R_{4D} \otimes K_{\Sigma
}^{1/2} )^{\vee},
\end{equation}
where $R_{4D}$ is the $U(1)$ R-symmetry bundle of the $4D$ $\mathcal{N}=1$
theory and $K_{\Sigma}$ is the canonical bundle on $\Sigma$. The tangent
bundle on $T=T_{8}$ splits like
\begin{equation}
T_{8}=T_{6} \oplus T\Sigma,
\end{equation}
so that the second Chern class of the 6D R-symmetry, and the Pontryagin
classes of the tangent bundle become
\begin{subequations}
\label{eq:RandTsplit}%
\begin{align}
&  c_{2}(R_{6D})=-\left(  c_{1}(R_{4D})+\frac{1}{2}c_{1}(K_{\Sigma})\right)
^{2}\\
&  p_{1}(T_{8})= p_{1}(T_{6}) + p_{1}(T\Sigma)\\
&  p_{2}(T_{8})=p_{1}(T_{6}) p_{1}(T\Sigma).
\end{align}
\end{subequations}
The infrared R-symmetry, $R_{4D}$, is determined by extremization, where any abelian $U(1)$ flavor symmetry, if present, can mix with the UV R-symmetry. The parameter $\epsilon$ takes into account this mixing and the infrared R-symmetry is given by,
\begin{equation}
R_{4D}=R_{I_3}-\frac{1}{2} t_{K_{\Sigma}}+ \epsilon \, t_{U(1)}
\end{equation}
where $t_{K_{\Sigma}}$ is the generator of the $U(1)$ structure group of $\Sigma$, which is associated to the canonical bundle $K_{\Sigma}$,  $t_{U(1)}$ is the generator of the $U(1)$ flavor symmetry, and $R_{I_3}$ is the Cartan generator of the 6D $SU(2)$ R-symmetry.

We now turn on an abelian flavor flux for the flavor symmetry $G_{R}$ with
corresponding background field strength $F_{R}$, which correspond to turning
on a line bundle $\mathcal{L}_{\Sigma}$ on the Riemann surface. $G_{R}$
decomposes now into $H_{R} \times U(1) \subset G_{R}$, where $F_{H_{R}}$ is
the field strength of the non-abelian subgroup of $G_{R}$, and $F_{U(1)}$ is
the field strength of the leftover $U(1)$ flavor symmetry. A
representation $\rho$ of $G_{R}$ decomposes in the following way
\begin{align} \label{eq:gtdec}
G_{R}  &  \supset H_{R} \times U(1)\\
\rho(G_{R})  &  =\bigoplus_{i} \rho_{i}(H_{R}) \otimes\rho_{i}(U(1))=\rho
_{i}(H_{R})_{q_{i}} ,
\end{align}
where the $q_{i}$ charges label the $U(1)$ representations.  Under $H_{R} \times U(1) \subset G_{R}$ the Chern character splits as follows
\begin{equation} \label{eq:chcldec}
\mathrm{ch}(G_{R})_{\rho}= \sum_{i} \mathrm{ch}(H_{R})_{\rho_{i}}
\wedge\mathrm{ch}(U(1))_{q_{i}}.
\end{equation}
The Chern character expansion reads\footnote{The standard definition of the Chern character is $\mathrm{ch}(F)= {\rm Tr}\left( {\rm exp}\left(\frac{iF}{2\pi }\right) \right)$ and in order to avoid carrying factors of $2 \pi $ around we substitute $\frac{F}{2\pi} \rightarrow F $.}
\begin{equation}
\mathrm{ch}(G)_{\rho}= d_{\rho}+i \mathrm{tr}_{\rho}(F_{G}) -\frac
{\mathrm{tr}_{\rho}(F^{2}_{G})}{2}- i \frac{\mathrm{tr}_{\rho}(F^{3}_{G})}{6}+
\frac{\mathrm{tr}_{\rho}(F^{4}_{G})}{24} + \ldots
\end{equation}
where $d_{\rho}$ is the dimension of the representation, and for our $U(1)$ factor:
\begin{equation}
\mathrm{ch}( {U(1)})_{q}= 1+q c_1({U(1)})+ \frac{q^2}{2}c_1({U(1)})^2+ \frac{q^3}{6}c_1({U(1)})^3+ \frac{q^4}{24}c_1({U(1)})^4+\ldots
\end{equation}
and the $U(1)$ flavor symmetry splits into 4D part, $\mathcal L_{U(1)}$, the flux on $\Sigma$, $\mathcal L_{\Sigma}$,  and mixing with R-symmetry part,
\begin{equation}
c_1(U(1))= c_1(\mathcal{L}_{U(1)})+ c_1(\mathcal L_{\Sigma}) + \epsilon c_1(R_{4D}).
\end{equation}
where we recall that $\epsilon$ is the mixing parameter.

In order to convert the trace in a representation to the normalized trace ${\Tr}$ (i.e. such that the integral of ${\Tr F^2}/4$ is the instanton number) we need to introduce the following
\begin{equation}
\mathrm{tr}_{\rho}(F_{G}^{2})=\mathrm{Ind}(\rho(G)) \mathrm{Tr}(F_{G}^2),
\end{equation}
where $\mathrm{Ind}$ is the index of the representation, and in particular for adjoint and fundamental representations:
\begin{equation}
\mathrm{Ind}(\mathrm{adj}(G))=h^{\vee}_{G},\qquad\mathrm{Ind}(\mathrm{fund}%
(G))=s_{G}.
\end{equation}

The flavor symmetry curvatures in $I_{8}$ generically split,
\begin{subequations}
\label{eq:FRLsplit}%
\begin{align}
\mathrm{Tr}(F_{G_R}^{2})=  &  \mathrm{Tr}(F_{H_{R}}^{2}) - x (c_1(\mathcal{L}_{U(1)})
+ c_{1}(\mathcal{L}_{\Sigma}) + \epsilon c_{1}(R_{4D}) )^{2},\\
\mathrm{tr_{fund}}(F_{G_R}^{4})=  &  \mathrm{tr_{fund}}(F_{H_{R}}^{4})+y_1 (c_1(\mathcal{L}_{U(1)})
+ c_{1}(\mathcal{L}_{\Sigma}) + \epsilon c_{1}(R_{4D})
)^{4}+\\
&  + y_{2} \mathrm{Tr}(F_{H_{R}}^{2}) ( c_1(\mathcal{L}_{U(1)}) + c_{1}%
(\mathcal{L}_{\Sigma}) + \epsilon c_{1}(R_{4D}) )^{2}+\nonumber\\
&  + 6 y_{3} {\rm ch}_3(H_R)_{\mathrm{fund}} ( c_1(\mathcal{L}_{U(1)}) +
c_{1}(\mathcal{L}_{\Sigma}) + \epsilon c_{1}(R_{4D}) ) ,\nonumber
\end{align}
\end{subequations}
where given a general group theory decomposition \eqref{eq:gtdec} and its related Chern character splitting formula \eqref{eq:chcldec}  $x,y_{1},y_{2},y_{3}$ read
\begin{align}
&x= \frac{1}{{\rm Ind}(\rho(G))} \sum_{i} d_i q_i^2\\
&y_1= \sum_{i} d_i q_i^4\\
& y_2 = - 6 \sum_{i} {\rm Ind}(\rho(H_i)) \, q_i^2 \\
& y_3 =  \sum_{i} 4 q_i^3,
\end{align}
where the sum is over the various representations of the branching rule of $G_R \supset H_R \times U(1)$.

We now minimally embed $U(1)$ inside $G_{R}$ considering the following breaking patterns
\begin{itemize}
\item for $SU(k)$ with $H=SU(k-1)$: $\mathbf{k}=\mathbf{(k-1)}_{-1}%
\oplus\mathbf{1}_{-k+1}$,

\item for $SO(2k)$ with $H=SO(2k-2)$: $\mathbf{2k}=\mathbf{(2k-2)}_{0}%
\oplus\mathbf{1}_{2}\oplus\mathbf{1}_{-2}$,

\item for $E_{6}$ with $H=SO(10)$: $\mathbf{78}=\mathbf{45}_{0}\oplus
\mathbf{16}_{3}\oplus\mathbf{\overline{16}}_{-3}\oplus\mathbf{1}_{0}$,

\item for $E_{7}$ with $H=E_6$: $\mathbf{133}=\mathbf{78}_{0}%
\oplus\mathbf{27}_{2}\oplus\overline{\mathbf{27}}_{-2}\oplus\mathbf{1}_{0}$,

\item for $E_{8}$ with $H=E_{7}$: $\mathbf{248}=\mathbf{133}_{0}%
\oplus\mathbf{1}_{2}\oplus\mathbf{1}_{0}\oplus\mathbf{1}_{-2}\oplus
\mathbf{56}_{1}\oplus\mathbf{56}_{-1}$,
\end{itemize}
The explicit values of the constants $x,y_{1}%
,y_{2},y_{3},y_{4}$ are tabulated in table \ref{tab:decconst} for the group decomposition listed above, where we do not
need to specify $y_{1},y_{2},y_{3}$ when $t_{G}=0$, i.e. for $SO(8)$ and
$E_{6,7,8}$, because the terms proportional to $\mathrm{tr_{fund}}(F^{4})$
vanish in the anomaly polynomial.
\begin{table}[t!]
\centering
\begin{tabular}
[c]{|c||c|c|c|c|c|}\hline
$G$ & $SU(k) $ & $SO(2k) $ & $E_{6}$ & $E_{7}$ & $E_{8}$\\\hline\hline
$H$ & $SU(k-1)$ & $SO(2k-2)$ & $SO(10)$ & $E_{6}$ & $E_{7}$\\\hline
$x$ & $2k(k-1)$ & $8$ & 24 & 12 & 4\\\hline
$y_{1}$ & $k(k-1)(k^{2}-3k+3)$ & $32$ & $/$ & $/$ & $/$\\\hline
$y_{2}$ & $-3$ & $0$ & $/$ & $/$ & $/$\\\hline
$y_{3}$ & $-4$ & $0$ & $/$ & $/$ & $/$\\\hline
\end{tabular}
\caption{Curvature splitting constants for minimal embedding of $U(1)$ into
$G$. }%
\label{tab:decconst}%
\end{table}

Having decomposed all the Chern classes and curvatures for the global symmetry of the 6D anomaly polynomial, we are now ready to compute the 4D anomaly polynomial and central charges by
integrating $I_{8}$ on $\Sigma$ by plugging into \eqref{eq:6Dap} the bundle decompositions
\eqref{eq:RandTsplit} and \eqref{eq:FRLsplit}. The 4D anomaly polynomial is then:
\begin{align}
\label{eq:I6}I_{6}=  &  \frac{k_{RRR}}{6} c_{1}(R)^{3}- \frac{k_{R}}{24}c_1(R)%
p_{1}(T) +k_{RH_LH_L}\frac{\mathrm{Tr}(F^{2}_{H_{L}})}{4} c_{1}(R) + k_{RH_RH_R}%
\frac{\mathrm{Tr}(F^{2}_{H_{R}})}{4} c_{1}(R) +\nonumber\\
&  %
+ k_{U(1)H_LH_L} \frac{\mathrm{Tr}(F^{2}_{H_{L}})}{4}
c_1(\mathcal L_{U(1)})+k_{U(1)H_RH_R} \frac{\mathrm{Tr}(F^{2}_{H_{R}})}{4}
c_1(\mathcal L_{U(1)})+\nonumber\\
&  + \frac{k_{U(1)U(1)U(1)}}{6}c_1(\mathcal L_{U(1)})^{3} + \frac{k_{U(1)RR}}{2}c_1(\mathcal L_{U(1)})
c_{1}(R)^{2}+ \frac{k_{U(1)U(1)R}}{2}c_1(\mathcal L_{U(1)})^{2} \, c_{1}(R) \nonumber  \\
&+ \frac{k_{U(1)}}{24}c_1(\mathcal L_{U(1)})p_1(T)+k_{H_L H_L H_L} {\rm ch}_3(H_L)_{\mathrm{fund}}+ k_{H_R H_R H_R} {\rm ch}_3(H_R)_{\mathrm{fund}}
\end{align}
where the coefficients of the anomaly polynomial are all computed with respect to the IR R-symmetry $R_{4D}$, and for convenience we drop the subscript, such that $R=R_{4D}$ and
$T=T_{6}$. The central charges are then given by
\begin{align}
&  a=\frac{9}{32}k_{RRR}- \frac{3}{32}k_{R}\\
&  c=\frac{9}{32}k_{RRR}- \frac{5}{32}k_{R}.
\end{align}
Some of the coefficients of \eqref{eq:I6} will now depend on the mixing parameter $\epsilon$.
$a$ and $c$ then read
\begin{align}
&  a(\epsilon)= \frac{9}{64} \left(  x l \epsilon\left(  6 \xi_{R}-8
\kappa_{R}+3 \nu_{R} x \epsilon^{2}\right)  +3 l y_{1} \omega_{R} \epsilon
^{3}+2(1-g)\left(  -24 \alpha+16 \beta-3\xi_{R} x \epsilon^{2}\right)  \right)
\label{eq:aeps}\\
&  c(\epsilon) = \frac{3}{64} \left(  x l \epsilon\left(  18 \xi_{R}-40
\kappa_{R}+9 \nu_{R} x \epsilon^{2}\right)  +9 l y_{1} \omega_{R} \epsilon
^{3}+2(1-g) \left(  -72 \alpha+80 \beta-9\xi_{R} x \epsilon^{2}\right)
\right)  \label{eq:ceps}%
\end{align}
where
\begin{equation}
\int_{\Sigma} c_{1}(K_{\Sigma}) = 2(g-1),\qquad\int_{\Sigma} c_{1}%
(\mathcal{L}_{\Sigma}) =l,
\end{equation}
with $g$ the genus of $\Sigma$. We now maximize $a(\epsilon)$, so that
a-maximization \cite{Intriligator:2003jj} fixes the value of $\epsilon$
\begin{equation}
\epsilon= \frac{2(1-g) \xi_{R} x -\sqrt{z}}{3 l\left(  \nu_{R} x^{2}+y_{1}
\omega_{R}\right)  }%
\end{equation}
where
\begin{equation} \label{eq:z}
z=x \left(  4(1-g)^{2}\xi_{R}^{2} x-2 l^{2} (3\xi_{R}-4 \kappa_{R}) \left(
\nu_{R} x^{2}+y_{1} \omega_{R} \right)  \right)  ,
\end{equation}
We plug the fixed value of $\epsilon$ into \eqref{eq:aeps} and \eqref{eq:ceps} to get the values
of $a$ and $c$. When we gauge the left flavor symmetry we will need to compute the contribution to the
coefficient of the beta function as follow
\begin{equation}
b^{\mathrm{matter}}_L = \frac{3}{2}k_{RH_LH_L}
\end{equation}
and in terms of the 6D anomaly polynomial coefficients it explicitly reads
\begin{align}
&  b^{\mathrm{matter}}_{L}=\frac{\chi x \left(  \sqrt{z}+2 (g-1) \xi_{R} x\right)  }{ 4\left(
\nu_{R} x^{2}+y_{1}\omega_{R} \right)  }- 3(g-1)\xi_{L}
\end{align}
and the other coefficients in \eqref{eq:I6} are
\label{eq:APcoeff4D}%
\begin{align}
&  k_{H_LH_LH_L}=0,\quad k_{H_RH_RH_R}= 3 \frac{l \omega_{R} y_{3}}{8}, \quad k_{U(1)H_LH_L}%
=-l x \chi,\quad k_{U(1)H_RH_R}=y_{2} \omega_{R}-2 l x \nu_{R},\\
&  k_{U(1)U(1)U(1)}=3\frac{l(x^{2}\nu_{R}+y_{1}\omega_{R})}{2}, \quad k_{U(1)RR}%
=\frac{8lx\kappa_{R}}{3}, \quad k_{U(1)U(1)R}=-\sqrt{z}, \quad
k_{U(1)}=-12 x l \kappa_R\\
&  k_{RH_RH_R}=\frac{(2 \nu_{R} x-y_{2}\omega_{R}) \sqrt{z}-2 (g-1) \xi_{R} \left(
4 \nu_{R} x^{2}+x y_{2} \omega_{R} +6y_{1} \omega_{R}\right)  }{ 6\left(
\nu_{R} x^{2}+y_{1} \omega_{R} \right)  }%
\end{align}

\subsubsection{Absence of $U(1)$ Flux}

In case there is no $U(1)$ flux on $\Sigma$ the mixing parameter is trivial
$\epsilon=0$ and $a$ and $c$ read
\begin{subequations}
\begin{align}
&  a(l=0)= \frac{9}{4} \left(  (1-g)\left(  -3 \alpha+2\beta\right)  \right)
\label{eq:anf}\\
&  c(l=0) = \frac{3}{4} \left(  (1-g) \left(  -9 \alpha+10 \beta\right)
\right)  \label{eq:cnf}%
\end{align}
\end{subequations}
whereas the values of the beta functions and the other anomaly polynomial
coefficients can be computed from the one of the with $U(1)$ flux by setting
$l=0$. In this case we also have to assume that $g\neq0$, otherwise the
anomaly polynomial coefficients must be computed in another way, and not by
reduction of the 6D anomaly polynomial.

\subsubsection{Rank 1 Conformal Matter}
We now restrict to the case where $Q=1$ M5 brane probe an ADE singularity. The
values of $a$ for $ SO(8), E_{6}, E_{7} E_{8}$ are tabulated in
\ref{tab:aQ1CM}. \begin{table}[h]
\centering
\begin{tabular}
[c]{|c||c|}\hline
$G$ & $a(g,l)$\\\hline
$SO(8)$ & $\frac{2 (g-1)^2 \left(\sqrt{(g-1)^2+8 l^2}-g+1\right)+l^2 \left(16 \sqrt{(g-1)^2+8 l^2}+51 g-51\right)}{16 l^2}$\\\hline
$E_{6}$ & $-\frac{9 (g-1)^{3}}{4 l^{2}}+\frac{3 (g-1)^{2} \sqrt{9 (g-1)^{2}+60
l^{2}}}{4 l^{2}}+5 \sqrt{9 (g-1)^{2}+60 l^{2}}+\frac{1209 (g-1)}{16}$\\\hline
$E_{7}$ & $-\frac{125 (g-1)^{3}}{4 l^{2}}+\frac{25 (g-1)^{2} \sqrt{25
(g-1)^{2}+48 l^{2}}}{4 l^{2}}+12 \sqrt{25 (g-1)^{2}+48 l^{2}}+\frac{3225
(g-1)}{8}$\\\hline
$E_{8}$ & $-\frac{3645 (g-1)^{3}}{4 l^{2}}+\frac{405 (g-1)^{2} \sqrt{81
(g-1)^{2}+28 l^{2}}}{4 l^{2}}+35 \sqrt{81 (g-1)^{2}+28 l^{2}}+\frac{49887
(g-1)}{16}$\\\hline
\end{tabular}
\caption{Value of $a$ depending on the genus $g$ and the $U(1)$ flux $l$ on
$\Sigma$. }%
\label{tab:aQ1CM}%
\end{table}

For $SO(2k)$ the value of $a$ generalizes
\begin{footnotesize}
\begin{align}
a(g,l)_{SO(2k)}=  &\frac{1}{48 (5-2 k)^2 l^2}\left(18 (g-1)^2 (k-3)^2 \left(\sqrt{9 (g-1)^2 (k-3)^2+4 (2 k-5) (5 k-14) l^2}-3 (g-1) (k-3)\right)+\right. \nonumber\\
&(2 k-5)l^2\left(9 (g-1) (k-3) (2 k (16 k-89)+251)-112 \sqrt{9 (g-1)^2 (k-3)^2+4 (2 k-5) (5 k-14) l^2}+\right. \nonumber \\
&\left.\left.40 k \sqrt{9 (g-1)^2 (k-3)^2+4 (2 k-5) (5 k-14) l^2}\right)\right)
\end{align}
\end{footnotesize}
The values of $c$ are in table \ref{tab:cQ1CM}. \begin{table}[h]
\centering
\begin{tabular}
[c]{|c||c|}\hline
$G$ & $c(g,l)$\\\hline
$SO(8)$ & $\frac{(g-1)^2 \left(\sqrt{(g-1)^2+8 l^2}-g+1\right)+l^2 \left(10 \sqrt{(g-1)^2+8 l^2}+29 g-29\right)}{8 l^2}
$\\\hline
$E_{6}$ & $-\frac{9 (g-1)^{3}}{4 l^{2}}+\frac{3 (g-1)^{2} \sqrt{9 (g-1)^{2}+60
l^{2}}}{4 l^{2}}+\frac{11}{2} \sqrt{9 (g-1)^{2}+60 l^{2}}+\frac{637 (g-1)}{8}%
$\\\hline
$E_{7}$ & $-\frac{125 (g-1)^{3}}{4 l^{2}}+\frac{25 (g-1)^{2} \sqrt{25
(g-1)^{2}+48 l^{2}}}{4 l^{2}}+\frac{51}{4} \sqrt{25 (g-1)^{2}+48 l^{2}}+415
(g-1)$\\\hline
$E_{8}$ & $-\frac{3645 (g-1)^{3}}{4 l^{2}}+\frac{405 (g-1)^{2} \sqrt{81
(g-1)^{2}+28 l^{2}}}{4 l^{2}}+\frac{145}{4} \sqrt{81 (g-1)^{2}+28 l^{2}}%
+\frac{25269 (g-1)}{8}$\\\hline
\end{tabular}
\caption{Value of $c$ depending on the genus $g$ and the $U(1)$ flux $l$ on
$\Sigma$. }%
\label{tab:cQ1CM}%
\end{table}

The value of $c$ for $SO(2k)$ generalizes the one for $SO(8)$ as follows
\begin{footnotesize}
\begin{align}
c(g,l)_{SO(2k)}=  &\frac{1}{24 (5-2 k)^2 l^2}\left(9 (g-1)^2 (k-3)^2 \left(\sqrt{9 (g-1)^2 (k-3)^2+4 (2 k-5) (5 k-14) l^2}-3 g (k-3)+3 k-9\right)+\right.
\nonumber\\
& (2 k-5)l^2\left( (g-1)(-3339 +3552 k-1263 k^2+150 k^3)-58 \sqrt{9 (g-1)^2 (k-3)^2+4 (2 k-5) (5 k-14) l^2}+\right. \nonumber \\
&  \left.\left.22 k \sqrt{9 (g-1)^2 (k-3)^2+4 (2 k-5) (5 k-14) l^2}\right)\right)
\end{align}
\end{footnotesize}
The values of the contributions to the beta function coefficient $b^{\mathrm{matter}}_{L}$ are given in
table \ref{tab:bLQ1CM} \begin{table}[h]
\centering
\begin{tabular}
[c]{|c||c|}\hline
$G$ & $b^{\mathrm{matter}}_{L}(g,l)$\\\hline
$SO(2k)$ & $\frac{6(g-1) (k-3)^2+ \left(\sqrt{9 (g-1)^2 (k-3)^2+4 (2 k-5) (5 k-14) l^2}\right)}{2 k-5}$\\\hline
$E_{6}$ & $ 3\sqrt{(g-1)^{2}+\frac{20 l^{2}}{3}}+15 g-15$\\\hline
$E_{7}$ & $ \left(  \sqrt{25 (g-1)^{2}+48 l^{2}}+40 g-40\right)
$\\\hline
$E_{8}$ & $\sqrt{81 (g-1)^{2}+28 l^{2}}+126 g-126$\\\hline
\end{tabular}
\caption{Value of $b^{\mathrm{matter}}_{L}$ depending on the genus $g$ and the $U(1)$ flux $l$
on $\Sigma$. }%
\label{tab:bLQ1CM}%
\end{table}

The values of the other coefficients in the anomaly polynomial can
be simply read of from \eqref{eq:APcoeff4D} with the substitution of the 6D
anomaly polynomial coefficients \eqref{eq:APGT} and the group theory data in
table \ref{tab:groupconst}. We display in table \ref{tab:bRQ1CM} the values for $k_{RH_RH_R}$.
\begin{table}[h]
\centering
\begin{tabular}
[c]{|c||c|}\hline
$G$ & $k_{RH_RH_R}$\\\hline
$SO(2k)$ & $ \frac{6 (g-1) (k-3)^2+2 \left(\sqrt{9 (g-1)^2 (k-3)^2+4 (2 k-5) (5 k-14) l^2}\right)}{6 k-15} $\\\hline
$E_{6}$ & $ 4 \sqrt{(g-1)^2+\frac{20 l^2}{3}}+8 g-8$\\\hline
$E_{7}$ & $2 \left(\sqrt{25 (g-1)^2+48 l^2}+10 g-10\right) $\\\hline
$E_{8}$ & $\frac{10}{3} \sqrt{81 (g-1)^2+28 l^2}+60 (g-1)$\\\hline
\end{tabular}
\caption{Value of $k_{RH_RH_R}$ depending on the genus $g$ and the $U(1)$ flux $l$
on $\Sigma$. }%
\label{tab:bRQ1CM}%
\end{table}T

In the case where there is no $U(1)$ flux for the flavor symmetry $G_{R}$ the values
of $a,c,b^{\mathrm{matter}}_{L}$ are tabulated in table \ref{tab:abcnoflux}.
\begin{table}[h!]
\centering
\begin{tabular}
[c]{|c||c|c|c|}\hline
$G$ & $a(g)$ & $c(g)$ & $b^{\mathrm{matter}}_{L}(g)$\\\hline
$SO(2k)$ & $\frac{3}{16} (g-1) (k-3) (16 k-39)$ & $\frac{1}{8} (g-1) (k-3) (25 k-57)$ & $ 3(g-1) (k-3)$\\\hline
$E_{6}$ & $\frac{1569 (g-1)}{16}$ & $\frac{829 (g-1)}{8}$ & $18 (g-1)$\\\hline
$E_{7}$ & $\frac{3945 (g-1)}{8}$ & $\frac{2035 (g-1)}{4}$ & $45 (g-1)$\\\hline
$E_{8}$ & $\frac{57447 (g-1)}{16}$ & $\frac{29139 (g-1)}{8}$ & $135
(g-1)$\\\hline
\end{tabular}
\caption{Values of $a,c,b^{\mathrm{matter}}_{L}$ depending on the genus $g$, with no flavor
flux, $l=0$. }%
\label{tab:abcnoflux}%
\end{table}

As a crosscheck we also compare our values of the central charge and flavor central charge to the one in section 2 of \cite{Kim:2018bpg} for $(SO(2k), SO(2k))$  conformal matter on a $T^2$ with flavor symmetry fluxes. We can check that the our results agree with the one in \cite{Kim:2018bpg} by plugging into our formulas $k=N+3$ and a different normalization for the $U(1)$ flavor flux $l=-2z$, where $z$ is their quantized flavor flux on $\Sigma$ not to be confused with \eqref{eq:z}, (because they use a different branching rule, $\mathbf{4N+12}=\mathbf{(4N+10)}_{0}
\oplus\mathbf{1}_{1}\oplus\mathbf{1}_{-1}$, w.r.t. the one we used here and we introduced above).

For $Q>1$ the values of the central charges and beta functions are more
complicated, especially in the case with $U(1)$ flux, but can be recovered
from the formulas at the beginning of this section by keeping $Q$ general.

\subsubsection{Gauging and Conformal Windows}

Now that we have constructed generalized conformal matter blocks, we can couple
them to a gauge theory sector. To do so we need to gauge the left flavor
symmetry (i.e. the one which has not been broken by a $U(1)$ flux), but in order to have a IR fixed point this
procedure gives a constraint on the values of the UV beta functions coefficient,
$b^{\mathrm{matter}}_{L}$, i.e. a conformal window,
\begin{equation}
\frac{3}{2} h^{\vee}_{G} \lesssim b^{\mathrm{matter}}_{L}(g,l) \leq 3 h^{\vee}_{G}.
\end{equation}
We also require that $a,c > 0$, namely we satisfy various
coarse conditions to have an SCFT. The conformal
windows for each of the above cases are then as follows:

\newpage

\begin{itemize}

\item For $SO(2k)$ there are various complicated solutions, we explicitly
write here the one for $SO(8)$:

\begin{enumerate}
\item $g=0$ and $(4\leq | l |\leq 7)$ ,
\item $g=1$ and $(4\leq |l |\leq6)$ ,
\item $g=2$ and $(3\leq |l| \leq5)$ ,
\item $g=3$ and $(2\leq |l| \leq4)$,
\item $g=4$ and $(-4\leq l \leq4)$ ,
\item $g=5$ and $(-3\leq l \leq3)$,
\item $g=6$ and $(-2\leq l \leq2)$,
\item $g=7$ and $l=0$.
\end{enumerate}
\item For $E_{6}$ we have for

\begin{enumerate}
\item $g=0$ and $(5\leq |l| \leq6)$,

\item $g=1$ and $(3\leq |l| \leq4)$

\item $g=2$ and $(-2\leq l \leq2)$,

\item $g=3$ and $l=0$.
\end{enumerate}

\item For $E_{7}$ we have for

\begin{enumerate}
\item $g=0$ and $(12\leq |l| \leq14)$,

\item $g=1$ and $(2\leq l \leq4)$.
\end{enumerate}

\item For $E_{8}$ we have for

\begin{enumerate}
\item $g=0$ and $(33\leq | l| \leq40)$,

\item $g=1$ and $(9\leq |l| \leq17)$.
\end{enumerate}
\end{itemize}

In principle we can add conformal matter with $Q>1$, so that $Q$ enters and
complicates the bounds.  Alternatively we can couple $M$, $Q=1$ conformal matter theories to a $\mathcal N=1$ vector,
such that the bound generalizes as follows:
\begin{equation}
\frac{3}{2}h_{G}^{\vee} \lesssim \sum_{i=1}^{M}b^{\mathrm{matter}}_{L}(g_{i},l_{i})\leq 3 h_{G}^{\vee}.
\end{equation}
and the bounds will be expressed in terms of the choices of genera $\{g_{i}\}$
and $U(1)$ fluxes for $G_{R}$, $\{l_{i}\}$.

\newpage

\newpage

\bibliographystyle{utphys}
\bibliography{6D4Dglue}

\end{document}